\documentclass[preprintnumbers,nofootinbib,noshowpacs,eqsecnum,prd,superscriptaddress,letterpaper]{revtex4-2}
\pdfoutput=1
\usepackage{graphicx}
\usepackage[fleqn]{amsmath}
\usepackage{amssymb}
\usepackage{url}
\usepackage{array,multirow}
\usepackage{hyperref,color}
\usepackage[normalem]{ulem}
\usepackage{slashed}
\usepackage{rotating}
\usepackage{afterpage}
\usepackage{ifthen}
\usepackage{mciteplus}
\usepackage{color}
\usepackage{adjustbox}

\newcommand{\hc}{\mathrm{h.c.}}

\newcommand{\Sla}[1]{/\!\!\!\!#1}

\def\lsim{\raise0.3ex\hbox{$\;<$\kern-0.75em\raise-1.1ex\hbox{$\sim\;$}}}
\def\gsim{\raise0.3ex\hbox{$\;>$\kern-0.75em\raise-1.1ex\hbox{$\sim\;$}}}

\newcommand{\Dfb}{\mbox{$\raisebox{1mm}{\boldmath ${}^\leftrightarrow$} \hspace{-4mm} D$}}

\begin{document}

\preprint{YITP-SB-2024-04, UWThPh 2024-7}

\title{Dimension-eight Operator Basis for Universal Standard Model Effective Field Theory}
\author{Tyler Corbett}
\email{corbett.t.s@gmail.com}
\affiliation{Faculty of Physics, University of Vienna, Boltzmanngasse 5, A-1090 Wien, Austria}
\author{Jay Desai}
\email{jay.desai@stonybrook.edu}
\affiliation{C.N. Yang Institute for Theoretical Physics,
  Stony Brook University, Stony Brook New York 11794-3849, USA}
\author{O.\ J.\ P.\ \'Eboli}
\email{eboli@if.usp.br}
\affiliation{Instituto de F\'{\i}sica, 
Universidade de S\~ao Paulo, S\~ao Paulo -- S\~ao Paulo 05580-090, Brazil.}
\author{M.~C.~Gonzalez-Garcia,}
\email{maria.gonzalez-garcia@stonybrook.edu}
\affiliation{C.N. Yang Institute for Theoretical Physics,
  Stony Brook University, Stony Brook New York 11794-3849, USA}
\affiliation{Departament de Fis\'{\i}ca Qu\`antica i
  Astrof\'{\i}sica and Institut de Ciencies del Cosmos, Universitat de
  Barcelona, Diagonal 647, E-08028 Barcelona, Spain}
\affiliation{Instituci\'o Catalana de Recerca i Estudis
  Avan\c{c}ats (ICREA), Passeig Lluis Companys 23, 08010 Barcelona,
  Spain.}
\begin{abstract}  
  We present the basis of dimension-eight operators associated with
  universal theories. We first derive a complete list of independent
  dimension-eight operators formed with the Standard Model bosonic
  fields characteristic of such universal new physics
  scenarios. Without imposing C or P symmetries the basis contains 175 operators 
  -- that is, the assumption of Universality reduces the number of 
  independent SMEFT coefficients at dimension eight from 44807 to 175.
  89 of the 175 universal operators are included in the general 
  dimension-eight operator basis in the literature.  The 86 additional
  operators involve higher derivatives of the Standard Model bosonic 
  fields and can be rotated in favor of operators involving fermions
  using the Standard Model equations of motion for the bosonic fields.
  By doing so we obtain the allowed fermionic operators generated in this 
  class of models which we map into the corresponding 86 independent
  combinations of operators in the dimension-eight basis of
  Ref.~\cite{Murphy:2020rsh}.
\end{abstract}

\maketitle

\section{Introduction}
\label{sec:intro}

The Standard Model (SM) based on the
$SU(3)_C \otimes SU(2)_L \otimes U(1)_Y$ gauge symmetry has been
extensively tested at the large hadron collider (LHC) and so far, no
deviation of its predictions~\cite{Pasztor:2713476} or new heavy state
have been observed~\cite{Yuan:2020fyf}.  The natural conclusion is
that there must be a mass gap between the electroweak scale and the
beyond the Standard Model (BSM) physics required to address the
well-known shortcomings of the SM.  In this scenario, precision
measurements of SM processes are an important tool to probe BSM
physics and Effective Field Theory (EFT)~\cite{Weinberg:1978kz,
  Georgi:1985kw, Donoghue:1992dd} has become the standard tool
employed to search for hints of new physics. \medskip

The paradigmatic advantage of EFTs for BSM searches is its
model--independence since they are based exclusively on the low-energy
accessible states and symmetries.  Assuming that the scalar particle
observed in 2012~\cite{ Aad:2012tfa, Chatrchyan:2012xdj} belongs to an
electroweak doublet, the $SU(2)_L \otimes U(1)_Y$ gauge symmetry can
be realized linearly at low energies. The resulting model is the
so-called Standard Model EFT (SMEFT) which can be written as
\begin{eqnarray}
{\cal L}_{\rm eff} = {\cal L}_{\rm SM} + \sum_{j=1} \sum_n \frac{f^{(j)}_n}{\Lambda^j} {\cal O}^{(j)}_n \;\; ,
\label{l:eff}
\end{eqnarray}
where the higher-dimension operators ${\cal O}^{(j)}_n$ involve
gauge--boson, Higgs--boson and/or fermionic fields with Wilson
coefficients $f_n$ and $\Lambda$ is a characteristic scale.\medskip

There is a plethora of analyses of the LHC data in terms of the SMEFT
up to dimension-six; see for instance~\cite{ Corbett:2012dm,
  Corbett:2012ja, Ellis:2014jta, Corbett:2015ksa, Butter:2016cvz,
  Baglio:2017bfe, Aguilar-Saavedra:2018ksv, Ellis:2018gqa,
  daSilvaAlmeida:2018iqo, Brivio:2019ius, Ellis:2020unq,
  Dawson:2020oco, Ethier:2021bye, Almeida:2021asy} and references
therein.  In order to assess the importance of the different
contributions in the $1/\Lambda$ expansion in such analysis, as well
as avoid the appearance of phase space regions where the cross section
is negative~\cite{Baglio:2017bfe}, one is required in many cases to
perform the full calculation at order $1/\Lambda^4$. As is well known
the consistent calculation at order $1/\Lambda^4$ requires the
introduction of the contributions stemming from dimension-eight
operators. \medskip

At this point the advantage of the model-independent approach
mentioned above becomes a limitation due to the large number of Wilson
coefficients.  Already at dimension-six there are 2499 possible
operators when taking flavor into account~\cite{Alonso:2013hga,
  Henning:2015alf}. At dimension-eight the number grows to
44,807~\cite{Murphy:2020rsh,Li:2020gnx}. Clearly such large number of
operators precludes a complete general analysis at any order beyond
$1/\Lambda$ and we are forced to reintroduce some model dependent
hypothesis. In this realm, identifying physically motivated hypothesis
able to capture a large class of BSM theories becomes the new
paradigm. \medskip

One such well-motivated hypothesis is that of {\sl Universality},
which in brief refers to BSM scenarios where the new physics (NP)
dominantly couples to the gauge bosons of the Standard Model. It was
first put forward in the context of the analysis of electroweak
precision data from LEP and low energy experiments, with the
introduction of the oblique parameters $S, T, U$~\cite{Peskin:1990zt,
  Peskin:1991sw} (or
$\epsilon_1, \epsilon_2, \epsilon_3$~\cite{Altarelli:1991fk}) which
captured the dominant NP effects in the observables.  In the context
of the SMEFT, Universality formally refers to BSM models for which the
low-energy effects can be parametrized in terms of operators involving
exclusively the SM bosons, hereon referred to as {\sl bosonic
  operators}~\cite{Wells:2015uba}.  Ultraviolet (UV) completions
that satisfy this
specific definition of universal theories include theories in which
the new states couple only to the bosonic sector, as in composite
Higgs models~\cite{Panico:2015jxa}, as well as models where the SM
fermions are coupled to new states via SM-like
currents~\cite{Barbieri:2004qk, Marzocca:2020jze} { like in type I
  two-Higgs-doublet models~\cite{Dawson:2023ebe}.}

In the  EFT framework not all operators at a given order are
independent as operators related by local changes of variables in
quantum field theories possessing a jacobian determinant equal to one
at the origin exhibit the same $S$--matrix
elements~\cite{Chisholm:1961tha, Kamefuchi:1961sb}. In particular,
operators connected by the use of the classical equations of motion
(EOM) of the SM fields lead to the same $S$--matrix
elements~\cite{Politzer:1980me, Georgi:1991ch, Arzt:1993gz,
  Simma:1993ky}\footnote{{When considering higher orders in the
    $1/\Lambda$ expansion one needs to take care when applying the
    EOM. While they are consistent when at the highest order in the
    expansion considered, at lower orders one needs to include terms
    ``beyond linear order''.  Alternatively, the application of field
    redefinitions is always
    consistent~\cite{Criado:2018sdb,Ellis:2023zim}.}}. In general, a
given SMEFT basis trades some of the bosonic operators for other
bosonic operators and operators involving fermions, hereon called {\sl
  fermionic operators}, in order to keep only independent
operators. Therefore, the action of a rotated operator is equivalent
to a relation between the Wilson coefficients in the basis. These
relations for universal dimension-six operators were obtained in
Ref.~\cite{Wells:2015uba}. \medskip

This work represents the next step in the exploration of the BSM
effects for universal theories by presenting the SMEFT operator basis
and relations implied by the Universality hypothesis at
dimension-eight. As a first step we search for a complete list of
independent dimension-eight operators composed exclusively with SM
bosons before the use of EOM.  A large fraction of these operators
involve higher derivatives of the gauge bosons and/or the Higgs field
and therefore, in the existing dimension-eight
basis~\cite{Murphy:2020rsh,Li:2020gnx}, they have been generically
eliminated in favor of fermionic operators.  Consequently, in
universal theories only a subset of the fermionic operators of the
general dimension-eight operator basis are generated and, furthermore,
their Wilson coefficients are related.  In this work we use, for
concreteness, the basis presented by Murphy in
Ref.~\cite{Murphy:2020rsh} which we refer to as M8B.  Thus, the
program at hand is first to identify a suitable basis of independent
bosonic operators at dimension-eight and then by application of EOM to
identify the combination of fermionic operators of M8B associated with
universal theories. \medskip

{The relevance of constructing the most general EFT
  within a minimal set of assumptions -- such as that of Universality-- is precisely to provide a tool for phenomenological studies as model independent
  as possible within that assumption. 
  On this front, it is important to stress that the universality assumption
  allows us to perform detailed studies at $1/\Lambda^4$ without resorting to
  very simplified hypothesis where just one dimension-eight operator is
  considered, or to specific UV completions.
  For instance, working in the framework of universal
  models, Ref.~\cite{Corbett:2023qtg} studies the impact of
  dimension-eight operators on the experimental analysis of anomalous
  triple gauge couplings by combining the available electroweak
  precision data and electroweak diboson ($W^+ W^-$,$W^\pm Z$,
  $W^\pm \gamma$) productions. It is interesting to notice that the
  inclusion of dimension-eight operators breaks the relation
  $\lambda_\gamma = \lambda_Z$ that holds for the dimension-six
  operators.  Another possible application is the complete
  $1/\Lambda^4$ analysis of Drell-Yan processes~\cite{NosPrep} that
  goes beyond the $S$, $T$, $W$ and $Y$ oblique parameter
  analysis~\cite{Farina:2016rws} with the introduction of further
  contributions to the electroweak gauge boson propagators. }\medskip

{  For the sake of illustration  we also present in
  Section~\ref{sec:uv} a few simple UV completions
  of the SM that give rise only to bosonic operators when heavy states
  are integrated out at tree level.  As expected, once a specific UV model
  is specified, only a subset of the possible dimension-eight universal
  operators is generated, and its  number grows with
  the complexity of the UV completion and its mass spectrum.
  Thus the results in this paper can be generically utilized in two different
  approaches. Firstly, as mentioned above,  it allows to perform a
  $1/\Lambda^4$ complete
  analysis in a totally model agnostic way by considering all universal
  dimension-six and -eight operators which contribute to the process of
  interest. Alternatively, it can be of practical use when working within
  a specific universal UV completion matched to the SMEFT by  integrating out
  the heavy states to obtain the generated bosonic 
  effective operators up to dimension eight. In this case the results
  in appendix~\ref{app:rotbosfer} can be
  used to rotate these generated bosonic operators to M8B without having
  to do each time the exercise of applying the equivalence of operators by
  integration by parts, Fierz identities or equations of motion because
  it has been already taken care of.}\medskip

The work is organized as follows. Sec.~\ref{sec:not} contains
our notation and framework. In Section~\ref{sec:not} we present
our notation and framework. Section~\ref{sec:boson} is dedicated to
presenting our basis of independent dimension-eight universal bosonic
operators while in Sec.~\ref{sec:fstruct} we construct the Lorentz
structures involving fermions associated with the product of SM
currents, which are used in Sec.~\ref{sec:ferm} to obtain the basis of
universal fermionic operators.  { In Section~\ref{sec:uv} we introduce a
few simple bosonic UV completions and the corresponding low-energy operators, while we present our final remarks in
Sec.~\ref{sec:rem}.} The work is complemented with three appendices.
The full explicit expressions of the relations between the bosonic and
fermionic operators for universal theories are presented in
Appendix~\ref{app:rotbosfer}. For convenience we include in
Appendix~\ref{sec:relations} a compilation of the relations more
frequently employed, and we reproduce in Appendix~\ref{app:ferM8B} the
subset of M8B operators which appear in the universal
operators. \medskip

\section{Notation and Framework}
\label{sec:not}

Our conventions are such that the SM lagrangian reads
\begin{eqnarray}
  {\cal L}_{\rm SM} = &&-\frac{1}{4} G^A_{\mu\nu} G^{A\mu\nu}
  -\frac{1}{4} W^a_{\mu\nu} W^{a\mu\nu} -\frac{1}{4} B_{\mu\nu}
                         B^{\mu\nu} + | D_\mu H|^2 + \lambda v^2 |H|^2 - \lambda |H|^4
  \nonumber\\  
                      && + \sum_{f\in \{q, \ell, u, d, e\}} i \bar{f}
                         \Sla{D} f    { -} \left[
                         \left( \widetilde{H}^\dagger \bar{u} y^{u\dagger} q
                         +\bar{q} y^d d H + \bar{\ell} y^e e H + {\rm h.c.}
                         \right)  \right] \;,
\end{eqnarray}
where $G^A_{\mu\nu}, W^a_{\mu\nu}, B_{\mu\nu}$ stand for the field
strength tensors of $SU(3)_c, SU(2)_L, U(1)_Y$ respectively.  We
denoted the quark and lepton doublets by $q$ and $\ell$ while the
$SU(2)_L$ singlets are $u$, $d$ and $e$ and the respective Yukawa
couplings are $y^{u,d,e}$.  We also define
$\widetilde{H}_j = \epsilon_{jk} H^{k\dagger}$ with
$\epsilon_{12} = +1$\footnote{It should be noted, with our conventions
  for $\tilde H_j$ and $\epsilon_{jk}$ that assuming $y^{u\dagger}$ is
  diagonal will result in a wrong sign for the up-quark
  mass. Therefore if one neglects CKM considerations $y^{u\dagger}$
  should be assumed to be proportional to
  $-{\rm diag}(m_u,m_c,m_t)$.}.  The covariant derivative for objects
in the fundamental representation reads
$D_\mu = \partial_\mu - i g_s T^A G^A_\mu - i g \frac{\tau^a}{2}
W^a_\mu - i g^\prime Y B_\mu$ where $Y$ is the hypercharge of the
particle, $T^A$ are the $SU(3)_c$ generators and $\tau^a$ stands for
the Pauli matrices. On the other hand, the covariant derivatives for
the field strengths are
\begin{equation}
  D_\rho B^{\mu\nu} = \partial_\rho B^{\mu\nu} 
  \;\;\;,\;\;\;
   D_\rho W^{a\mu\nu} = \partial_\rho W^{a\mu\nu}  + g
     \epsilon^{abc} W^b_\rho W^{c \mu \nu}  
  \;\;\;,\;\;\;
   D_\rho G^{A\mu\nu} = \partial_\rho G^{A\mu\nu} + g_s
     f^{ABC} G^B_\rho G^{C \mu \nu}  \;,
\end{equation}
where $f^{ABC}$ are the $SU(3)_c$ structure constants. We denote the
$SU(3)_c$ completely symmetric constants by $d^{ABC}$. \medskip

As mentioned above the first step in the program is to obtain the
basis of independent dimension-eight operators consisting only of SM
bosons. In order to do so we first obtained the number of independent
operators belonging to each of the different bosonic classes before
applying the EOM using available packages like {\rm
  BASISGEN}~\cite{Criado:2019ugp}, a modified version of {\rm
  ECO}~\cite{Marinissen:2020jmb} given in Ref.~\cite{Kondo:2022wcw}
and {\rm GrIP}~\cite{Banerjee:2020bym}. Next, we wrote down all
possible operators satisfying the SM gauge symmetry and Lorentz
invariance. In this process, we worked with the irreducible Lorentz
representation of the field strengths
\begin{eqnarray}
  X^{\mu\nu}_{L,R} = \frac{1}{2} \left( X^{\mu\nu} \mp i
    \widetilde{X}^{\mu\nu}  \right) 
  \;\;\;\;\hbox{ with }\;\;\;\;
  \widetilde{X}^{\mu\nu} = \frac{1}{2} \epsilon^{\mu\nu\rho\sigma}
    X_{\rho\sigma} \;,
\end{eqnarray}
where we defined the Levi-Civita totally antisymmetric tensor
$\epsilon_{0123} = - \epsilon^{0123} = + 1$.  The transformation
properties of these fields under the Lorentz group are simple,
$X_L \sim (1,0)$ and $X_R \sim (0,1)$ under $SU(2)_L \otimes SU(2)_R$.
The Bianchi identity reads $ D_\mu \widetilde{X}^{\mu\nu} =0$ implying
that $ D_\mu X^{\mu\nu}_L = D_\mu X^{\mu\nu}_R$ . At this point, we
obtained all possible linear relations between our set of operators
using $SU(3)$ and $SU(2)$ Fierz transformations~\cite{Nishi:2004st,
  Pal:2007dc, Liao:2012uj} summarized in
Appendix~\ref{sec:relations}.\medskip

Further, linear relations between the effective operators in a given
class can be obtained using integration by parts (IBP) for which we
follow a procedure similar to the one described in
Ref.~\cite{Hays:2018zze}.  In brief, given the field content and
number of derivatives in a given class we obtain all operators
invariant under gauge and Lorentz transformations.  To obtain the
relations among them implied by IBP we write all the vector structures
$y^\nu_j$ that contain one less derivative than the operator class
under consideration, then the IBP relations are obtained by setting
$D_\nu y^\nu_j =0$.  At this point, we consider the Fierz and IBP
linear relations and eliminate as many operators as there are
independent relations.  In order to apply the EOM more easily, we then
express the final set of operators in terms of the field strengths
$X^{\mu\nu}$ and their duals.\medskip

As illustration of the above procedure, let us consider the
$D^2 B_L H^4$ operator class that contains eight
members\footnote{Terms like $D_\mu D_\nu B^{\mu\nu}_L$ give rise to
operators in the $X_L B_L H^4$ class and were not considered for
simplicity.} \footnote{Hereon $D^\nu H^\dagger$ stands for
  $(D^\nu H)^\dagger$ for the sake of simplicity.}
:
\begin{eqnarray}
  && x_1 = B_L^{\mu\nu} (D_\mu H^\dagger D_\nu H )(H^\dagger H) \;,
  \\
  && x_2 = B_L^{\mu\nu} (D_\mu H^\dagger H )( H^\dagger D_\nu H) \;,
  \\
  && x_3 = (D_\mu B_L^{\mu\nu}) (D_\nu H^\dagger H)( H^\dagger  H) \;, 
  \\
  && x_4 = (D_\mu B_L^{\mu\nu}) (H^\dagger D_\nu H)( H^\dagger  H) \;,
  \\
  && x_5 = B_L^{\mu\nu} (D_\mu H^\dagger  \tau^I D_\nu H)( H^\dagger \tau^I H) \;,
  \\
  && x_6 = B_L^{\mu\nu} (D_\mu H^\dagger \tau^I  H) (H^\dagger \tau^I D_\nu H) \;,
  \\
  && x_7 = (D_\mu B_L^{\mu\nu}) (D_\nu H^\dagger \tau^I  H)( H^\dagger \tau^I  H) \;, 
  \\
  && x_8 = (D_\mu B_L^{\mu\nu}) (H^\dagger \tau^I  D_\nu H) (H^\dagger \tau^I  H) \;.
\end{eqnarray}
At this stage, we consider operators and their hermitian conjugates as
different structures.  In this example, linear Fierz relations can be
obtained using Eq.~(\ref{eq:fierz2}) leading to
\begin{eqnarray}
  &&  x_5 = 2 x_2 - x_1 \;, 
     \\
  &&  x_6 = 2 x_1 - x_2 \;, 
     \\
  && x_7 =  x_3 \;, 
     \\
  && x_8 = x_4 \;.
\end{eqnarray}
We can see clearly from these relations that we can trade $(x_5, x_6,
x_7, x_8)$ for $(x_1, x_2, x_3, x_4)$. Therefore, we focus on the
latter operator set when obtaining the IBP relations which are derived
from the following vector operators 
\begin{eqnarray}
  && y^\nu_1 =  B_L^{\mu\nu} (D_\mu H^\dagger  H)( H^\dagger H) \;,
  \\
 && y^\nu_2 =  B_L^{\mu\nu} (H^\dagger  D_\mu H) (H^\dagger H) \;,
  \\
&&   y^\nu_3 =  (D_\mu B_L^{\mu\nu})( H^\dagger H)^2 \;.
\end{eqnarray}
The IBP relations are, then, derived from $D_\nu y^\nu_j =0$ and they
read
\begin{eqnarray}
  &&  x_1 + x_2 - x_3 =0 \; ,
     \\
  &&  x_1 + x_2 + x_4 =0 \; ,
  \\
    &&  x_3 + x_4 =0 \; .
\end{eqnarray}
Just two of the last relations are independent, so we have two
independent operators that we can choose to be ${x_1}$ and ${x_3}$ since
this choice renders the rotations of these operators into M8B
straightforward. \medskip

Once the set of independent bosonic operators have been identified
we apply the EOM to those with one or more derivatives acting on the
gauge strength tensors and two or more acting on the Higgs field.
With our conventions the EOM read
\begin{eqnarray}
    D_{\mu}G^{A\mu\nu} &=& -J^{A\nu}_G\;, \nonumber \\
    D_{\mu}W^{I\mu\nu} &=& -\frac{ig}{2}H^\dagger \overleftrightarrow{D}^{I\nu} H -
    J^{I\nu}_W\;, \label{eq:eom1}
 \\
    D_{\mu}B^{\mu\nu} &=& -\frac{ig'}{2}H^\dagger \overleftrightarrow{D}^{\nu} H -J^\nu_B\;, \nonumber \\
    (D^2H^\dagger)^j &=& \lambda v^2 H^{\dagger j} - 2\lambda (H^\dagger H)H^{\dagger j} - J^j_H\;, \nonumber
\end{eqnarray}
where
$H^\dagger \overleftrightarrow{D}^{I\nu} H = H^\dagger \tau^I D^\nu H
- D^\nu H^\dagger \tau^I H $ and we have defined the fermionic
``currents''
  \begin{eqnarray}
    J^{A\mu}_{G} &=& g_s \displaystyle\sum_{f \in \{q,u,d\}}
    \sum_a \overline{f}_a\gamma^\mu T^A f_a\;, \nonumber   \\
    J^{I\mu}_{W} &=& \frac{g}{2} \displaystyle\sum_{f \in \{q,l\}}
    \sum_a \overline{f}_a\gamma^\mu \tau^I f_a\;, \nonumber \\
    J_B^\mu &= &g' \displaystyle\sum_{f \in \{q,l,u,d,e\}} \sum_a Y_f
    \overline{f}_a\gamma^\mu f_a\; ,
\label{eq:fcurrents}  
    \\
    J^j_H &=& {\displaystyle \sum_{ab}}\, \left\{
        y_{ab}^{u\dagger}
    \, (\overline{u}_a\,q_{bk})\, \epsilon^{ jk }\, +
\, y^d_{ab},(
    \,\overline{q}_a^j  
    \, d_b)\, + \,  y^e_{ab}\,(\overline{l}_a^j \,  e_b)\;\right\}\;, \nonumber \\
    {J^\dagger_H}_j &=& {\displaystyle \sum_{ab}}\, \left\{
       y^u_{ab}
    \, (\overline{q}^k_a\,u_b)\, { \epsilon_{kj}}\, +
\,y^{d\dagger}_{ab},(
    \,\overline{d}_a  
    \, q_{bj})\, + \, y^{e\dagger}_{ab}\,(\overline{e}_a \,  l_{bj})\;
    \right\} .\nonumber
\end{eqnarray}
$Y_f$ are the fermionic hypercharges,
$\{Y_q,Y_l,Y_u,Y_d,Y_e\} =
\{\frac{1}{6},-\frac{1}{2},\frac{2}{3},-\frac{1}{3},-1\}$ and
$J^j _H$, does not contain the CKM matrix because the fermion fields
in these equations are in gauge eigenstates (labeled with the latin
indexes $a$, $b$ or $c$) and so are the Yukawa matrices $y^f$. In
addition, we denote the $SU(2)_L$ indices as $ijk$.\medskip

Expressing the fermionic operators generated by products of these
currents and their derivatives in terms of operators in the M8B basis
requires in some cases trivial but lengthy field manipulations which
make use of identities involving the $SU(2)$ and $SU(3)$ generators as
well as Fierz field rearrangements~\cite{Nishi:2004st, Pal:2007dc,
  Liao:2012uj}; see Appendix~\ref{sec:relations} for the more
frequently employed relations. In addition, the simplification also
involves the equations of motion for the fermions which in our
notation read
\begin{eqnarray}
     i\slashed{D}{l_a}_j  &=& {\displaystyle\sum_b}\,  y^e_{ab}\,e_b\,H_j \;,
     \hspace*{2.4cm}
     i\slashed{D}e_a = {\displaystyle\sum_b}\,
     y^{e\dagger}_{ab}\,{l_b}_j \,H^{\dag,j }  \;, 
\nonumber \\
    i\slashed{D}d_a &=& {\displaystyle\sum_b}\, y^{d\dagger}_{ab}\,
    q_{bj} \, H^{\dag,j }\;,
\hspace*{2cm}    i\slashed{D}u_a =  {\displaystyle\sum_b}\, 
   y^{u\dagger}_{ab}\,q_{bj} \,\widetilde{H}^{\dag,j }\;,
   \nonumber  \\
    i\slashed{D}{q_a}_j  &=& {\displaystyle\sum_b}\, \left[\,
     y^d_{ab}\,d_b\,H_j  +  y^u_{ab}\, u_b\,
    \widetilde{H}_j \, \right]\, \;,
  \label{eq:eom2}
\end{eqnarray}
together with the covariant conservation of the gauge currents which
imply that
\begin{eqnarray}
D_\mu J_B^{\mu} =  -i\,\frac{g'}{2}\, D_\mu(H^\dagger
  \overleftrightarrow{D}^{\mu} H)\;,
  &\hspace*{1cm}& 
D_\mu J_W^{I,\mu} = -i\, \frac{g}{2}\,D_\mu(H^\dagger \overleftrightarrow{D}^{I\mu} H)\;,
\end{eqnarray}
and the commutators of the covariant derivatives of the gauge currents
are
\begin{eqnarray}
  {[D_{\alpha},D_{\beta}]}J_B^\mu=0\;,\, \hspace*{1cm}  {[D_{\alpha},D_{\beta}]}J_W^{I\nu}=
\,g\,\epsilon^{IJK} \,W_{\alpha\beta}^J \,J_W^{ K\nu}\;,
\hspace*{1cm} {[D_{\alpha},D_{ \beta}]}J_G^{A\nu}
=\,g_s\,f^{ABC} \,G_{\alpha\beta}^B \,J_G^{ C\nu}\;.
\end{eqnarray}  
%

\section{Independent bosonic operators}
\label{sec:boson}

The building blocks of the operator basis for universal theories are
the Higgs field $H$, the SM field strengths
($X^{\mu\nu}_{L,R} \sim B^{\mu\nu}_{L,R}, W^{a\mu\nu}_{L,R},
G^{A\mu\nu}_{L,R}$) and covariant derivatives $D$.  As mentioned above
we obtain the number of independent operators with this field content
using the packages {\rm BASISGEN} ~\cite{Criado:2019ugp} and {\rm
  ECO}~\cite{Marinissen:2020jmb,Kondo:2022wcw}. Doing so one finds,
prior to the application of the EOM and without imposing $C$ and $P$
symmetries, there are 175 independent bosonic operators at
dimension-eight. Of those, 89 can be chosen to be those included in
M8B, and which, for convenience, we list in
Table~\ref{tab:murbos}. They include all independent operators without
derivatives acting on the gauge strength tensors and with up to one
derivative acting on each Higgs field.  They lead to a rich and
well-known phenomenology.  For example, the operators in the classes
$X^4$, $X^3 X^\prime$ and $X^2X^{\prime\;2}$ generate anomalous
quartic and higher gauge self-couplings that have no triple gauge
vertex associated to them~\cite{Eboli:2006wa, Eboli:2016kko}. The
operator in the $H^8$ class modifies the Higgs self-couplings and the
operatores in the $X^3H^2$ class give rise to multi
$H$~\cite{Binoth:2006ym, Gomez-Ambrosio:2022why, Binoth:2006ym,Botella:2009pq, Kilian:2018bhs} and gauge
boson~\cite{Chivukula:2005ji, Green:2016trm} vertices, {\em e.g.}
anomalous triple gauge couplings~\cite{Degrande:2013kka,Corbett:2023qtg}. Furthermore, the operators in class $X^2 H^4$
class give finite renormalization to the SM input
parameters~\cite{Corbett:2023qtg} and they also generate multi Higgs
and gauge boson vertices~\cite{Corbett:2023yhk,
  Corbett:2021iob}. \medskip

\begin{table}[t!]
  \begin{tabular}[t]{|l|l||l|l||l|l|}
\hline    
  \multicolumn{2}{|c||}{\boldmath $ 1:X^4,\, X^3 X^{\prime}$} &
  \multicolumn{2}{c||}
              {\boldmath  $1:X^2 X^{\prime 2}$}
&
              \multicolumn{2}{c|}{\boldmath$2:H^8$}
\\
\hline\hline
$Q_{G^4}^{(1)}$ & $ (G_{\mu\nu}^A G^{A\mu\nu}) (G_{\rho\sigma}^B G^{B\rho\sigma})$ 
&
$Q_{G^2W^2}^{(1)} $ & $ (W_{\mu\nu}^I W^{I\mu\nu}) (G_{\rho\sigma}^A
G^{A\rho\sigma})$
&
$Q_{H^8} $ & $ (H^\dag H)^4$
  \\\cline{5-6}\cline{5-6}
$Q_{G^4}^{(2)}$ & $ (G_{\mu\nu}^A \widetilde{G}^{A\mu\nu}) (G_{\rho\sigma}^B \widetilde{G}^{B\rho\sigma})$ 
  & $Q_{G^2W^2}^{(2)}$ & $ (W_{\mu\nu}^I \widetilde{W}^{I\mu\nu})
  (G_{\rho\sigma}^A \widetilde{G}^{A\rho\sigma})$
  &
 \multicolumn{2}{c|}{\boldmath$3:H^6D^2$}
 \\\cline{5-6}
 \cline{5-6}
$Q_{G^4}^{(3)} $ & $ (G_{\mu\nu}^A G^{B\mu\nu}) (G_{\rho\sigma}^A
  G^{B\rho\sigma})$
  & $Q_{G^2W^2}^{(3)} $ & $ (W_{\mu\nu}^I G^{A\mu\nu}) (W_{\rho\sigma}^I
  G^{A\rho\sigma})$
&$Q_{H^6}^{(1)}$  & $(H^{\dag} H)^2 (D_{\mu} H^{\dag} D^{\mu} H)$ 
  \\
$Q_{G^4}^{(4)}$ & $ (G_{\mu\nu}^A \widetilde{G}^{B\mu\nu}) (G_{\rho\sigma}^A \widetilde{G}^{B\rho\sigma})$  
   &$Q_{G^2W^2}^{(4)} $ & $ (W_{\mu\nu}^I \widetilde{G}^{A\mu\nu})
  (W_{\rho\sigma}^I \widetilde{G}^{A\rho\sigma})$
& $Q_{H^6}^{(2)}$  & $(H^{\dag} H) (H^{\dag} \tau^I H) (D_{\mu} H^{\dag} \tau^I D^{\mu} H)$
  \\\cline{5-6}\cline{5-6}
$Q_{G^4}^{(5)} $ & $ (G_{\mu\nu}^A G^{A\mu\nu}) (G_{\rho\sigma}^B
  \widetilde{G}^{B\rho\sigma})$
  &$Q_{G^2W^2}^{(5)} $ & $ (W_{\mu\nu}^I \widetilde{W}^{I\mu\nu})
                         (G_{\rho\sigma}^A G^{A\rho\sigma})$
&  \multicolumn{2}{c|}{\boldmath$4:H^4D^4$}
  \\\cline{5-6}\cline{5-6}
$Q_{G^4}^{(6)} $ & $ (G_{\mu\nu}^A G^{B\mu\nu}) (G_{\rho\sigma}^A
  \widetilde{G}^{B\rho\sigma})$
  &$Q_{G^2W^2}^{(6)} $ & $ (W_{\mu\nu}^I W^{I\mu\nu}) (G_{\rho\sigma}^A
    \widetilde{G}^{A\rho\sigma})$
& $Q_{H^4}^{(1)}$  &  $(D_{\mu} H^{\dag} D_{\nu} H) (D^{\nu} H^{\dag} D^{\mu} H)$ 
    \\
$Q_{G^4}^{(7)} $ & $ d^{ABE} d^{CDE} (G_{\mu\nu}^A G^{B\mu\nu})
  (G_{\rho\sigma}^C G^{D\rho\sigma})$
  &$Q_{G^2W^2}^{(7)} $ & $ (W_{\mu\nu}^I G^{A\mu\nu}) (W_{\rho\sigma}^I
    \widetilde{G}^{A\rho\sigma})$
&%
$Q_{H^4}^{(2)}$  &  $(D_{\mu} H^{\dag} D_{\nu} H) (D^{\mu} H^{\dag} D^{\nu} H)$ \\ 
$Q_{G^4}^{(8)} $ & $ d^{ABE} d^{CDE} (G_{\mu\nu}^A
  \widetilde{G}^{B\mu\nu}) (G_{\rho\sigma}^C
  \widetilde{G}^{D\rho\sigma})$ 
&  $Q_{G^2B^2}^{(1)} $ & $ (B_{\mu\nu} B^{\mu\nu}) (G_{\rho\sigma}^A G^{A\rho\sigma})$
&
$Q_{H^4}^{(3)}$  &  $(D^{\mu} H^{\dag} D_{\mu} H) (D^{\nu} H^{\dag} D_{\nu} H)$  
\\\cline{5-6}\cline{5-6}
$Q_{G^4}^{(9)} $ & $ d^{ABE} d^{CDE} (G_{\mu\nu}^A G^{B\mu\nu})
      (G_{\rho\sigma}^C \widetilde{G}^{D\rho\sigma})$ 
&$Q_{G^2B^2}^{(2)} $ & $ (B_{\mu\nu} \widetilde{B}^{\mu\nu})
(G_{\rho\sigma}^A \widetilde{G}^{A\rho\sigma})$
& \multicolumn{2}{c|}{\boldmath$5:X^3H^2$}
\\\cline{5-6}\cline{5-6}
$Q_{W^4}^{(1)} $ & $ (W_{\mu\nu}^I W^{I\mu\nu}) (W_{\rho\sigma}^J W^{J\rho\sigma})$ 
&$Q_{G^2B^2}^{(3)} $ & $ (B_{\mu\nu} G^{A\mu\nu}) (B_{\rho\sigma} G^{A\rho\sigma})$
 & $Q_{G^3H^2}^{(1)} $ & $ f^{ABC} (H^\dag H) G_{\mu}^{A\nu}
    G_{\nu}^{B\rho} G_{\rho}^{C\mu}$    
\\
$Q_{W^4}^{(2)} $ & $ (W_{\mu\nu}^I \widetilde{W}^{I\mu\nu})
  (W_{\rho\sigma}^J \widetilde{W}^{J\rho\sigma})$
&  $Q_{G^2B^2}^{(4)} $ & $ (B_{\mu\nu} \widetilde{G}^{A\mu\nu})
(B_{\rho\sigma} \widetilde{G}^{A\rho\sigma})$
& $Q_{G^3H^2}^{(2)} $ & $ f^{ABC} (H^\dag H) G_{\mu}^{A\nu} G_{\nu}^{B\rho} \widetilde{G}_{\rho}^{C\mu}$   
\\
$Q_{W^4}^{(3)} $ & $ (W_{\mu\nu}^I W^{J\mu\nu}) (W_{\rho\sigma}^I
  W^{J\rho\sigma})$ 
&$Q_{G^2B^2}^{(5)} $ & $ (B_{\mu\nu} \widetilde{B}^{\mu\nu})
    (G_{\rho\sigma}^A G^{A\rho\sigma})$
  &$Q_{W^3H^2}^{(1)} $ & $\epsilon^{IJK} (H^\dag H) W_{\mu}^{I\nu}
    W_{\nu}^{J\rho} W_{\rho}^{K\mu}$
\\
$Q_{W^4}^{(4)} $ & $ (W_{\mu\nu}^I \widetilde{W}^{J\mu\nu})
  (W_{\rho\sigma}^I \widetilde{W}^{J\rho\sigma})$ 
& $Q_{G^2B^2}^{(6)} $ & $ (B_{\mu\nu} B^{\mu\nu}) (G_{\rho\sigma}^A
  \widetilde{G}^{A\rho\sigma})$
& $Q_{W^3H^2}^{(2)} $ & $ \epsilon^{IJK} (H^\dag H) W_{\mu}^{I\nu}
    W_{\nu}^{J\rho} \widetilde{W}_{\rho}^{K\mu}$
  \\
  $Q_{W^4}^{(5)} $ & $ (W_{\mu\nu}^I W^{I\mu\nu}) (W_{\rho\sigma}^J
  \widetilde{W}^{J\rho\sigma})$ 
&$Q_{G^2B^2}^{(7)} $ & $ (B_{\mu\nu} G^{A\mu\nu}) (B_{\rho\sigma}
  \widetilde{G}^{A\rho\sigma})$
&$Q_{W^2BH^2}^{(1)}$ & $ \epsilon^{IJK} (H^\dag \tau^I H) B_{\mu}^{\,\nu}
     W_{\nu}^{J\rho} W_{\rho}^{K\mu}$  
  \\
$Q_{W^4}^{(6)} $ & $ (W_{\mu\nu}^I W^{J\mu\nu}) (W_{\rho\sigma}^I
  \widetilde{W}^{J\rho\sigma})$ 
& $Q_{W^2B^2}^{(1)} $ & $ (B_{\mu\nu} B^{\mu\nu}) (W_{\rho\sigma}^I
  W^{I\rho\sigma})$
&$Q_{W^2BH^2}^{(2)} $ & $ \epsilon^{IJK} (H^\dag \tau^I H)
  (\widetilde{B}^{\mu\nu} W_{\nu\rho}^J W_{\mu}^{K\rho}$
  \\
$Q_{B^4}^{(1)} $ & $ (B_{\mu\nu} B^{\mu\nu}) (B_{\rho\sigma} B^{\rho\sigma})$
&$Q_{W^2B^2}^{(2)} $ & $ (B_{\mu\nu} \widetilde{B}^{\mu\nu})
  (W_{\rho\sigma}^I \widetilde{W}^{I\rho\sigma})$
&  
&  $+ B^{\mu\nu}W_{\nu\rho}^J \widetilde{W}_{\mu}^{K\rho})$
 \\\cline{5-6} \cline{5-6}   
$Q_{B^4}^{(2)} $ & $ (B_{\mu\nu} \widetilde{B}^{\mu\nu}) (B_{\rho\sigma}
  \widetilde{B}^{\rho\sigma})$ 
&$Q_{W^2B^2}^{(3)} $ & $ (B_{\mu\nu} W^{I\mu\nu}) (B_{\rho\sigma}
  W^{I\rho\sigma})$
& \multicolumn{2}{c|}{\boldmath $6:X^2H^4$}  
\\\cline{5-6}\cline{5-6}
$Q_{B^4}^{(3)} $ & $ (B_{\mu\nu} B^{\mu\nu}) (B_{\rho\sigma} \widetilde{B}^{\rho\sigma})$
  &$Q_{W^2B^2}^{(4)} $ & $ (B_{\mu\nu} \widetilde{W}^{I\mu\nu})
    (B_{\rho\sigma} \widetilde{W}^{I\rho\sigma})$
  &$Q_{G^2H^4}^{(1)} $ & $ (H^\dag H)^2 G^A_{\mu\nu} G^{A\mu\nu}$
\\  
$Q_{G^3B}^{(1)} $ & $ d^{ABC} (B_{\mu\nu} G^{A\mu\nu}) (G_{\rho\sigma}^B
  G^{C\rho\sigma})$
& $Q_{W^2B^2}^{(5)} $ & $ (B_{\mu\nu} \widetilde{B}^{\mu\nu})
  (W_{\rho\sigma}^I W^{I\rho\sigma})$
&    $Q_{G^2H^4}^{(2)} $ & $ (H^\dag H)^2 \widetilde G^A_{\mu\nu}
      G^{A\mu\nu}$
\\
$Q_{G^3B}^{(2)} $ & $ d^{ABC} (B_{\mu\nu} \widetilde{G}^{A\mu\nu})
  (G_{\rho\sigma}^B \widetilde{G}^{C\rho\sigma})$  
&$Q_{W^2B^2}^{(6)} $ & $ (B_{\mu\nu} B^{\mu\nu}) (W_{\rho\sigma}^I
  \widetilde{W}^{I\rho\sigma})$
  &$Q_{W^2H^4}^{(1)} $ & $ (H^\dag H)^2 W^I_{\mu\nu} W^{I\mu\nu}$  
  \\
$Q_{G^3B}^{(3)} $ & $ d^{ABC} (B_{\mu\nu} \widetilde{G}^{A\mu\nu})
  (G_{\rho\sigma}^B G^{C\rho\sigma})$  
& $Q_{W^2B^2}^{(7)} $ & $ (B_{\mu\nu} W^{I\mu\nu}) (B_{\rho\sigma}
  \widetilde{W}^{I\rho\sigma})$
  &$Q_{W^2H^4}^{(2)}$ &  $(H^\dag H)^2 \widetilde W^I_{\mu\nu}
      W^{I\mu\nu}$  
  \\
$Q_{G^3B}^{(4)} $ & $ d^{ABC} (B_{\mu\nu} G^{A\mu\nu}) (G_{\rho\sigma}^B
  \widetilde{G}^{C\rho\sigma})$
&&
   &$Q_{W^2H^4}^{(3)}$ & $(H^\dag \tau^I H) (H^\dag \tau^J H)
    W^I_{\mu\nu} W^{J\mu\nu}$  
  \\
\cline{1-4}
\cline{1-4}
\multicolumn{4}{|c||}{\boldmath$7:X^2H^2D^2$} 
  &$Q_{W^2H^4}^{(4)}$ &
$(H^\dag \tau^I H) (H^\dag \tau^J H) \widetilde
    W^I_{\mu\nu} W^{J\mu\nu}$\\  
\cline{1-4}
\cline{1-4}
$Q_{G^2H^2D^2}^{(1)}$  &  $(D^{\mu} H^{\dag} D^{\nu} H) G_{\mu\rho}^A G_{\nu}^{A \rho}$ 
&
$Q_{B^2H^2D^2}^{(1)}$  &  $(D^{\mu} H^{\dag} D^{\nu} H) B_{\mu\rho} B_{\nu}^{\,\,\,\rho}$ 
&$Q_{B^2H^4}^{(1)} $ & $  (H^\dag H)^2 B_{\mu\nu} B^{\mu\nu}$
\\
$Q_{G^2H^2D^2}^{(2)}$  &  $(D^{\mu} H^{\dag} D_{\mu} H) G_{\nu\rho}^A G^{A \nu\rho}$ 
&
$Q_{B^2H^2D^2}^{(2)}$  &  $(D^{\mu} H^{\dag} D_{\mu} H) B_{\nu\rho} B^{\nu\rho}$ 
&$Q_{WBH^4}^{(1)}$ &  $(H^\dag H) (H^\dag \tau^I H) W^I_{\mu\nu}
B^{\mu\nu}$
\\
$Q_{G^2H^2D^2}^{(3)}$  &  $(D^{\mu} H^{\dag} D_{\mu} H) G_{\nu\rho}^A \widetilde{G}^{A \nu\rho}$ &
$Q_{B^2H^2D^2}^{(3)}$  &  $(D^{\mu} H^{\dag} D_{\mu} H) B_{\nu\rho} \widetilde{B}^{\nu\rho}$
&
 $Q_{WBH^4}^{(2)} $ & $ (H^\dag H) (H^\dag \tau^I H) \widetilde
   W^I_{\mu\nu} B^{\mu\nu}$
\\
$Q_{W^2H^2D^2}^{(1)}$  &  $(D^{\mu} H^{\dag} D^{\nu} H) W_{\mu\rho}^I W_{\nu}^{I \rho}$ &
$Q_{WBH^2D^2}^{(1)}$  &  $(D^{\mu} H^{\dag} \tau^I D_{\mu} H) B_{\nu\rho} W^{I \nu\rho}$
&$Q_{B^2H^4}^{(2)}$ &  $(H^\dag H)^2 \widetilde B_{\mu\nu}
     B^{\mu\nu}$\\\cline{5-6}\cline{5-6}
$Q_{W^2H^2D^2}^{(2)}$  &  $(D^{\mu} H^{\dag} D_{\mu} H) W_{\nu\rho}^I W^{I \nu\rho}$ &
$Q_{WBH^2D^2}^{(2)}$  &  $(D^{\mu} H^{\dag} \tau^I D_{\mu} H) B_{\nu\rho} \widetilde{W}^{I \nu\rho}$
&\multicolumn{2}{|c|}{\boldmath$8:XH^4D^2$} \\\cline{5-6}\cline{5-6}
$Q_{W^2H^2D^2}^{(3)}$  &  $(D^{\mu} H^{\dag} D_{\mu} H) W_{\nu\rho}^I \widetilde{W}^{I \nu\rho}$ &
$Q_{WBH^2D^2}^{(3)}$  &  $i (D^{\mu} H^{\dag} \tau^I D^{\nu} H) (B_{\mu\rho} W_{\nu}^{I \rho}$
&$Q_{WH^4D^2}^{(1)}$  & $(H^{\dag} H) (D^{\mu} H^{\dag} \tau^I D^{\nu} H) W_{\mu\nu}^I$
\\
$Q_{W^2H^2D^2}^{(4)}$  &  $i \epsilon^{IJK} (D^{\mu} H^{\dag} \tau^I D^{\nu} H) W_{\mu\rho}^J W_{\nu}^{K \rho}$ &
&\hspace{2cm}  $- B_{\nu\rho} W_{\mu}^{I\rho})$ &
$Q_{WH^4D^2}^{(2)}$  & $(H^{\dag} H) (D^{\mu} H^{\dag} \tau^I D^{\nu} H) \widetilde{W}_{\mu\nu}^I$
\\
$Q_{W^2H^2D^2}^{(5)}$  &  $\epsilon^{IJK} (D^{\mu} H^{\dag} \tau^I D^{\nu} H) (W_{\mu\rho}^J \widetilde{W}_{\nu}^{K \rho}$& 
$Q_{WBH^2D^2}^{(4)}$  &  $(D^{\mu} H^{\dag} \tau^I D^{\nu} H) (B_{\mu\rho} W_{\nu}^{I \rho}$ &
$Q_{WH^4D^2}^{(3)}$  & $\epsilon^{IJK} (H^{\dag} \tau^I H) (D^{\mu} H^{\dag} \tau^J D^{\nu} H) W_{\mu\nu}^K$
\\
&\hspace*{2.5cm} $- \widetilde{W}_{\mu\rho}^J W_{\nu}^{K \rho})$ &
&     \hspace{2cm}    $+ B_{\nu\rho} W_{\mu}^{I\rho})$  &
$Q_{WH^4D^2}^{(4)}$  & $\epsilon^{IJK} (H^{\dag} \tau^I H) (D^{\mu} H^{\dag} \tau^J D^{\nu} H) \widetilde{W}_{\mu\nu}^K$
\\
$Q_{W^2H^2D^2}^{(6)}$  &  $i \epsilon^{IJK} (D^{\mu} H^{\dag} \tau^I D^{\nu} H) (W_{\mu\rho}^J \widetilde{W}_{\nu}^{K \rho}$ & 
$Q_{WBH^2D^2}^{(5)}$  &  $i (D^{\mu} H^{\dag} \tau^I D^{\nu} H) (B_{\mu\rho} \widetilde{W}_\nu^{^I \rho}$ &
$Q_{BH^4D^2}^{(1)}$  & $(H^{\dag} H) (D^{\mu} H^{\dag} D^{\nu} H) B_{\mu\nu}$
\\
& \hspace*{2.5cm}$+ \widetilde{W}_{\mu\rho}^J W_{\nu}^{K \rho})$ &
& \hspace{2cm} $- B_{\nu\rho} \widetilde{W}_\mu^{^I \rho})$ 
&$Q_{BH^4D^2}^{(2)}$  & $(H^{\dag} H) (D^{\mu} H^{\dag} D^{\nu} H) \widetilde{B}_{\mu\nu}$
\\
&&
$Q_{WBH^2D^2}^{(6)}$  &  $(D^{\mu} H^{\dag} \tau^I D^{\nu} H) (B_{\mu\rho} \widetilde{W}_\nu^{^I \rho}$ &
&
\\
&&
& \hspace{2cm}$+ B_{\nu\rho} \widetilde{W}_\mu^{^I \rho})$ &&\\
\hline
\end{tabular}
  \caption{Independent bosonic operators belonging to  M8B.}
\label{tab:murbos}
\end{table}

The first task at hand is, therefore, to identify a suitable set for
the remaining 86 operators following the procedure sketched in the
previous section. Since our final objective is to find the
corresponding combinations of fermionic operators generated after
application of the EOM, we select the 86 operators for which the
transformation can be more directly implemented. With this in mind, we
make the following choice of operators.\medskip

\subsection{Operators with Higgs  fields and two or more derivatives}

Prior to applying the EOM, the classes $H^6 D^2$, $H^4 D^4$ and
$H^2 D^6$ contain 18 independent bosonic operators of which five are
those included in the corresponding classes in
Table~\ref{tab:murbos}. As for the remaining 13 independent bosonic
operators, 2 of them are in the class $H^6D^2$ and we chose them as
\begin{eqnarray}
\begin{array}{lll}
  R^{(1)}_{H^6D^2}=   (D^2 H^\dagger H)( H^\dagger H)( H^\dagger H)\;,
  &\hspace*{1cm} & R^{(2)}_{H^6D^2}=   (H^\dagger D^2H) (H^\dagger H)( H^\dagger H)\; .  
\end{array}
\label{eq:86bos1}
\end{eqnarray}

In addition, there are 10 independent operators in the class $H^4D^4$
selected to be
\begin{eqnarray}
\begin{array}{lll}
R^{(1)}_{H^4D^4}=    (D^2 H^\dagger \tau^I  H) (D^\mu H^\dagger \tau^I D_\mu  H)\;,
&\hspace*{1cm}&
R^{(2)}_{H^4D^4}=    (D^2 H^\dagger D_\mu  H)  ( H^\dagger D^\mu  H) \;, \\[+0.1cm] 
R^{(3)}_{H^4D^4}=     (D_\mu H^\dagger D^2  H) (D^\mu H^\dagger  H)\;,
&\hspace*{1cm}&
R^{(4)}_{H^4D^4}=      (H^\dagger \tau^I D^2  H)  (D_\mu H^\dagger \tau^I  D^\mu  H) \;, \\[+0.1cm]
R^{(5)}_{H^4D^4}=     (D^2 H^\dagger   H) (D_\mu  H^\dagger D^\mu  H ) \;, 
&\hspace*{1cm}&
R^{(6)}_{H^4D^4}=    ( H^\dagger D^2 H)  (D^\mu H^\dagger D_\mu H) \;, \\[+0.1cm]
R^{(7)}_{H^4D^4}=      (D^2 H^\dagger D^2  H)( H^\dagger  H)\;,
&\hspace*{1cm}&
R^{(8)}_{H^4D^4}=    (D^2 H^\dagger  H)(D^2 H^\dagger H) \;, \\[+0.1cm]
R^{(9)}_{H^4D^4}=     (D^2 H^\dagger  H)  ( H^\dagger D^2  H) \;,
&\hspace*{1cm}&
R^{(10)}_{H^4D^4}=     ( H^\dagger D^2  H)  ( H^\dagger D^2  H)\;,
\end{array}
\label{eq:86bos2}
\end{eqnarray}
while there is only one in the class $H^2D^6$
\begin{eqnarray}
  R^{(1)}_{H^2D^6}=  (D^\mu D^2  H^\dagger D^\mu D^2  H)\;.
\label{eq:86bos3}
\end{eqnarray}  

As we will see upon application of EOM they generate combinations of
fermionic operators with two fermions of classes $\psi^2 H^5$ and
$\psi^2 H^3 D^2$, and operators with four fermions in classes
$\psi^4 H^2$ and $\psi^4 D^2$ with chiralities
$(\overline L L)(\overline R R)$, $(\overline L R)(\overline L R)$ and
$(\overline L R)(\overline R L)$, with related Wilson
coefficients. Explicit expressions for the relations can be found in
Eqs.~\eqref{eq:r1}--\eqref{eq:r13} of
Appendix~\ref{app:rotbosfer}. \medskip

\subsection{Operators with gauge field strengths and derivatives}

There are 19 independent operators in classes $X^3 D^2$,
$X^2 X^\prime D^2$ and $X^2 D^4$ none of which is included in
M8B. Four involve three powers of the $W$ field strength tensor and
another four three powers of the $G$ tensor and we selected them to be
\begin{eqnarray}
\begin{array}{lll}  
  R^{(1)}_{W^3D^2}=       {W^I_{\mu\nu} (D_{\alpha}W^{J,\alpha \mu})(D_{\beta}W^{K,\beta \nu})\epsilon^{IJK}}\;, 
 &\hspace*{0.5cm}& 
R^{(1)}_{G^3D^2}  =     \; G^A_{\mu\nu} (D_{\alpha}G^{B,\alpha
  \mu})(D_{\beta}G^{C,\beta \nu}) f^{ABC}
\;,  \\[+0.1cm]
R^{(2)}_{W^3D^2} =    {\widetilde{W}^I_{\mu\nu} (D_{\alpha}W^{J,\alpha \mu})(D_{\beta}W^{K,\beta \nu})\epsilon^{IJK}}
\;,
&\hspace*{0.5cm}& 
R^{(2)}_{G^3D^2}  =  {\widetilde{G}^A_{\mu\nu} (D_{\alpha}G^{B,\alpha
  \mu})(D_{\beta}G^{C,\beta \nu}) f^{ABC}}
  \;,  \\[+0.1cm]
      R^{(3)}_{W^3D^2}  =    {W^I_{\mu \nu} W^{J,\nu}_{\rho} (D^{\mu}D_{\alpha}W^{K,\alpha \rho}) \epsilon^{IJK}}
      \;,
      &\hspace*{0.5cm}&
R^{(3)}_{G^3D^2} =   {G^A_{\mu \nu} G^{B,\nu}_{\rho} (D^{\mu}D_{\alpha}G^{C,\alpha \rho}) f^{ABC}}
  \;,  \\[+0.1cm]      
  R^{(4)}_{W^3D^2} =  {W^I_{\mu \nu} \widetilde{W}^{J,\nu}_{\rho} (D^{\mu}D_{\alpha}W^{K,\alpha \rho} - D^{\rho}D_{\alpha}W^{K,\alpha \mu}) \epsilon^{IJK}} \;,
      &\hspace*{0.5cm}&
R^{(4)}_{G^3D^2} =    {G^A_{\mu \nu} \widetilde{G}^{B,\nu}_{\rho} (D^{\mu}D_{\alpha}G^{C,\alpha \rho} - D^{\rho}D_{\alpha}G^{C,\alpha \mu}) f^{ABC}} \; .
\end{array}
\label{eq:86bos4}
\end{eqnarray}
Eight operators contain two powers of $W^{\mu\nu}$ or $G^{\mu\nu}$
together with $B^{\mu\nu}$ which can be chosen as
\begin{eqnarray}
\begin{array}{lll}  
  R^{(1)}_{BW^2D^2} =      (D^\mu B_{\mu\nu}) W^{I,\nu\rho} (D^\alpha W^{I}_{\rho\alpha}) \; ,
      &\hspace*{0.5cm}&
  R^{(1)}_{BG^2D^2} =     {G^A_{\mu\nu} (D_{\alpha}G^{A,\alpha
  \mu})(D_{\beta}B^{\beta \nu})} \; , \\[+0.1cm]
  R^{(2)}_{BW^2D^2} =      (D^\mu B_{\mu\nu}) \widetilde{W}^{I,\nu\rho} (D^\alpha W^{I}_{\rho\alpha}) \; ,
  &\hspace*{0.5cm}&
 R^{(2)}_{BG^2D^2} = 
   {\widetilde{G}^A_{\mu\nu} (D_{\alpha}G^{A,\alpha
  \mu})(D_{\beta}B^{\beta \nu})} 
  \; , \\[+0.1cm]  
R^{(3)}_{BW^2D^2}  =   {B_{\mu \nu} W^{I,\nu}_{\rho} (D^{\mu}D_{\alpha}W^{I,\alpha \rho} - D^{\rho}D_{\alpha}W^{I,\alpha \mu})}
\; ,
      &\hspace*{0.5cm}&
R^{(3)}_{BG^2D^2} =     {B_{\mu \nu} G^{A,\nu}_{\rho} (D^{\mu}D_{\alpha}G^{A,\alpha \rho} - D^{\rho}D_{\alpha}G^{A,\alpha \mu})}
    \; , \\[+0.1cm]
R^{(4)}_{BW^2D^2} =     {B_{\mu \nu} \widetilde{W}^{I,\nu}_{\rho} (D^{\mu}D_{\alpha}W^{I,\alpha \rho} - D^{\rho}D_{\alpha}W^{I,\alpha \mu})}\; , 
      &\hspace*{0.5cm}&
    R^{(4)}_{BG^2D^2} =   {B_{\mu \nu} \widetilde{G}^{A,\nu}_{\rho} (D^{\mu}D_{\alpha}G^{A,\alpha \rho} - D^{\rho}D_{\alpha}G^{A,\alpha \mu})}\;.
\end{array}
\label{eq:86bos5}
\end{eqnarray}
These operators modify the triple (multi) gauge couplings. Upon
application of the EOM they will lead to combinations of two-fermion
operators in the classes $\psi^2 H^5$, $\psi^2 H^4 D$,
$\psi^2 X H^2 D$, $\psi^2 X^2 H$, and $\psi^2 X^2 D$, and uniquely
generate four-fermion operators in the class $\psi^4 X$ (see
Eqs.~\eqref{eq:r14}--\eqref{eq:r29}). \medskip

Finally, there are three operators in $X^2 D^4$, one per gauge boson,
\begin{eqnarray}
\begin{array}{lllll}  
R^{(1)}_{B^2D^4}=  
 D^\rho D^{\alpha}B_{\alpha\mu} D_\rho D^{\beta}B_{\beta}^{\mu}  \; , & \hspace*{0.1cm} &
R^{(1)}_{W^2D^4} =   D^\rho D^{\alpha}W^I_{\alpha\mu} D_\rho D^{\beta}W_{\beta}^{I,\mu} \; , & \hspace*{0.1cm} &
R^{(1)}_{G^2D^4}  =     D^\alpha D^\mu G^A_{\mu\nu} 
 D_\alpha D^\rho G^{A,\;\nu}_{\rho}   \;.
\end{array}
\label{eq:86bos6}
\end{eqnarray}
They affect the gauge boson propagators and can give rise to
ghosts~\cite{Coleman:1969xz} in addition to anomalous multi gauge
boson vertices. Equations of motion rotate these three operators to
combinations of two-fermion operators in classes $\psi^2 H^5$,
$\psi^2 H^4 D$, $\psi^2 H^2 D^3$, and $\psi^2 X H^2 D$ as well as
four-fermion operators in classes $\psi^4 H^2$ -- with chiralities
$(\overline L L)(\overline R R)$, $(\overline L R)(\overline L R)$ and
$(\overline L R)(\overline R L)$ -- and $\psi^4 D^2$ with chiralities
$(\overline L L)(\overline R R)$, $(\overline L R)(\overline L R)$,
$(\overline L R)(\overline R L)$, and $(\overline R R)(\overline R R)$
which can be found in Eqs.~\eqref{eq:r30}--\eqref{eq:r32}. \medskip

\subsection{Operators with field strengths, Higgs fields and derivatives}

There are 62 independent bosonic operators in the class $X^2 H^2 D^2$
prior the use of EOM.  M8B contains 18 operators in this class; see
Table~\ref{tab:murbos}. There are, therefore, 44 additional
independent bosonic operators in class $X^2 H^2 D^2$ of which 9
contain two powers of the hypercharge field strength tensor and
another 9 contain two powers of the gluon field strength tensor
\begin{eqnarray}
\begin{array}{lll}  
R^{(1)}_{B^2H^2 D^2} = 
B_{\mu\nu} B^{\mu\nu} (D^2H^\dagger H)\; ,
&\hspace*{1cm}&
R^{(1)}_{G^2H^2 D^2} = 
     G^{A}_{\mu\nu} G^{A\mu\nu} (D^2H^\dagger H) \; , \\[+0.1cm]
R^{(2)}_{B^2H^2 D^2} = 
B_{\mu\nu} B^{\mu\nu} (H^\dagger D^2H) \; ,
&\hspace*{1cm}&
R^{(2)}_{G^2H^2 D^2} = 
     G^{A}_{\mu\nu} G^{A\mu\nu} (H^\dagger D^2H) \; , \\[+0.1cm]  
R^{(3)}_{B^2H^2 D^2} = 
B_{\mu\nu} \widetilde{B}^{\mu\nu} (D^2H^\dagger H)\; ,
&\hspace*{1cm}&
R^{(3)}_{G^2H^2 D^2} = 
     G^{A}_{\mu\nu} \widetilde{G}^{A\mu\nu} (D^2H^\dagger H) \; , \\[+0.1cm]
R^{(4)}_{B^2H^2 D^2} = 
B_{\mu\nu} \widetilde{B}^{\mu\nu} (H^\dagger D^2H) \; ,
&\hspace*{1cm}&
R^{(4)}_{G^2H^2 D^2} = 
     G^{A}_{\mu\nu} \widetilde{G}^{A\mu\nu} (H^\dagger D^2H) \; , \\[+0.1cm] 
R^{(5)}_{B^2H^2 D^2} = 
(D^\mu B_{\mu\nu}) B^{\alpha\nu}  (D_\alpha H^\dagger H)\; ,
&\hspace*{1cm}&
R^{(5)}_{G^2H^2 D^2} = 
     (D^\mu G^{A}_{\mu\nu}) G^{A\alpha\nu}  (D_\alpha H^\dagger H) \; , \\[+0.1cm]
R^{(6)}_{B^2H^2 D^2} = 
(D^\mu B_{\mu\nu}) B^{\alpha\nu}  (H^\dagger D_\alpha H) \; ,
&\hspace*{1cm}&
R^{(6)}_{G^2H^2 D^2} = 
     (D^\mu G^{A}_{\mu\nu}) G^{A\alpha\nu}  (H^\dagger D_\alpha H) \; , \\[+0.1cm]
R^{(7)}_{B^2H^2 D^2} = 
(D^\mu B_{\mu\nu}) \widetilde{B}^{\alpha\nu}  (D_\alpha H^\dagger H)\; ,
&\hspace*{1cm}&
R^{(7)}_{G^2H^2 D^2} = 
     (D^\mu G^{A}_{\mu\nu}) \widetilde{G}^{A\alpha\nu}  (D_\alpha H^\dagger H) \; , \\[+0.1cm]
R^{(8)}_{B^2H^2 D^2} = 
(D^\mu B_{\mu\nu}) \widetilde{B}^{\alpha\nu}  (H^\dagger D_\alpha H)\; ,
&\hspace*{1cm}&
R^{(8)}_{G^2H^2 D^2} = 
     (D^\mu G^{A}_{\mu\nu}) \widetilde{G}^{A\alpha\nu}  (H^\dagger D_\alpha H) \; , \\[+0.1cm]
R^{(9)}_{B^2H^2 D^2} = 
(D^\mu B_{\mu\alpha})(D_\nu  B^{\nu\alpha})  (H^\dagger H)\; ,
&\hspace*{1cm}&
R^{(9)}_{G^2H^2 D^2} =     
(D^\mu G^{A}_{\mu\alpha})(D_\nu  G^{A\nu\alpha})  (H^\dagger H)\;,
\end{array}
\label{eq:86bos7}
\end{eqnarray}
while 13 contain two powers of the $W$ field strength tensor and another 13
contain the product of the hypercharge and $W$ field strength tensors
\begin{eqnarray}
\begin{array}{lll}  
R^{(1)}_{W^2H^2 D^2} = 
W^I_{\mu\nu} W^{I,\mu\nu} (D^2H^\dagger H)\; ,&\hspace*{1cm} &
R^{(1)}_{BWH^2 D^2} = 
 B_{\mu\nu} W^{I,\mu\nu} (H^\dagger \tau^I D^2H) \; , \\[+0.1cm]
R^{(2)}_{W^2H^2 D^2} = 
W^I_{\mu\nu} W^{I,\mu\nu} (H^\dagger D^2H) \; ,&\hspace*{1cm} &
R^{(2)}_{BWH^2 D^2} = 
 B_{\mu\nu} W^{I,\mu\nu} (D^2 H^\dagger \tau^I H) \; , \\[+0.1cm]
R^{(3)}_{W^2H^2 D^2} = 
W^I_{\mu\nu} \widetilde{W}^{I,\mu\nu} (D^2H^\dagger H)\; ,&\hspace*{1cm} &
R^{(3)}_{BWH^2 D^2} = 
 B_{\mu\nu} \widetilde{W}^{I,\mu\nu} (H^\dagger \tau^I D^2H) \; , \\[+0.1cm]
R^{(4)}_{W^2H^2 D^2} = 
W^I_{\mu\nu} \widetilde{W}^{I,\mu\nu} (H^\dagger D^2H) \; ,  &\hspace*{1cm} &
R^{(4)}_{BWH^2 D^2} = 
 B_{\mu\nu} \widetilde{W}^{I,\mu\nu} (D^2 H^\dagger \tau^I H) \; , \\[+0.1cm]
R^{(5)}_{W^2H^2 D^2} = 
(D^\mu W^I_{\mu\nu}) W^{I,\alpha\nu}  (D_\alpha H^\dagger H)\; , &\hspace*{1cm} &
R^{(5)}_{BWH^2 D^2} = 
 (D^\mu B_{\mu\alpha}) W^{I,\alpha\nu}  (D_\nu H^\dagger \tau^I H)\; , \\[+0.1cm]
R^{(6)}_{W^2H^2 D^2} = 
(D^\mu W^I_{\mu\nu}) W^{I,\alpha\nu}  (H^\dagger D_\alpha H)\; , &\hspace*{1cm} &
R^{(6)}_{BWH^2 D^2} = 
 (D^\mu B_{\mu\alpha}) W^{I,\alpha\nu}  (H^\dagger \tau^I D_\nu H)\; , \\[+0.1cm]
R^{(7)}_{W^2H^2 D^2} = 
(D^\mu W^I_{\mu\nu}) \widetilde{W}^{I,\alpha\nu}  (D_\alpha H^\dagger H)\; ,
&\hspace*{1cm} &
R^{(7)}_{BWH^2 D^2} = 
 (D^\mu B_{\mu\alpha}) \widetilde{W}^{I,\alpha\nu}  (D_\nu H^\dagger \tau^I H)\; , \\[+0.1cm]
R^{(8)}_{W^2H^2 D^2} = 
(D^\mu W^I_{\mu\nu}) \widetilde{W}^{I,\alpha\nu}  (H^\dagger D_\alpha H)\; , &\hspace*{1cm} &
R^{(8)}_{BWH^2 D^2} = 
 (D^\mu B_{\mu\alpha}) \widetilde{W}^{I,\alpha\nu}  (H^\dagger \tau^I D_\nu H)\; , \\[+0.1cm]
R^{(9)}_{W^2H^2 D^2} = 
(D^\mu W^I_{\mu\alpha}) (D_\nu W^{I,\nu\alpha})  (H^\dagger H)\; , &\hspace*{1cm} &
R^{(9)}_{BWH^2 D^2} = 
 (D^\mu  W^I_{\mu\nu}) B^{\nu\alpha} (D_\alpha H^\dagger\tau^I H)\; , \\[+0.1cm]
R^{(10)}_{W^2H^2 D^2} = 
\epsilon^{IJK} (D^\mu W^I_{\mu\nu}) W^{J,\rho\nu} (D_\rho H^\dagger \tau^K H) \; ,
&\hspace*{1cm} &
R^{(10)}_{BWH^2 D^2} = 
 (D^\mu  W^I_{\mu\nu}) B^{\nu\alpha} (H^\dagger\tau^I D_\alpha H)\; , \\[+0.1cm]
R^{(11)}_{W^2H^2 D^2} = 
 \epsilon^{IJK} (D^\mu W^I_{\mu\nu}) W^{J,\rho\nu} (H^\dagger \tau^K 
 D_\rho H)\; ,&\hspace*{1cm} &
R^{(11)}_{BWH^2 D^2} = 
 (D^\mu  W^I_{\mu\nu}) \widetilde{B}^{\nu\alpha} (D_\alpha H^\dagger \tau^I H)\; , \\[+0.1cm]
 R^{(12)}_{W^2H^2 D^2} = 
 \epsilon^{IJK} (D^\mu W^I_{\mu\nu}) \widetilde{W}^{J,\rho\nu} (D_\rho H^\dagger \tau^K H) \; 
, &\hspace*{1cm} &
 R^{(12)}_{BWH^2 D^2} = 
 (D^\mu  W^I_{\mu\nu}) \widetilde{B}^{\nu\alpha} (H^\dagger\tau^I D_\alpha H)\; , \\[+0.1cm]
R^{(13)}_{W^2H^2 D^2} = 
\epsilon^{IJK} (D^\mu W^I_{\mu\nu}) \widetilde{W}^{J,\rho\nu} (H^\dagger
\tau^K D_\rho H)  \; ,
&\hspace*{1cm} &
R^{(13)}_{BWH^2 D^2} = 
 (D^\mu B_{\mu\alpha})(D_\nu  W^{I,\nu\alpha})  (H^\dagger\tau^I H)\; .\\[+0.2cm]
\end{array}
\label{eq:86bos8}
\end{eqnarray}

Generically, operators in this class modify the gauge couplings of the
Higgs boson and vertices with two scalars and two or more gauge
bosons.  As we will see upon application of EOM they generate
combinations of fermionic operators with two fermions belonging to the
classes $\psi^2 H^5$, $\psi^2 H^4 D$, $\psi^2 X^2 H$, and
$\psi^2 X H^2 D$, and also operators with four fermions in classes
$\psi^4 H^2$ involving chiralities $(\overline L L)(\overline R R)$,
$(\overline L R)(\overline L R)$, $(\overline L R)(\overline R L)$,
and $(\overline R R)(\overline R R)$.  Explicit expressions for the
relations can be found in Eqs.~\eqref{eq:r33}--\eqref{eq:r76} of
Appendix~\ref{app:rotbosfer}. \medskip

Class $X H^4 D^2$ contains 10 independent operators, six
of them in M8B and another four which we chose as
\begin{eqnarray}
\begin{array}{lll}  
R^{(1)}_{BH^4D^2} = 
    (D_{\alpha}B^{\alpha \mu})(H^\dagger \overleftrightarrow{D}_{\mu} H) (H^\dagger H)
  \; ,  &\hspace*{1cm} & 
R^{(1)}_{WH^4D^2} =  (D^\mu W^I_{\mu\nu})
  (H^\dag \overleftrightarrow{D}^{I \nu} H)  (H^\dag H)
  \; , \\[+0.1cm]
R^{(2)}_{WH^4D^2} = 
    \epsilon^{IJK}(H^\dag \tau^I H)(D^{\nu}H^\dag \tau^{J} H) (D^{\mu}W^K_{\mu\nu}) 
  \; , &\hspace*{1cm} &
R^{(3)}_{WH^4D^2} = 
    \epsilon^{IJK} (H^\dag \tau^I H) (H^\dag \tau^{J}D^\nu H) (D^{\mu}W^K_{\mu\nu})  \;.
\\[+0.2cm]
\end{array}
\label{eq:86bos9}
\end{eqnarray}
As seen in Eqs.~\eqref{eq:r77}--\eqref{eq:r80}, these four bosonic
operators are rotated by EOM to combinations of two-fermion operators
in classes $\psi^2 H^5$ and $\psi^2 H^4 D$.  \medskip

Finally, there are six independent operators in class $X H^2 D^4$,
none of which are in M8B, and that we write as
\begin{eqnarray}
\begin{array}{lllll}  
  R^{(1)}_{BH^2D^4} =& 
     { (D^\mu H^\dagger D^2H) (D^\nu B_{\mu\nu} ) }
     \; ,
     &\hspace*{0.2cm} &
R^{(1)}_{WH^2D^4} =& 
     { (D^\mu H^\dagger \tau^I D^2H) (D^\nu W^I_{\mu\nu} ) }
  \; , \\[+0.1cm]     
R^{(2)}_{BH^2D^4} = &
   { (D^2H^\dagger { D^\mu H}) (D^\nu B_{\mu\nu} )  }
   \; ,
   &\hspace*{0.2cm} &
R^{(2)}_{WH^2D^4} =& 
   { (D^2H^\dagger \tau^I D^\mu H) (D^\nu W^I_{\mu\nu} )  }
  \; , \\[+0.1cm]   
R^{(3)}_{BH^2D^4} =& 
   { (D^\mu H^\dagger D^\alpha H - D^\alpha H^\dagger D^\mu H)
    (D_\alpha D^\nu B_{\mu\nu} ) }
   \; ,
&\hspace*{0.2cm} &
R^{(3)}_{WH^2D^4} =& 
   {
    (D^\mu H^\dagger\tau^I D^\alpha H-D^\alpha H^\dagger\tau^I D^\mu H)
    (D_\alpha D^\nu W^I_{\mu\nu} ) }    \;.\\[+0.2cm]
\end{array}
\label{eq:86bos10}
\end{eqnarray}
Application of EOM on these six operators will give two-fermion
operators in classes $\psi^2 H^5$, $\psi^2 H^4 D$ and
$\psi^2 H^3 D^2$, and four-fermion operators in classes $\psi^4 H^2$
and $\psi^4 H D$. \medskip

We finish this section by pointing out that an alternative basis of 86
dimension-eight purely bosonic operators has been presented in
Refs.~\cite{Chala:2021pll,Chala:2021cgt} motivated by the study of
off-shell Green's functions.  The universal basis presented here and
that in these references are related by IBP and Bianchi identities. As
mentioned above the basis of bosonic operators presented in this
section was selected with the aim of allowing for a more direct
implementation of the EOM and a more transparent identification of the
resulting Lorentz structures involving fermions and the corresponding
fermionic operator combinations associated with universal theories, as
we discuss next. \medskip

\section{Products of fermionic currents}
\label{sec:fstruct}

In universal theories, fermionic operators are either generated
involving the SM fermionic currents or originate through the use of
EOM for the bosonic fields on purely bosonic operators.  As such the
only possible fermionic Lorentz structures are those listed in
Eq.~(\ref{eq:fcurrents}).  Consequently, the Wilson coefficients of
the possible fermionic operators in universal theories have well
defined relations. At this point, it is interesting to identify the
possible current combinations which are generated by the application
of the EOM to the bosonic operators listed in
Sec.~\ref{sec:boson}. These combinations contain two and four fermion
fields. \medskip


Most of operators exhibiting two fermionic fields originate from
direct contraction of the gauge and Higgs currents in
Eq.~\eqref{eq:fcurrents} with dimension-five bosonic structures.  In
addition, some two-fermion operators contain derivatives of the
fermionic currents in Eq.~\eqref{eq:fcurrents} contracted with
dimension-four bosonic structures. The generated structures are:
\begin{flalign}  
    ({\rm D}{\Psi}^2_+)_B^{\mu\nu}\equiv& D^\mu J_B^{\nu}+D^\nu J_B^{\mu}=
    g'\,\Big\{ \sum_a
    \sum_{f \in \{q,l,u,d,e\}} Y_f
    \left[D^\mu(\overline{f}_a\gamma^\nu  f_a)+ D^\nu(\overline{f}_a\gamma^\mu  f_a)\right]\Big\}\;,
\label{eq:fd1}
\end{flalign}
\begin{flalign}  
    ({\rm D}{\Psi}^2_+)_W^{K\mu\nu}\equiv& D^\mu J_W^{K\nu}+D^\nu J_W^{K\mu}=
   \frac{g}{2}\,\Big\{ \sum_a
    \sum_{f \in \{q,l\}} 
    \left[D^\mu(\overline{f}_a\gamma^\nu \tau^K f_a)+ D^\nu(\overline{f}_a\gamma^\mu \tau^K f_a)\right]\Big\}\;,
\label{eq:fd2}    
\end{flalign}
\begin{flalign}  
    ({\rm D}{\Psi}^2_-)_W^{K\mu\nu} \equiv& D^\mu J_W^{K\nu}-D^\nu J_W^{K\mu}=
  \frac{g}{2}\,\Big\{\,-\,  i\, \epsilon^{\mu\nu\sigma\alpha}
  \sum_{f \in \{q,l\}} \sum_a
  \overline{f}_a\gamma^\sigma
  \overleftrightarrow{D}^{K\alpha} f_a
   \nonumber
  \\  &
 +\sum_{ab}\left[ y^e_{ab} \,(\bar l_a\, \sigma^{\mu\nu}\, e_b)\, \tau^K \, H \,+\,
     y^d_{ab}\, (\overline{q}_a\, \sigma^{\mu\nu}\,d_b)\, \tau^K \, H \, +\,  
   y^u_{ab}\,(\overline{q}_a\, \sigma^{\mu\nu}\, u_b)\, \tau^K\,
     \widetilde H \, +\, {\rm h.c.} \right]\Big\} \;,
\label{eq:fd3}
\end{flalign}
\begin{flalign}  
  ({\rm D}{\Psi}^2_-)_G^{A\mu\nu}\equiv& D^\mu J_G^{A\nu}-D^\nu J_G^{A\mu}=
  g_s\,\Big\{ \, i\, \epsilon^{\mu\nu\sigma\alpha}\,\sum_a
  \left[\
    \overline{u}_a\gamma^\sigma \overleftrightarrow{D}^\alpha  T^A u_a
   + \overline{d}_a\gamma^\sigma \overleftrightarrow{D}^\alpha  T^A d_a
   - \overline{q}_a\gamma^\sigma \overleftrightarrow{D}^\alpha  T^A q_a\right]\nonumber
  \\
  & +2\,\sum_{ab}\left[
     y^d_{ab}\, (\overline{q}_a\, \sigma^{\mu\nu}\, T^A\, d_b)\,  H \, +\,  
     y^u_{ab}\, (\overline{q}_a\, \sigma^{\mu\nu}\, T^A\, u_b)\, \widetilde H\,
     +\, {\rm h.c.} \right]\Big\}\;.
 \label{eq:fd4} 
\end{flalign}
In order to facilitate the comparison with M8B we have transformed the
last two equations using the relations in the
Appendix~\ref{sec:relations}. In principle the same procedure could
have been applied to the first two relations, however, we kept the
form used in M8B. \medskip

Conversely, most operators containing four fermion fields originate
from the product of two currents in Eq.~\eqref{eq:fcurrents}
contracted with a field strength tensor, two Higgs fields or the
derivative of a Higgs field. The operator rotations to M8B require the
knowledge of sixteen current products. There are three structures
coming from the product of two scalar $J_H$'s:
\begin{flalign}  
  ({\Psi}^4)^{jk}_{HH}\equiv&  J_H^j\,J_H^k\,
    = \sum_{a,b,c,d}\Big\{ y^{u\dagger}_{ab}\,y^{u\dagger}_{cd}
(\overline{u}_a\, q_{bn})\, \epsilon^{ jn}\,
(\overline{u}_c\, q_{dm})\, \epsilon^{ km}\,
+\,   y^d_{ab}\,  y^d_{cd}
  \,(\overline{q}_a^j  \, d_b)\,
  (\overline{q}_c^k \, d_d)\,
\nonumber\\ &
  +\,
     y^e_{ab}\,  y^e_{cd} \,(\overline{l}_a^j \, e_b)\,\,(\overline{l}_c^k\, e_d)\,     
    +\,\Big[
       y^e_{ab}\, y^{u\dagger}_{cd} \,(\overline{l}_a^j \, e_b)\,
(\overline{u}_c\, q_{dm})\, \epsilon^{ km }\,+\;
 y^d_{ab}\, y^{u\dagger}_{cd} \,(\overline{q}_a^j \, d_b)\,
(\overline{u}_c\, q_{dm})\, \epsilon^{ km }\,\nonumber \\&+
       y^e_{ab}\,  y^d_{cd} \,(\overline{l}_a^j \, e_b)\,
      \,(\overline {q}_c^k\, d_d) \;+\; j\leftrightarrow k \;\Big]\,
    \Big\} \;,
\label{eq:4fhh1}    
\end{flalign}
\begin{flalign}      
        ({\Psi}^4)_{HH}
        \equiv&  J_H^{j}\,{J^\dagger_H}_j\,
        = \sum_{a,b,c,d}\Big\{
 - \sum_{f\in\{ u,d\} }             y^{f\dagger}_{ab}\,  y^f_{cd}\,
\left[\frac{1}{6}\,
  (\overline{f}_a\,\gamma^\mu\, {f_d}) \,
  (\overline{q}_c\,\gamma^\mu\, {q_b}) \,+\,
   (\overline{f}_a\,\gamma^\mu\, T^A\, {f_d}) \,
  (\overline{q}_c\,\gamma^\mu\, T^A\,{q_b}) \right] 
                 \nonumber
  \\
                &         \,-\,
        \frac{1}{2}\,y^{e\dagger}_{ab}\,  y^e_{cd}\,
  (\overline{e}_a\,\gamma^\mu\, {e_d}) \,
(\overline{l}_c\,\gamma^\mu\, {l_b})
        \;+\;\Big[  \,  y^e_{ab}\,  y^u_{cd}
  \,(\overline{l}_a^j \, e_b)\,
    { \epsilon_{kj}}\,(\overline{q}^k_c\,u_d)
                   \nonumber
  \\
& 
\,{ +}\,
 y^u_{ab}\,  y^d_{cd} \,(\overline{q}_a^j \, u_b)\,
\epsilon_{jk}\,(\overline{q}^k_c\,d_d)
\,+\,
 y^e_{ab}\, y^{d\dagger}_{cd} \,(\overline{l}_a\, e_b)\,
\,(\overline{d}_c\,q_d)\, +\, {\rm h.c.} \Big] 
        \Big\}\;,
\label{eq:4fhh2}            
\end{flalign}
\begin{flalign}  
{[({\Psi}^4)_{HH}]}^j _{k}\equiv&  J_H^j\,{J^\dagger_H}_k\,
                             \,=\,\,\frac{1}{2}\,\delta^j_k \, ({\Psi}^4)_{HH}
                             \nonumber
  \\
 & { +}\,\,\frac{1}{2}\,(\tau^I)^j_k\, 
\sum_{a,b,c,d}\Big\{
{                   y^{u\dagger}_{ab}\,  y^u_{cd}\,
\left[\frac{1}{6}\,
  (\overline{u}_a\,\gamma^\mu\, {u_d}) \,
  (\overline{q}_c\,\gamma^\mu\, \tau^I\,{q_b}) \,+\,
   (\overline{u}_a\,\gamma^\mu\, T^A\, {u_d}) \,
  (\overline{q}_c\,\gamma^\mu\, \tau^I\,T^A\,{q_b}) \right] }
    \nonumber
  \\ &
        {  -      y^{d\dagger}_{ab}\,  y^d_{cd}\,
 \left[\frac{1}{6}\,
   (\overline{d}_a\,\gamma^\mu\, {d_d}) \,
   (\overline{q}_c\,\gamma^\mu\, \tau^I\,{q_b}) \,+\,
    (\overline{d}_a\,\gamma^\mu\, T^A\, {d_d}) \,
   (\overline{q}_c\,\gamma^\mu\, \tau^I\,T^A\,{q_b}) \right] }
     \nonumber
  \\ &        \,-\,
        \frac{1}{2}\,y^{e\dagger}_{ab}\,  y^e_{cd}\,
  (\overline{e}_a\,\gamma^\mu\, {e_d}) \,
(\overline{l}_c\,\gamma^\mu\, \tau^I\,{l_b})
\;+\;\Big[ { -
   y^e_{ab}\,  y^u_{cd}
  \,(\overline{l}_a^m \, e_b)\,
  (\tau^I\epsilon)_{mn}\,(\overline{q}^n_c\,u_d)}
\nonumber \\
&  
\,  {  -\,
 y^u_{ab}\,  y^d_{cd} \, (\overline{q}^m_c\,d_d) \,
(\tau^I\epsilon)_{mn}\, (\overline{q}_a^n \, u_b) }
\,+\,
 y^e_{ab}\, y^{d\dagger}_{cd} \,(\overline{l}_a\, e_b)\,\tau^I
\,(\overline{d}_c\,q_d)\, +\, {\rm h.c.} \Big] 
\Big\} \;,
\label{eq:4fhh3}    
\end{flalign}  
where, in writing the right-hand side of the above equations, we have
made again use of the relations listed in Appendix~\ref{sec:relations}
to express the fermion currents in the combinations appearing in
M8B. \medskip

The product of two gauge currents $J^\mu_{B,W,G}$ gives rise to
Lorentz scalar and tensor structures. The tensor ones related to
bosonic operators are
\begin{flalign}      
  ({\Psi}^4)^{I\mu\nu}_{WW}\equiv& \epsilon^{IJK} \, J_W^{J\mu}\,J_W^{K\nu}
  =\frac{g^2}{4}\sum_{a,b}\Big\{2\,\epsilon^{IJK}
    (\overline{q}_a\gamma^\mu\tau^J q_a)
      (\bar l_b\gamma^\nu\tau^K l_b)\,
     \nonumber 
   \\
  &{ +}\,\epsilon^{\mu\nu\rho\sigma}
  \left[
    (\bar l_a\gamma_\rho\tau^I l_b)(\bar l_b\gamma_\sigma l_a)
    +\frac{1}{3}
    (\overline{q}_a\gamma_\rho\tau^I q_b)(\overline{q}_b\gamma_\sigma q_a)
    +2 (\overline{q}_a\gamma_\rho\tau^I T^Aq_b)(\overline{q}_b\gamma_\sigma T^Aq_a)
    \right]
  \Big\}\;,\label{eq:4ft1}
\end{flalign}
\begin{flalign}      
  ({\Psi}^4)^{A\mu\nu}_{GG}\equiv& f^{ABC}\, J_G^{B\mu}\,J_G^{C\nu}
  =g_s^2\sum_{a,b}\Big\{ f^{ABC} \sum_{f \in \{q,u,d\}}
  \sum_{f' \in \{q,u,d\},f'\neq f}
  (\overline{f}_a\gamma^\mu T^B f_a)(\overline{f}'_b\gamma^\nu T^C f'_b)
      \nonumber
  \\
  & { -} \frac{1}{2}\,\epsilon^{\mu\nu\rho\sigma}
  \Big[
    (\overline{u}_a\gamma_\rho T^A u_b)(\overline{u}_b\gamma_\sigma u_a)
    + (\overline{d}_a\gamma_\rho T^A d_b)(\overline{d}_b\gamma_\sigma d_a)
    \nonumber\\
&    -\frac{1}{2} (\overline{q}_a\gamma_\rho T^A q_b)(\overline{q}_b\gamma_\sigma q_a)
    -\frac{1}{2} (\overline{q}_a\gamma_\rho\tau^I T^A q_b)
    (\overline{q}_b\gamma_\sigma\tau^I\, q_a)
    \Big]
  \Big\}\;,\label{eq:4ft2}
\end{flalign}
\begin{flalign}        
  ({\Psi}^4)^{A\mu\nu}_{GB}\equiv &J_G^{A\mu}\,J_B^\nu
  =g_s\, g' \sum_{a,b}\sum_{f \in \{q,u,d\}}
  \sum_{f' \in \{q,l,u,d,e\}} 
  Y_{f'} (\overline{f}_a\gamma^\mu T^A f_a)(\overline{f}'_b\gamma^\nu f'_b)\;,
\end{flalign}
\begin{flalign}        
  ({\Psi}^4)^{I\mu\nu}_{WB}\equiv& J_W^{I\mu}\,J_B^\nu
  =\frac{g}2\, g' \sum_{a,b}\sum_{f \in \{q,l\}}
  \sum_{f' \in \{q,l,u,d,e\}} 
  Y_{f'} (\overline{f}_a\gamma^\mu \tau^I f_a)(\overline{f}'_b\gamma^\nu f'_b)\;,
\label{eq:4ft3}
\end{flalign}
where we have Fierz transformed the first two equations above for
later convenience.  On the other hand, the generated Lorentz scalar
structures are

\begin{flalign}        
({\Psi}^4)_{BB}\equiv& J_B^{\mu}\,J_{B\mu}  
  ={g'}^2 \sum_{a,b} \sum_{f,f' \in \{q,l,u,d,e\}} Y_f Y_{f'}
  (\overline{f}_a\gamma^\mu  f_a)(\overline{f}'_b\gamma_\mu f'_b) \;,
\label{eq:4fs1}  
\end{flalign}
\begin{flalign}          
  ({\Psi}^4)_{WW}\equiv& \sum_I J_W^{I\mu}\,J_{W\mu}^I\, 
  =\frac{g^2}{4}\sum_{a,b}\Big\{
(\overline{q}_a\gamma^\mu\tau^I q_a)
  (\overline{q}_b\gamma_\mu\tau^I q_b)\,+ 2\,
  (\overline{q}_a\gamma^\mu\tau^I q_a)
  (\bar l_b\gamma_\mu\tau^I l_b)\,
  \nonumber \\ &\,
    +2  (\bar l_a\gamma^\mu l_b)(\bar l_b\gamma_\mu l_a)
-  (\bar l_a\gamma^\mu l_a)(\bar l_b\gamma_\mu l_{ b})
\Big\}\;,
\label{eq:4fs2}  
\end{flalign}
\begin{flalign}        
  ({\Psi}^4)_{GG}\equiv &\sum_A J_G^{A\mu}\,J_{G\mu}^{A}\, 
  =g_s^2\sum_{a,b} \Big\{
  \sum_{f \in \{q,u,d\}} 
  \sum_{f' \in \{q,u,d\},f'\neq f}
  (\overline{f}_a\gamma^\mu T^A f_a)(\overline{f}'_b\gamma_\mu T^A f'_b)\nonumber\\
  &
  + \frac{1}{2}
 \sum_{f \in \{u,d\}}\, 
    (\overline{f}_a\gamma^\mu  f_b)(\overline{f}_b\gamma_\mu  f_a)
    - \frac{1}{6}
\sum_{f \in \{q,u,d\}}\, 
(\overline{f}_a\gamma^\mu  f_a)(\overline{f}_b\gamma_\mu  f_b)\nonumber\\
&  \mathbf{ +}  \frac{1}{4} (\overline{q}_a\gamma^\mu  q_b)(\overline{q}_b\gamma_\mu  q_a)
+    \frac{1}{4} (\overline{q}_a\gamma^\mu\,\tau^I  q_b)(\overline{q}_b\gamma_\mu\,\tau^I  q_a)  
\Big\}\;,
\label{eq:4fs3}  
\end{flalign}
\begin{flalign}        
 ({\Psi}^4)^I_{WB}\equiv& J_W^{I\mu}\,J_{B\mu}\, 
 =\frac{gg'}{2}\sum_{a,b} \sum_{f' \in \{q,l,u,d,e\}} \sum_{f \in
   \{q,l\} } Y_{ f^\prime}
 (\overline{f}_a\gamma^\mu \tau^I f_a)(\overline{f}'_b\gamma_\mu f'_b)
 \;.
\label{eq:4fs4}  
\end{flalign}

There are only two products of the scalar current $J_H$ with a gauge
current that are generated
\begin{flalign}        
  ({\Psi}^4)_{BH}^{\mu j}\equiv& J_B^\mu\,J_H^j
  =\,g'\,  \sum_{a,b,c} \sum_{f \in \{q,l,u,d,e\}}\times\nonumber\\ 
&\Big\{ Y_f\, y^{u\dagger}_{ab}\,
  (\overline{f}_c\,\gamma^\mu\, f_c)
  \, (\overline{u}_a\,q_{bk})\,
    \epsilon^{ j k  }\, + \,Y_f\,
    \, y^d_{ab}\, (\overline{f}_c\,\gamma^\mu\, f_c)
    ( \,\overline{q}_a^j  
    \, d_b)\, + \, Y_f\,  y^e_{ab}\,
(\overline{f}_c\,\gamma^\mu\, f_c)
    \,(\overline{l}_a^j \,  e_b)\Big\}\; ,
\label{eq:4fhb}
\end{flalign}
\begin{flalign}        
  ({\Psi}^4)_{WH}^{I\mu j}\equiv& J_W^{I\mu}\,J_H^j
  =\,\frac{g}{2}\,  \sum_{a,b,c} \sum_{f \in \{q,l\}}\times\nonumber\\ 
&\Big\{ y^{u\dagger}_{ab}\,
  (\overline{f}_c\,\gamma^\mu\, \tau^I\,f_c)
  \, (\overline{u}_a\,q_{bk})\,
    \epsilon^{j k}\, +
    \, y^d_{ab}\, (\overline{f}_c\,\gamma^\mu\,\tau^I\, f_c)
    ( \,\overline{q}_a^j  
    \, d_b)\, +   y^e_{ab}\,
(\overline{f}_c\,\gamma^\mu\,\tau^I f_c)
    \,(\overline{l}_a^j \,  e_b)\Big\}\;
    .
\label{eq:4fhw}    
\end{flalign}

Finally, some operators with four fermions come from direct contraction of
derivatives of two currents. They are:
\begin{flalign}        
  ({\rm D}{\Psi}^4)_{BB} \equiv& D^\alpha  J_B^\mu D_\alpha J_{B\mu}=
  {g'}^2 \sum_{a,b} \sum_{f,f' \in \{q,l,u,d,e\}} Y_f Y_{f'}
  D^\alpha (\overline{f}_a\gamma^\mu f_a)\, D_\alpha(\overline{f}'_b\gamma_\mu f'_b)\;,
\label{eq:4fd1}
\end{flalign}
\begin{flalign}          
  ({\rm D}{\Psi}^4)_{GG}\equiv& \sum_A D^\alpha  J_G^{A\mu}\,D_\alpha J_{G\mu}^{A}\, 
  =\,g_s^2\sum_{a,b}\Big\{
  \sum_{f \in \{q,u,d\}} \sum_{f' \in \{q,u,d\},f'\neq f}
  D^\alpha (\overline{f}_a\gamma^\mu T^A f_a)\,D_\alpha(\overline{f}'_b\gamma^\nu T^A f'_b)\nonumber\\
  &
  +
\frac{1}{2}  \sum_{f\in\{u,d\}} 
  D^\alpha  (\overline{f}_a\gamma^\mu  f_b)\, D_\alpha(\overline{f}_b\gamma_\mu  f_a)
  -\frac{1}{6}
\sum_{f\in\{q,u,d\}} 
D^\alpha(\overline{f}_a\gamma^\mu  f_a)\, D_\alpha(\overline{f}_b\gamma_\mu  f_b) \nonumber
\\&
+\frac{1}{4} D^\alpha  (\overline{q}_a\gamma^\mu  q_b)\, D_\alpha(\overline{q}_b\gamma_\mu  q_a) 
+\frac{1}{4} D^\alpha  (\overline{q}_a\gamma^\mu\,\tau^I  q_b)\, D_\alpha(\overline{q}_b\gamma_\mu\,\tau^I  q_a)
\Big\}\;,
\label{eq:4fd2}
\end{flalign}
\begin{flalign}        
  ({\rm D}{\Psi}^4)_{WW}\equiv& \sum_I D^\alpha J_W^{I\mu}\, D_\alpha J_{W\mu}^I\, 
  =\frac{g^2}{4}\sum_{a,b}\Big\{
D^\alpha(\overline{q}_a\gamma^\mu\tau^I q_a)
\, D^\alpha (\overline{q}_b\gamma_\mu\tau^I q_b)\,+ 2\,
  D^\alpha(\overline{q}_a\gamma^\mu\tau^I q_a)\,
  D_\alpha(\bar l_b\gamma_\mu\tau^I l_b)\,
  \nonumber \\ &\,
    +2\,  D^\alpha(\bar l_a\gamma^\mu l_b)\, D_\alpha(\bar l_b\gamma_\mu l_a)
\,-\, D^\alpha (\bar l_a\gamma^\mu l_a)\, D_\alpha(\bar l_b\gamma_\mu
l_{ b})
\Big\}\;,
\label{eq:4fd3}
\end{flalign}
\begin{flalign}        
({\rm D}{\Psi}^4)_{HH}\equiv& D^\alpha J_H^{j}\, D_\alpha {J^\dagger_H}_j\, 
=\sum_{abcd}\Big\{\Big[
\, y^e_{ab} \,  y^u_{cd}\,
D^\mu(\overline{l}^j_a\,e_b)\,\epsilon_{ k j}\,
 D_\mu(\overline{q}^k_c\, u_d)  \,{ +}\,
  y^u_{ab}\,  y^d_{cd} \, 
 \, D^\mu(\overline{q}^j_a\,u_b)
 \,\epsilon_{jk}\,
 D_\mu(\overline{q}^k_{ c}\,d_{ d})
\nonumber
\\ & +\,
 y^e_{ab} \, y^{d\dagger}_{cd}\, D^\mu(\overline{l}^j_a\,e_b)\,
 D_\mu(\overline{d}_c\, q_d)\,
 \;+\;{\rm h.c.}\Big]\;
 -\,\frac{1}{2}\, y^e_{ab}\, y^{e\dagger}_{cd}\,
D^\mu(\overline{l}_a\,\gamma^\nu\, l_d)
\,D_\mu(\overline{e}_c\,\gamma_\nu\, e_b)\, \nonumber \\
&  -\sum_{f\in\{u,d\}}
    \, y^f_{ab}\, y^{f\dagger}_{cd}\,
  \left[\,\frac{1}{6}\,
  D^\mu(\overline{q}_a\,\gamma^\nu\, q_d)\,D_\mu(\overline{f}_c\,\gamma_\nu\, f_b)\,
    +\,
    D^\mu( \overline{q}_a\,\gamma^\nu\,T^A\, q_d)\,D_\mu(\overline{f}_c\,\gamma_\nu\,T^A\, f_b)\right] \Big\}\;.
\label{eq:4fd4}
\end{flalign}
Notice that these last structures do not need any further
simplification as their present form appear in M8B. \medskip

\section{Fermionic operators for universal theories}
\label{sec:ferm}

We are now in position to present the combination of dimension-eight
fermionic operators that are associated with universal theories. We
call such combinations {\sl universal fermionic operators} since they
are the ones with independent couplings. That is, in universal
theories the couplings of the fermionic operators must be linear
combinations of the 86 independent couplings of the universal
fermionic operators listed here. \medskip

The 86 universal fermionic operators are formed by the contraction of
the fermionic Lorentz structures listed in Sec.~\ref{sec:fstruct} with
the remaining bosonic pieces of the universal operators listed in
Sec.~\ref{sec:boson}.  For convenience, we express them in terms of
the fermionic operators in M8B and we employ the M8B naming and
numbering of the operator classes.  Also for convenience, we reproduce
in Appendix~\ref{app:ferM8B} the subset of M8B operators which appear
in the universal operators listed here.  In addition, we have included
a factor $i$ to make the operators hermitian whenever
possible. \medskip

The full relation between the 86 bosonic operators in
Sec.~\ref{sec:boson}, the universal fermionic operators, and the
bosonic operators in M8B can be found in
Appendix~\ref{app:rotbosfer}. \medskip

\subsection{Two-Fermion Operators}

There are 62 independent universal combinations of two-fermion
operators in the following classes:

\begin{itemize}

\item {\bf Class} {\boldmath$9:\psi^2X^2H +\hc$:} there are 16
  universal operators in this class arising from the direct
  contraction of the Higgs fermionic current Eq.~\eqref{eq:fcurrents}
  with two gauge boson strength tensors
\begin{flalign}
    Q^{(1)}_{\psi^2 B^2 H} \equiv (J_H H) B_{\mu\nu} B^{\mu\nu}=&    
      {\displaystyle \sum_{pr}}\, \Big[
         y^{u\dagger}_{rp} \, Q^{\dagger(1)}_{quB^2H}\, +\,
        \, y^d_{pr}\, Q^{(1)}_{qdB^2H}\, 
        + \,  y^e_{pr}\, Q^{(1)}_{leB^2H}\,\Big] \;,
\nonumber
\end{flalign}\vspace*{-0.6cm}
\begin{flalign}  
      Q^{(2)}_{\psi^2 B^2 H} \equiv (J_H H) B_{\mu\nu} \widetilde B^{\mu\nu}=&
     {\displaystyle \sum_{pr}}\, \Big[
         y^{u\dagger}_{rp} \, Q^{\dagger(2)}_{quB^2H}\, +\,
        \, y^d_{pr}\, Q^{(2)}_{qdB^2H}\, 
        + \,  y^e_{pr}\, Q^{(2)}_{leB^2H}\,\Big] \;,
\nonumber
\end{flalign}\vspace*{-0.6cm}
\begin{flalign}  %
     Q^{(1)}_{\psi^2 W^2 H} \equiv (J_H H) W^I_{\mu\nu} W^{I\mu\nu}=&
     {\displaystyle \sum_{pr}}\, \Big[
         y^{u\dagger}_{rp} \, Q^{\dagger(1)}_{quW^2H}\, +\,
        \, y^d_{pr}\, Q^{(1)}_{qdW^2H}\, 
        + \,  y^e_{pr}\, Q^{(1)}_{leW^2H}\,\Big] \;,
\nonumber
\end{flalign}\vspace*{-0.6cm}
\begin{flalign}  
     Q^{(2)}_{\psi^2 W^2 H} \equiv (J_H H) W^I_{\mu\nu} \widetilde W^{I\mu\nu}=&
     {\displaystyle \sum_{pr}}\, \Big[
         y^{u\dagger}_{rp} \, Q^{\dagger(2)}_{quW^2H}\, +\,
        \, y^d_{pr}\, Q^{(2)}_{qdW^2H}\, 
        + \,  y^e_{pr}\, Q^{(2)}_{leW^2H}\,\Big] \;,
\label{eq:qclass9}      
\end{flalign}\vspace*{-0.6cm}
\begin{flalign}       
     Q^{(1)}_{\psi^2 G^2 H} \equiv (J_H H) G^A_{\mu\nu} G^{A\mu\nu}=&
     {\displaystyle \sum_{pr}}\, \Big[
         y^{u\dagger}_{rp} \, Q^{\dagger(1)}_{quG^2H}\, +\,
        \, y^d_{pr}\, Q^{(1)}_{qdG^2H}\, 
        + \,  y^e_{pr}\, Q^{(1)}_{leG^2H}\,\Big]\;,
        \nonumber
\end{flalign}\vspace*{-0.6cm}
\begin{flalign}  
    Q^{(2)}_{\psi^2 G^2 H} \equiv (J_H H) G^A_{\mu\nu} \widetilde G^{A\mu\nu} =&
     {\displaystyle \sum_{pr}}\, \Big[
         y^{u\dagger}_{rp} \, Q^{\dagger(2)}_{quG^2H}\, +\,
        \, y^d_{pr}\, Q^{(2)}_{qdG^2H}\, 
        + \,  y^e_{pr}\, Q^{(2)}_{leG^2H}\,\Big]\;,
        \nonumber
\end{flalign}\vspace*{-0.6cm}
\begin{flalign}  
     Q^{(1)}_{\psi^2 WB H} \equiv (J_H\tau^I H) B_{\mu\nu} W^{I\mu\nu}=&
     {\displaystyle \sum_{pr}}\, \Big[
       { -}\,
       y^{u\dagger}_{rp} \, Q^{\dagger(1)}_{quWBH}\, +\,
        \, y^d_{pr}\, Q^{(1)}_{qdWBH}\, 
        + \,  y^e_{pr}\, Q^{(1)}_{leWBH}\,\Big]  \;,   \nonumber
\end{flalign}\vspace*{-0.6cm}
\begin{flalign}  
    Q^{(2)}_{\psi^2 WB H} \equiv (J_H \tau^I H) B_{\mu\nu} \widetilde W^{I\mu\nu} =&
     {\displaystyle \sum_{pr}}\, \Big[
       { -}\,  y^{u\dagger}_{rp} \, Q^{\dagger(2)}_{quWBH}\, +\,
        \, y^d_{pr}\, Q^{(2)}_{qdWBH}\, 
        + \,  y^e_{pr}\, Q^{(2)}_{leWBH}\,\Big]    \;,    \nonumber
\end{flalign}
%
together with the hermitian conjugates of the above operators.  These
universal fermionic operators are generated when applying EOM to some
of the operators in class $X^2 H^2 D^2$ as can be seen in
Eqs.~\eqref{eq:r33}--\eqref{eq:r36},~\eqref{eq:r42}--\eqref{eq:r45},~\eqref{eq:r55}--\eqref{eq:r58},
and~\eqref{eq:r64}--\eqref{eq:r67}, \medskip

\item {\bf Hybrid class} {\boldmath$9\&14:\psi^2X^2H(D)$} contains 8
  operators exhibiting specific combinations of operators in classes
  $\psi^2X^2H$ and $\psi^2X^2D$ originated from contraction of the
  fermionic structures in Eq.~\eqref{eq:fd3} and~\eqref{eq:fd4} with
  two gauge strength tensors
\begin{flalign}  
    Q^{(1)}_{\psi^2 WB H(D)} \equiv& \,({\rm D}{\Psi}^2_-)_W^{I\mu\nu}B_{\mu\rho} W_\nu^{I\rho}
        = \frac{ g}{ 2}
        {\displaystyle \sum_{pr}}\Big\{\,{ -}
        \Big[
         y^u_{pr} \, Q^{(3)}_{quWBH}\, +\,
        \, y^d_{pr}\, Q^{(3)}_{qdWBH}\,
        + \,  y^e_{pr}\, Q^{(3)}_{leWBH}\,\,+\;{\rm h.c.}\,\Big]
    \nonumber
    \\&
    -\,i\,\left[Q^{(3)}_{q^2WBD}+Q^{(3)}_{l^2WBD}
    \right] \delta_{pr}\,\Big\} \;,
    \nonumber
\end{flalign}\vspace*{-0.6cm}
\begin{flalign}  
    Q^{(2)}_{\psi^2 WB H(D)} \equiv &\,({\rm D}{\Psi}^2_-)_W^{I\mu\nu}B_{\mu\rho} \widetilde W_\nu^{I\rho}
     = \frac{g}{2}
    {\displaystyle \sum_{pr}}\Big\{\, { -}\,\Big[
        i\, y^u_{pr} \, Q^{(3)}_{quWBH}\, +\,
        \,i\, y^d_{pr}\, Q^{(3)}_{qdWBH}\,
        + \, i\, y^e_{pr}\, Q^{(3)}_{leWBH}\,\,+\;{\rm h.c.}\,\Big]
   \nonumber \\&
    +\,i\,\left[Q^{(1)}_{q^2WBD}+Q^{(1)}_{l^2WBD}
  \right] \delta_{pr}\,\Big\} \;,
\nonumber
\end{flalign}\vspace*{-0.6cm}
\begin{flalign}  
    Q^{(1)}_{\psi^2 GB H(D)} \equiv& \,({\rm D}{\Psi}^2_-)_G^{A\mu\nu}B_{\mu\rho} G_\nu^{A\rho}
= g_s\,
    {\displaystyle \sum_{pr}}\Big\{\, { -} 2\,\Big[
         y^u_{pr} \, Q^{(3)}_{quGBH}\, +\,
        \, y^d_{pr}\, Q^{(3)}_{qdGBH}\,+\;{\rm h.c.}\,\Big]
   \nonumber\\&
     -\,i\,\big[Q^{(3)}_{q^2GBD}\,-\,Q^{(3)}_{u^2GBD}\,-\,
      Q^{(3)}_{d^2GBD}  \big] \delta_{pr}\,\Big\} \;,
\nonumber
\end{flalign}\vspace*{-0.6cm}
\begin{flalign}  
    Q^{(2)}_{\psi^2 GB H(D)} \equiv& \,({\rm D}{\Psi}^2_-)_G^{A\mu\nu}B_{\mu\rho} \widetilde G_\nu^{A\rho}
     = g_s\,
    {\displaystyle \sum_{pr}}\Big\{ \, { -} 2\,\Big[
        i\, y^u_{pr} \, Q^{(3)}_{quGBH}\, +\,
        i\, y^d_{pr}\, Q^{(3)}_{qdGBH}\,+\;{\rm h.c.}\,\Big]
   \nonumber \\&
   {  +}\,i\,\big[Q^{(1)}_{q^2GBD}\,-\,Q^{(1)}_{u^2GBD}\,-\,
      Q^{(1)}_{d^2GBD}  \big] \delta_{pr}\,\Big\} \;,
\nonumber
\end{flalign}\vspace*{-0.6cm}
\begin{flalign}  
    Q^{(1)}_{\psi^2 W^2 H(D)} \equiv& \,\epsilon^{IJK}({\rm D}{\Psi}^2_-)_W^{I\mu\nu}W^J_{\mu\rho} W_\nu^{K\rho}
     = \frac{g}{2}
    {\displaystyle \sum_{pr}}\Big\{\, \Big[
         y^u_{pr} \, Q^{(3)}_{quW^2H}\, +\,
        \, y^d_{pr}\, Q^{(3)}_{qdW^2H}\,
        + \,  y^e_{pr}\, Q^{(3)}_{leW^2H}\,\,+\;{\rm h.c.}\,\Big]
    \nonumber\\&
    -\,i\,\left[Q^{(4)}_{q^2W^2D}+Q^{(4)}_{l^2W^2D}
    \right] \delta_{pr}\,\Big\} \;,
    \label{eq:qclass9-14}
\end{flalign}\vspace*{-0.6cm}
\begin{flalign}      
Q^{(2)}_{\psi^2 W^2 H(D)} \equiv &\,\epsilon^{IJK}({\rm D}{\Psi}^2_-)_W^{I\mu\nu}W^J_{\mu\rho}
\widetilde W_\nu^{K\rho}
     = \frac{g}{2}
    {\displaystyle \sum_{pr}}\Big\{\, { +}\,\Big[
       i\,  y^u_{pr} \, Q^{(3)}_{quW^2H}\, +\,
       i \, y^d_{pr}\, Q^{(3)}_{qdW^2H}\,
        + \, i\, y^e_{pr}\, Q^{(3)}_{leW^2H}\,\,+\;{\rm h.c.}\,\Big]
    \nonumber\\&
    { +2 }\,i\,\left[Q^{(2)}_{q^2W^2D}+Q^{(2)}_{l^2W^2D}
    \right] \delta_{pr}\,\Big\} \;,
    \nonumber
\end{flalign}\vspace*{-0.6cm}
\begin{flalign}      
    Q^{(1)}_{\psi^2 G^2 H(D)} \equiv& \,f^{ABC}\,({\rm D}{\Psi}^2_-)_G^{A\mu\nu}G^B_{\mu\rho} G_\nu^{D\rho}
     = g_s\,
    {\displaystyle \sum_{pr}}\Big\{\, 2\,\Big[
         y^u_{pr} \, Q^{(5)}_{quG^2H}\, +\,
        \, y^d_{pr}\, Q^{(5)}_{qdG^2H}\,+\;{\rm h.c.}\,\Big]
    \nonumber\\&
    -\,i\,\big[Q^{(5)}_{q^2G^2D}\,-\,Q^{(5)}_{u^2G^2D}\,-\,
      Q^{(5)}_{d^2G^2D}  \big] \delta_{pr}\,\Big\} \;,
    \nonumber
\end{flalign}\vspace*{-0.6cm}
\begin{flalign}  
    Q^{(2)}_{\psi^2 G^2 H(D)} \equiv& \,f^{ABC}\,({\rm D}{\Psi}^2_-)_G^{A\mu\nu}G^B_{\mu\rho}
    \widetilde G_\nu^{D\rho}
     = g_s\,
    {\displaystyle \sum_{pr}}\Big\{{ +}\, 2\,\Big[
       i\,  y^u_{pr} \, Q^{(5)}_{quG^2H}\, +\,
        i\, y^d_{pr}\, Q^{(5)}_{qdG^2H}\,+\;{\rm h.c.}\,\Big]
    \nonumber\\&
    { +2 }\,i\,\big[Q^{(2)}_{q^2G^2D}\,-\,Q^{(2)}_{u^2G^2D}\,-\,
    Q^{(2)}_{d^2G^2D}  \big] \delta_{pr}\,\Big\} \;.
    \nonumber 
  \end{flalign}
  The universal fermionic operators in this hybrid class are generated
  when applying EOM to some of the operators in class $X^3 D^2$ as can
  be seen in
  Eqs.~\eqref{eq:r16}--\eqref{eq:r17},~\eqref{eq:r20}--\eqref{eq:r21},~\eqref{eq:r24}--\eqref{eq:r25},
  and~\eqref{eq:r28}--\eqref{eq:r29}.

\item {\bf Class} {\boldmath$11:\psi^2H^2D^3$:} the 2 operators in
  this class arise from the contraction of the structures in
  Eq.~\eqref{eq:fd1} and~\eqref{eq:fd2} with a current containing two
  symmetrized covariant derivatives acting on the Higgs field
\begin{flalign}  
      Q^{(1)}_{\psi^2H^2D^3} \equiv& \,
      i\,({\rm D}{\Psi}^2_+)_B^{\mu\nu} (D_{(\mu}D_{\nu)} H^\dagger H-H^\dagger D_{(\mu}D_{\nu)} H) 
\nonumber\\  &=\,2\,g'\, {\displaystyle \sum_{pr}}\,
      {\displaystyle \sum_{f\in \{q,l,u,d,e\}}} Y_f\,\Big[
        Q^{(1)}_{f^2 H^2 D^3}-Q^{(2)}_{f^2 H^2 D^3} \,{ +\, {\rm h.c.}}
        \Big]\,\delta_{pr}  \;,  \
\label{eq:qclass11}    
\end{flalign}\vspace*{-0.6cm}
\begin{flalign}  
      Q^{(2)}_{\psi^2H^2D^3} \equiv& \,
      i\,({\rm D}{\Psi}^2_+)_W^{I\mu\nu} (D_{(\mu}D_{\nu)} H^\dagger\tau^I H-H^\dagger \tau^ID_{(\mu}D_{\nu)} H) 
      \nonumber \\
      &=\,g \, {\displaystyle \sum_{pr}}\,
      {\displaystyle \sum_{f \in \{q,l\}}} \,\Big[
        Q^{(3)}_{f^2 H^2 D^3}-Q^{(4)}_{f^2 H^2 D^3} { +\, {\rm h.c.}}
        \Big]\,\delta_{pr} \;. \nonumber
\end{flalign}
These operators are generated directly from application of the EOM of
the Higgs field to the operators in class $X^2D^4$ as in
Eqs.~\eqref{eq:r30} and~\eqref{eq:r31} and class $XH^4 D^2$, see
Eqs.~\eqref{eq:r83} and~\eqref{eq:r86}.

\item {\bf Class} {\boldmath$12:\psi^2H^5 + \hc$} contains 2 operators
  originating from the contraction of the Higgs fermionic currents in
  Eq.~\eqref{eq:fcurrents} directly with Higgs fields:
\begin{flalign}    
      Q^{(1)}_{\psi^2H^5} \equiv \, (H^\dagger H)^2 (J_H H)=&
      {\displaystyle \sum_{pr}}\, \Big[
        y^{u\dagger}_{rp} \, Q^\dagger_{quH^5}\, +\,
        \, y^d_{pr}\, Q_{qdH^5}\, 
        + \,  y^e_{pr}\, Q_{leH^5}\,\Big]
\label{eq:qclass12}
\end{flalign}
and its hermitian conjugate.  These operators appear directly in the
application of the Higgs EOM to the two operators in class $H^6 D^2$
Eqs.~\eqref{eq:r1}--\eqref{eq:r2} but, as seen in
Appendix~\ref{app:rotbosfer}, they also arise in the rotation of a
large fraction of the 86 bosonic operators. This is so, because these
two operators in class $H^6 D^2$ are generically generated when
reducing the products of the Higgs-gauge currents introduced by the
gauge boson EOM to the bosonic operators in M8B.

\item {\bf Class} {\boldmath$13:\psi^2H^4D$:} there are four  universal
  fermionic operators in this class which appear in the contraction of
  the electroweak fermionic currents in Eq.~\eqref{eq:fcurrents} with
  the product of two Higgs pairs one of them containing one
  derivative.
\begin{flalign}
  Q^{(1)}_{\psi^2H^4D} \equiv \,
      i\,J_B^\mu (H^\dagger\overleftrightarrow{D}_{\mu} H) (H^\dagger H)
      &= \,g'\,  {\displaystyle \sum_{pr}}\,
      {\displaystyle \sum_{f \in \{q,l,u,d,e\}}} Y_f \,
        Q^{(1)}_{f^2 H^4 D} \delta_{pr} \;,
   \nonumber
\end{flalign}\vspace*{-0.6cm}
\begin{flalign}        
      Q^{(2)}_{\psi^2H^4D} \equiv \,
      i\,J_W^{I\mu}\left[ (H^\dagger\overleftrightarrow{D}^I_{\mu} H) (H^\dagger H)
        +(H^\dagger\overleftrightarrow{D}_\mu H) (H^\dagger \tau^I H)\right]   
      & =\,\frac{g}{2}\, {\displaystyle \sum_{pr}}\,
      {\displaystyle \sum_{f \in \{q,l\}}} \,
      Q^{(2)}_{f^2 H^4 D} \,\delta_{pr}   \;,
      \nonumber
\end{flalign}\vspace*{-0.6cm}
\begin{flalign}  
      Q^{(3)}_{\psi^2H^4D} \equiv \,
      i\,\epsilon^{IJK}\,J_W^{I\mu}(H^\dagger\overleftrightarrow{D}^J_{\mu} H)
      (H^\dagger \tau^K H)
      & =\,\frac{g}{2}\, {\displaystyle \sum_{pr}}\,
      {\displaystyle \sum_{f \in \{q,l\}}} \,
      Q^{(3)}_{f^2 H^4 D} \,\delta_{pr} \;,
      \label{eq:qclass13}    
\end{flalign}\vspace*{-0.6cm}
\begin{flalign}          
      Q^{(4)}_{\psi^2H^4D} \equiv \,
      \epsilon^{IJK}\,J_W^{I\mu}(H^\dagger\tau^J_{\mu} H) D_\mu(H^\dagger \tau^K H)
      & =\,\frac{g}{2}\, {\displaystyle \sum_{pr}}\,
      {\displaystyle \sum_{f \in \{q,l\}}} \,
      Q^{(4)}_{f^2 H^4 D} \,\delta_{pr} \;.
   \nonumber
\end{flalign}
Operators in this class are directly generated by application of the
gauge-boson EOM in operators in class $X H^4 D^2$ as seen in
Eqs.~\eqref{eq:r77}--\eqref{eq:r80}. They also arise in the complete
rotation to operators in M8B of some operators in classes $X^3 D^2$
(Eq.~\eqref{eq:r14}), $X^2 D^4$ (Eqs.~\eqref{eq:r30}--\eqref{eq:r31}),
$X^2 H^2 D^2$ (Eqs.~\eqref{eq:r37} ,~\eqref{eq:r38}, \eqref{eq:r46},
\eqref{eq:r47}, ~\eqref{eq:r51}, and ~\eqref{eq:r52}), and $X H^2 D^4$
(Eqs.~\eqref{eq:r83} and ~\eqref{eq:r86}). Notice that, for the sake
of simplicity in writing the expressions above, in operators
$Q^{(1)}_{f^2 H^4 D}$, where $f = u,d,e$, we have added a superscript
of $(1)$ to the M8B operators.  This minimal change of labeling is
reflected also when we list the operators in class 13 in Table
~\ref{tab:uniclass12-13-17}.

\item {\bf Class} {\boldmath$15:\psi^2XH^2D$:} It contains 24
  operators generated by the contraction of fermionic gauge currents
  in Eq.~\eqref{eq:fcurrents} with a gauge field strength tensors and a pair
  of Higgs bosons with one derivative.  In twelve operators the
  fermionic and Higgs currents are contracted with the $SU(2)_L$
  field strength tensor
\begin{flalign}
      Q^{(1)}_{\psi^2WH^2D} \equiv \,
      J^\nu_B D^\mu(H^\dagger\tau^I H)W^I_{\mu\nu}&=\,g'\,{\displaystyle \sum_{pr}}\,      
      {\displaystyle \sum_{ f \in \{ q,l,u,d,e\}}} Y_f \, Q^{(1)}_{f^2W H^2
        D}\,\delta_{pr} \;,
      \nonumber
\end{flalign}\vspace*{-0.6cm}
\begin{flalign}          
      Q^{(2)}_{\psi^2WH^2D} \equiv \,
      J^\nu_B D^\mu(H^\dagger\tau^I H)\widetilde W^I_{\mu\nu}&=\,g'\,{\displaystyle \sum_{pr}}\, 
     {\displaystyle \sum_{ f\in \{ q,l,u,d,e\}}} Y_f \, Q^{(2)}_{f^2W H^2
       D} \,\delta_{pr} \;,
     \nonumber
\end{flalign}\vspace*{-0.6cm}
\begin{flalign}          
      Q^{(3)}_{\psi^2WH^2D} \equiv \, \,{ i}\,
      J^\nu_B (H^\dagger
\overleftrightarrow{D}^I_{\mu} H)W^I_{\mu\nu}&= \,{ i}\,\,g'\,{\displaystyle \sum_{pr}}\,      
     {\displaystyle \sum_{ f\in \{ q,l,u,d,e\}}} Y_f \, Q^{(3)}_{f^2W H^2
       D} \, \delta_{pr}  \;,
     \nonumber
\end{flalign}\vspace*{-0.6cm}
\begin{flalign}         
 Q^{(4)}_{\psi^2WH^2D} \equiv \, \,{ i}\,
     J^\nu_B (H^\dagger
     \overleftrightarrow{D}^I_{\mu} H)\widetilde W^I_{\mu\nu}&=
\,{ i}\,
     \,g'\,{\displaystyle \sum_{pr}}\,
     {\displaystyle \sum_{ f\in \{ q,l,u,d,e\}}} Y_f \, Q^{(4)}_{f^2W H^2
       D}\,\delta_{pr} \;,
     \nonumber
\end{flalign}\vspace*{-0.6cm}
\begin{flalign}  
      Q^{(5)}_{\psi^2WH^2D} \equiv \,
      J^{I\nu}_W D^\mu(H^\dagger H)W^I_{\mu\nu}&=\,\frac{g}{2}
      \,{\displaystyle \sum_{pr}}\,      
        {\displaystyle \sum_{f \in \{q,l\}}}
        Q^{(5)}_{f^2 WH^2 D} \,\delta_{pr}    \;,
        \nonumber
\end{flalign}\vspace*{-0.6cm}
\begin{flalign}  
      Q^{(6)}_{\psi^2WH^2D} \equiv \,
      J^{I\nu}_W D^\mu(H^\dagger H)\widetilde W^I_{\mu\nu}&=\,\frac{g}{2}\,{\displaystyle \sum_{pr}}\, 
         {\displaystyle \sum_{f \in \{q,l\}}} \,
         Q^{(6)}_{f^2 W H^2 D} \,\delta_{pr} \;,
         \label{eq:qclass15-1}    
\end{flalign}\vspace*{-0.6cm}
\begin{flalign}          
      Q^{(7)}_{\psi^2WH^2D} \equiv \, \,{ i}\,
      J^{I\nu}_W (H^\dagger
      \overleftrightarrow{D}_{\mu} H)W^I_{\mu\nu}&= \,{ i}\,
      \, \frac{g}{2}\,{\displaystyle \sum_{pr}}\,      
        {\displaystyle \sum_{f \in \{q,l\}}}
        Q^{(7)}_{f^2 WH^2 D} \,\delta_{pr} \;,
        \nonumber
\end{flalign}\vspace*{-0.6cm}
\begin{flalign}         
 Q^{(8)}_{\psi^2WH^2D} \equiv \, \,{ i}\,
     J^{I\nu}_W (H^\dagger
     \overleftrightarrow{D}_{\mu} H)\widetilde W^I_{\mu\nu}&=\,{ i}\,
     \,\frac{g}{2}\,{\displaystyle \sum_{pr}}\,
        {\displaystyle \sum_{f \in \{q,l\}}}  \,
        Q^{(8)}_{f^2 WH^2 D} \,\delta_{pr} \;,
        \nonumber
\end{flalign}\vspace*{-0.6cm}
\begin{flalign}          
              Q^{(9)}_{\psi^2WH^2D} \equiv \,\epsilon^{IJK}\,
      J^{I\nu}_W D^\mu(H^\dagger\tau^J H)W^K_{\mu\nu}&=\,\frac{g}{2}
      \,{\displaystyle \sum_{pr}}\,      
        {\displaystyle \sum_{f \in \{q,l\}}}
        Q^{(9)}_{f^2 WH^2 D} \,\delta_{pr} \;,
        \nonumber
\end{flalign}\vspace*{-0.6cm}
\begin{flalign}  
      Q^{(10)}_{\psi^2WH^2D} \equiv \, \epsilon^{IJK}\,
      J^{I\nu}_W D^\mu(H^\dagger\tau^J H)\widetilde W^K_{\mu\nu}&=\,\frac{g}{2}\,{\displaystyle \sum_{pr}}\, 
         {\displaystyle \sum_{f \in \{q,l\}}} \,
         Q^{(10)}_{f^2 W H^2 D} \,\delta_{pr} \;,
         \nonumber
\end{flalign}\vspace*{-0.6cm}
\begin{flalign}  
      Q^{(11)}_{\psi^2WH^2D} \equiv \, \,{ i}\,\epsilon^{IJK}\,
      J^{I\nu}_W (H^\dagger
      \overleftrightarrow{D}^J_{\mu} H)W^K_{\mu\nu}&= \,{ i}\,
      \, \frac{g}{2}\,{\displaystyle \sum_{pr}}\,      
        {\displaystyle \sum_{f \in \{q,l\}}}
        Q^{(11)}_{f^2 WH^2 D} \,\delta_{pr} \;,
        \nonumber
\end{flalign}\vspace*{-0.6cm}
\begin{flalign}  
 Q^{(12)}_{\psi^2WH^2D} \equiv \, \,{ i}\, \epsilon^{IJK}\,
     J^{I\nu}_W (H^\dagger
     \overleftrightarrow{D}_{\mu}^J H)\widetilde W^K_{\mu\nu}&=\,{ i}\,
     \,\frac{g}{2}\,{\displaystyle \sum_{pr}}\,
        {\displaystyle \sum_{f \in \{q,l\}}}  \,
        Q^{(12)}_{f^2 WH^2 D} \,\delta_{pr}\;.
 \nonumber
\end{flalign}
They originate from direct application of EOM of the electroweak gauge
bosons in operators in classes $X^3 D^2$ (Eqs.~\eqref{eq:r14},
\eqref{eq:r15}, ~\eqref{eq:r22}, and ~\eqref{eq:r23}) as well as
$X^2 H^2 D^2$ (Eqs.~\eqref{eq:r46}-~\eqref{eq:r49} and
Eqs.~\eqref{eq:r51}-~\eqref{eq:r54}). They are also generated in the
rotation of operators $R^{(1)}_{W^2D^4}$~\eqref{eq:r31},
$R^{(3)}_{BH^2D^4}$~\eqref{eq:r83}, and
$R^{(3)}_{WH^2D^4}$~\eqref{eq:r86}. \medskip

In eight operators in this class the fermionic structures couple
to the hypercharge field strength tensor
\begin{flalign}
      Q^{(1)}_{\psi^2BH^2D} \equiv \,
      J^\nu_B D^\mu(H^\dagger H)B_{\mu\nu}&=\,g'\,{\displaystyle \sum_{pr}}\,      
     \Big[{\displaystyle \sum_{ f\in \{u,d,e\}}} Y_f \, Q^{(1)}_{f^2B H^2 D} +
        {\displaystyle \sum_{f \in \{q,l\}}} Y_f \,
        Q^{(5)}_{f^2 BH^2 D} \Big]\delta_{pr}\;,
        \nonumber
\end{flalign}\vspace*{-0.6cm}
\begin{flalign}          
      Q^{(2)}_{\psi^2BH^2D} \equiv \,
      J^\nu_B D^\mu(H^\dagger H)\widetilde B_{\mu\nu}&=\,g'\,{\displaystyle \sum_{pr}}\, 
     \Big[{\displaystyle \sum_{ f\in \{ u,d,e\}}} Y_f \, Q^{(2)}_{f^2B H^2 D} +
        {\displaystyle \sum_{f \in \{q,l\}}} Y_f \,
        Q^{(6)}_{f^2 BH^2 D} \Big]\delta_{pr}\;,
        \nonumber
\end{flalign}\vspace*{-0.6cm}
\begin{flalign}          
      Q^{(3)}_{\psi^2BH^2D} \equiv \, \,{ i}\,
      J^\nu_B (H^\dagger
\overleftrightarrow{D}_{\mu} H)B_{\mu\nu}&= \,{ i}\,\,g'\,{\displaystyle \sum_{pr}}\,      
     \Big[{\displaystyle \sum_{ f\in \{u,d,e\}}} Y_f \, Q^{(3)}_{f^2B H^2 D} +
        {\displaystyle \sum_{f \in \{q,l\}}} Y_f \,
        Q^{(7)}_{f^2 BH^2 D} \Big]\delta_{pr}\;,
        \nonumber
\end{flalign}\vspace*{-0.6cm}
\begin{flalign}          
 Q^{(4)}_{\psi^2BH^2D} \equiv \, \,{ i}\,
     J^\nu_B (H^\dagger
     \overleftrightarrow{D}_{\mu} H)\widetilde B_{\mu\nu}&=
\,{ i}\,
     \,g'\,{\displaystyle \sum_{pr}}\,
     \Big[{\displaystyle \sum_{ f\in \{u,d,e\}}} Y_f \, Q^{(4)}_{f^2B H^2 D} +
        {\displaystyle \sum_{f \in \{q,l\}}} Y_f \,
        Q^{(8)}_{f^2 BH^2 D} \Big]\delta_{pr}\;,
        \nonumber
\end{flalign}\vspace*{-0.6cm}
\begin{flalign}  
      Q^{(5)}_{\psi^2BH^2D} \equiv \,
      J^{I\nu}_W D^\mu(H^\dagger\tau^I H)B_{\mu\nu}&=\,\frac{g}{2}
      \,{\displaystyle \sum_{pr}}\,      
        {\displaystyle \sum_{f \in \{q,l\}}}
        Q^{(1)}_{f^2 BH^2 D} \,\delta_{pr} \;,
        \label{eq:qclass15-2}        
\end{flalign}\vspace*{-0.6cm}
\begin{flalign}          
      Q^{(6)}_{\psi^2BH^2D} \equiv \,
      J^{I\nu}_W D^\mu(H^\dagger\tau^I H)\widetilde B_{\mu\nu}&=\,\frac{g}{2}\,{\displaystyle \sum_{pr}}\, 
         {\displaystyle \sum_{f \in \{q,l\}}} \,
         Q^{(2)}_{f^2 B H^2 D} \,\delta_{pr}\;,
         \nonumber
\end{flalign}\vspace*{-0.6cm}
\begin{flalign}         
      Q^{(7)}_{\psi^2BH^2D} \equiv \, \,{ i}\,
      J^{I\nu}_W (H^\dagger
      \overleftrightarrow{D}^I_{\mu} H)B_{\mu\nu}&= \,{ i}\,
      \, \frac{g}{2}\,{\displaystyle \sum_{pr}}\,      
        {\displaystyle \sum_{f \in \{q,l\}}}
        Q^{(3)}_{f^2 BH^2 D} \,\delta_{pr} \;,
        \nonumber
\end{flalign}\vspace*{-0.6cm}
\begin{flalign}          
 Q^{(8)}_{\psi^2BH^2D} \equiv \, \,{ i}\,
     J^{I\nu}_W (H^\dagger
     \overleftrightarrow{D}^I_{\mu} H)\widetilde B_{\mu\nu}&=\,{ i}\,
     \,\frac{g}{2}\,{\displaystyle \sum_{pr}}\,
        {\displaystyle \sum_{f \in \{q,l\}}}  \,
        Q^{(4)}_{f^2 BH^2 D} \,\delta_{pr}\;.
          \nonumber
\end{flalign}
They are generated by direct application of EOM of the electroweak
gauge bosons in operators in class $X^2 H^2 D^2$
(Eqs.~\eqref{eq:r37}-~\eqref{eq:r40} and
Eqs.~\eqref{eq:r72}-~\eqref{eq:r75}). They are also generated in the
rotation of operators $R^{(3)}_{BH^2D^4}$~\eqref{eq:r83}, and
$R^{(3)}_{WH^2D^4}$~\eqref{eq:r86}. \medskip

And finally four operators  involve   the gluon strength tensor
\begin{flalign}
      Q^{(1)}_{\psi^2GH^2D} \equiv \,
      J^A_G D^\mu(H^\dagger H)G^A_{\mu\nu}&=\,g_s\,{\displaystyle \sum_{pr}}\,      
      \Big[{\displaystyle \sum_{ f\in \{u,d\}}}  \, Q^{(1)}_{f^2G H^2 D}\,
        +\,Q^{(5)}_{q^2G H^2 D}\,\Big]
        \delta_{pr}\;,
        \nonumber
\end{flalign}\vspace*{-0.6cm}
\begin{flalign}                 
      Q^{(2)}_{\psi^2GH^2D} \equiv \,
      J^A_G D^\mu(H^\dagger H)\widetilde G^A_{\mu\nu}&=\,g_s\,{\displaystyle \sum_{pr}}\, 
      \Big[{\displaystyle \sum_{ f\in \{u,d\}}}  \, Q^{(2)}_{f^2G H^2 D}\,
        +\,Q^{(6)}_{q^2G H^2 D}\,\Big]
        \,\delta_{pr}\;,
        \nonumber
\end{flalign}\vspace*{-0.6cm}
\begin{flalign}                
      Q^{(3)}_{\psi^2GH^2D} \equiv \, \,{ i}\,
      J^A_G (H^\dagger
\overleftrightarrow{D}_{\mu} H)G^A_{\mu\nu}&= \,{ i}\,\,g_s\,{\displaystyle \sum_{pr}}\,      \Big[
                   {\displaystyle \sum_{ f\in \{u,d\}}}  \, Q^{(3)}_{f^2G H^2 D} \,
                   +\,Q^{(7)}_{q^2G H^2 D}\,\Big]\delta_{pr} \;,
\label{eq:qclass15-3}        
\end{flalign}\vspace*{-0.6cm}
\begin{flalign}          
 Q^{(4)}_{\psi^2GH^2D} \equiv \, \,{ i}\,
     J^A_G (H^\dagger
     \overleftrightarrow{D}_{\mu} H)\widetilde G^A_{\mu\nu}&=
\,{ i}\,
     \,g_s\,{\displaystyle \sum_{pr}}\,
     \Big[    {\displaystyle \sum_{ f\in \{u,d\}}}  \, Q^{(4)}_{f^2G H^2 D}\,
+\,Q^{(8)}_{q^2G H^2 D}\,\Big]
     \delta_{pr} \;,
\nonumber
\end{flalign}
which stem from the direct application of the gluon EOM in operators
in class $X^2 H^2 D^2$, as in Eqs.~\eqref{eq:r59}--\eqref{eq:r62}, and
the operators $R^{(1)}_{B G^2D^2}$~\eqref{eq:r26} and
$R^{(2)}_{B G^2D^2}$~\eqref{eq:r27} of class $X^3D^2$.

\item {\bf Class} {\boldmath $17:\psi^2H^3D^2 + \hc$} is generated by
  direct contraction of the Higgs fermionic current in
  Eq.~\eqref{eq:fcurrents} with one Higgs field and two derivatives of
  Higgs fields. There are six independent such contractions
\begin{flalign}  
      Q^{(1)}_{\psi^2H^3D^2} \equiv \, (D_\mu H^\dagger D^\mu H) (J_H H)=&
      {\displaystyle \sum_{pr}}\, \Big[
         y^{u\dagger}_{rp} \, Q^{\dagger (1)}_{quH^3D^2}\, +\,
        \, y^d_{pr}\, Q^{(1)}_{qdH^3D^2}\,
        + \,  y^e_{pr}\, Q^{(1)}_{leH^3D^2}\,\Big] \;,\nonumber
\end{flalign}\vspace*{-0.6cm}
\begin{flalign}                 
      Q^{(2)}_{\psi^2H^3D^2} \equiv \, (D_\mu H^\dagger \tau^I D^\mu H) (J_H \tau^I H)=&
      {\displaystyle \sum_{pr}}\, \Big[
       { -}  y^{u\dagger}_{rp} \, Q^{\dagger (2)}_{quH^3D^2}\, +\,
        \, y^d_{pr}\, Q^{(2)}_{qdH^3D^2}\,
        + \,  y^e_{pr}\, Q^{(2)}_{leH^3D^2}\,\Big]\;,
        \label{eq:qclass17}    
\end{flalign}\vspace*{-0.6cm}
\begin{flalign}                 
      Q^{(3)}_{\psi^2H^3D^2} \equiv \, (H^\dagger D_\mu H) (J_H D^\mu H)=&      
      {\displaystyle \sum_{pr}}\, \Big[
         y^{u\dagger}_{rp} \, Q^{\dagger (5)}_{quH^3D^2}\, +\,
        \, y^d_{pr}\, Q^{(5)}_{qdH^3D^2}\, 
        + \,  y^e_{pr}\, Q^{(5)}_{leH^3D^2}\,\Big]\;,
\nonumber
\end{flalign}
and their hermitian conjugates.  These operators are generated
directly from applying the Higgs field EOM to the
operators in class $H^4 D^4$ and $H^2 D^6$ (see
Eqs.~\eqref{eq:r3}--\eqref{eq:r8} and~\eqref{eq:r13}). In addition
they also appear in the rotation of operators $X H^2 D^4$, as in
Eqs.~\eqref{eq:r81}--\eqref{eq:r82}
and~\eqref{eq:r84}--\eqref{eq:r85}, arising in the reduction of the
products of the Higgs gauge currents introduced by the gauge boson EOM
to the bosonic operators in M8B.

\end{itemize}

\subsection{Four-Fermion Operators}

We obtain 24 universal four-fermion operators in the following
classes:

\begin{itemize}

\item {\bf Class} {\boldmath$18:\psi^4H^2$:} contains eight universal
  fermionic operators obtained from the product of the four-fermion
  currents in Eqs.~\eqref{eq:4fhh1}--\eqref{eq:4fhh3} and
  Eqs.~\eqref{eq:4fs1}--\eqref{eq:4fs4} with a pair Higgs fields

\begin{flalign}  
 Q^{(1)}_{\psi^4H^2}\equiv &({\Psi}^4)_{HH} (H^\dagger H) = 
 {\displaystyle \sum_{prst}}\Big\{
 -\!\!\!\!{\displaystyle\sum_{f \in \{u,d\}}}
         y^{f\dagger}_{sr}\,  y^f_{pt}\,
        \left(\frac{1}{6}\,Q^{(1)}_{q^2f^2H^2}
        \,+\, Q^{(3)}_{q^2f^2H^2}\right)\, -\,
        \frac{1}{2}\, y^{e\dagger}_{sr}\,  y^e_{pt}\,
        Q^{(1)}_{l^2e^2H^2} \nonumber\\
& \;+\;\left[  \,{ - y^e_{pr}\,  y^u_{st} \, Q^{(1)}_{lequH^2}
\,+\,
   y^u_{pr}\,  y^d_{st} \, Q^{(1)}_{q^2udH^2}}
\,+\,
 y^e_{pr}\,  y^{d\dagger}_{ts} \, Q^{(1)}_{leqdH^2}\;+\,{\rm h.c.}\right]
        \Big\}\;,
\nonumber
\end{flalign}\vspace*{-0.6cm}
\begin{flalign}                         
  Q^{(2)}_{\psi^4H^2}\equiv& ({\Psi}^4)_{HH}^{jk}H_j H_k= 
  {\displaystyle \sum_{prst}}\Big\{
       y^{u\dagger}_{rp}\, y^{u\dagger}_{ts} \,
  Q^{\dagger(5)}_{q^2u^2H^2}
+\,   y^d_{pr}\,  y^d_{st}\, Q^{(5)}_{q^2d^2H^2}
+\, y^e_{pr}\,  y^e_{st} \, Q^{(3)}_{l^2e^2H^2} \nonumber\\
& +\,2\,\Big[ y^e_{pr}\,  y^{u\dagger}_{st}\, Q^{(5)}_{lequH^2}
  +\, y^d_{pr}\,  y^{u\dagger}_{st}\, Q^{(5)}_{q^2udH^2}
  +\, y^e_{pr}\,  y^d_{st}\, Q^{(3)}_{leqdH^2}\Big]\,\Big\} \;,
\nonumber
\end{flalign}\vspace*{-0.6cm}
\begin{flalign}                 
Q^{(3)}_{\psi^4H^2}
\equiv& {H^\dagger}^k[({\Psi}^4)_{HH}]^j_kH_j-\frac{1}{2}
Q^{(1)}_{\psi^4H^2}
=\nonumber\\
&
\frac{1}{2}\,
 {\displaystyle \sum_{prst}}\Big\{
 \!\!{ 
-y^{d\dagger}_{sr}\,  y^d_{pt}\,
        \left( \frac{1}{6}\,Q^{(2)}_{q^2d^2H^2}
        \,+\, Q^{(4)}_{q^2d^2H^2}\right)\,+\,
y^{u\dagger}_{sr}\,  y^u_{pt}\,
        \left(\frac{1}{6}\,Q^{(2)}_{q^2u^2H^2}
        \,+\, Q^{(4)}_{q^2u^2H^2}\right)}\, -\,        
\frac{1}{2}\, y^{e\dagger}_{sr}\,  y^e_{pt}\,
        Q^{(2)}_{l^2e^2H^2} \nonumber\\
& \;+\;\left[  \, { - y^e_{pr}\,  y^u_{st} \, Q^{(2)}_{lequH^2}
\, -\,
   y^u_{pr}\,  y^d_{st} \,\boldmath{ Q^{(2)}_{q^2udH^2}}}
\,+\,
 y^e_{pr}\,  y^{d\dagger}_{ts} \, Q^{(2)}_{leqdH^2}\;+\,{\rm h.c.}\right]
                                \Big\} \;,
\label{eq:qclass18}       
\end{flalign}\vspace*{-0.6cm}
\begin{flalign}                 
 Q^{(4)}_{\psi^4H^2} \equiv & ({\Psi}^4)_{BB} (H^\dagger H) =
 g'^2{\displaystyle \sum_{prst}}
\Big\{
    {\displaystyle\sum_{f\in \{q,l,u,d,e\}}}
    Y_f^2 Q^{(1)}_{f^4H^2}
        +2 {\displaystyle\sum_{f\in \{q,l\}}}{\displaystyle\sum_{f'\in \{u,d,e\}}}
      Y_f Y_f' Q^{(1)}_{f^2{f'}^2H^2}\nonumber \\
      &
+ 2\, Y_q\, Y_l\, Q^{(1)}_{l^2 q^2 H^2}
      +2\Big[ Y_e Y_u Q^{(1)}_{e^2u^2H^2} + Y_e Y_d Q^{(1)}_{e^2d^2H^2}+ Y_u Y_d Q^{(1)}_{u^2d^2H^2}\Big]
        \Big\}\delta_{pr}\delta_{st} \;,
\nonumber
\end{flalign}\vspace*{-0.6cm}
\begin{flalign}                 
 Q^{(5)}_{\psi^4H^2} \equiv & ({\Psi}^4)_{WW} (H^\dagger H) =
\frac{g^2}{4} {\displaystyle \sum_{prst}}
 \Big\{
\big[Q^{(3)}_{q^4H^2}
  + 2 Q^{(3)}_{l^2q^2H^2}\big]   
\delta_{pr}\delta_{st}
+ Q^{(1)}_{l^4H^2}\left(2\delta_{pt}\delta_{sr}-\delta_{pr}\delta_{st}\right)
\Big\} \;,
\nonumber
\end{flalign}\vspace*{-0.6cm}
\begin{flalign}                 
 Q^{(6)}_{\psi^4H^2} \equiv & ({\Psi}^4)_{GG} (H^\dagger H) =
g_s^2{\displaystyle \sum_{prst}} \Big\{
2 \big[Q^{(3)}_{q^2d^2H^2}+Q^{(3)}_{q^2u^2H^2} +Q^{(2)}_{u^2d^2H^2}\Big]
\delta_{pr}\delta_{st} \nonumber\\
&+
\frac{1}{2} \Big[Q_{u^4H^2} +Q_{d^4H^2}  
  \Big]\left(\delta_{pt}\delta_{sr}-\frac{1}{3}\delta_{pr}\delta_{st}\right)
+\frac{1}{2} Q^{(1)}_{q^4H^2}
\left(\frac{1}{2}\delta_{pt}\delta_{sr}
-\frac{1}{3}\delta_{pr}\delta_{st}\right)
+\frac{1}{4}  Q^{(3)}_{q^4H^2} \delta_{pt}\delta_{sr}
\Big\} \;,
\nonumber
\end{flalign}\vspace*{-0.6cm}
\begin{flalign}                 
  Q^{(7)}_{\psi^4H^2}
  \equiv & ({\Psi}^4)^I_{WB} (H^\dagger\tau^I H) =
 \frac{gg'}{2}
      {\displaystyle \sum_{prst}} \Big\{
      {\displaystyle\sum_{f\in \{q,e,u,d\}}} Y_f
        \Big[ Q^{(2)}_{l^2f^2H^2}+Q^{(2)}_{q^2f^2H^2}\Big]
        + Y_l \Big[Q^{(2)}_{l^4H^2}+Q^{(4)}_{l^2q^2H^2}\Big]
     \Big\}\delta_{pr}
     \delta_{st} \;,
\nonumber
\end{flalign}
together with the hermitian conjugate of $Q^{(2)}_{\psi^4 H^2}$.  The
operators $Q^{(1)}_{\psi^4H^2}$--$Q^{(3)}_{\psi^4H^2}$ are generated
by the use of the Higgs EOM directly in operators in class $H^4 D^4$
and $H^2 D^6$, see Eqs.~\eqref{eq:r9}--\eqref{eq:r13}, and in the
rotation of operators in class $X^2 D^4$, as in
Eqs.~\eqref{eq:r30}--\eqref{eq:r31}.
$Q^{(4)}_{\psi^4H^2}$--$Q^{(7)}_{\psi^4H^2}$ arise from direct
application of the EOM for the gauge bosons in operators of class
$X^2 H^2 D^2$, in particular the EOM of $B^{\mu\nu}$ in
$R^{(9)}_{B^2H^2D^2}$~\eqref{eq:r41}, the EOM of $W^{\mu\nu}$ in
$R^{(9)}_{W^2H^2D^2}$~\eqref{eq:r50}, the one for $G^{\mu\nu}$ in
$R^{(9)}_{G^2H^2D^2}$~\eqref{eq:r63}, and the EOM's for $B^{\mu\nu}$
and $W^{\mu\nu}$ in $R^{(13)}_{BWH^2D^2}$~\eqref{eq:r76}.
Here again, for convenience, when writing the expression of
$Q^{(4)}_{\psi^4H^2}$ in terms of operators in M8B,  we have added a superscript
$(1)$ in  $Q^{(1)}_{f^4 H^2}$,  where $f = u,d,e$; and in
$Q^{(1)}_{e^2u^2H^2}$ and $Q^{(1)}_{e^2d^2H^2}$.
This minimal change of labeling is reflected also when we
list the operators in class 18 in Table ~\ref{tab:uniclass18}.


\item {\bf Class} {\boldmath$19:\psi^4X$:} the eight universal
  operators in this class are formed by the contraction of the
  four-fermion tensor currents in Eqs.~\eqref{eq:4ft1}--\eqref{eq:4ft3}
  with a gauge strength tensor
\begin{flalign}  
  Q^{(1)}_{\psi^4W}\equiv ({\Psi}^4)^{I\mu\nu}_{WW} W_{\mu\nu}^I =&
  \frac{g^2}{2} {\displaystyle \sum_{prst}}
  \Big\{ Q^{(5)}_{l^2q^2W}\,\delta_{pr}  \delta_{st} \,{ -}\,
    \Big[Q^{(2)}_{l^4 W}+ \frac{1}{3} Q^{(2)}_{q^4 W} + 2 Q^{(4)}_{q^4 W}\Big]
  \delta_{pt}\delta_{rs} \Big\} \;,
\nonumber
\end{flalign}\vspace*{-0.6cm}
\begin{flalign}                         
Q^{(2)}_{\psi^4W}\equiv ({\Psi}^4)^{I\mu\nu}_{WW} \widetilde W_{\mu\nu}^I =&
  \frac{g^2}{2} {\displaystyle \sum_{prst}}
  \Big\{ Q^{(6)}_{l^2q^2W}\,\delta_{pr}  \delta_{st} \,{ +}\,
    \Big[Q^{(1)}_{l^4 W}+ \frac{1}{3} Q^{(1)}_{q^4 W} + 2 Q^{(3)}_{q^4 W}\Big]
  \delta_{pt}\delta_{rs} \Big\} \;,
\nonumber
\end{flalign}\vspace*{-0.6cm}
\begin{flalign}                         
Q^{(3)}_{\psi^4W}\equiv ({\Psi}^4)^{I\mu\nu}_{WB} W_{\mu\nu}^I =&
  \frac{g g'}{2} {\displaystyle \sum_{prst}}
  \Big\{
  -Y_l\Big[Q^{(1)}_{l^4W}+Q^{(1)}_{l^2 q^2W}\Big] -Y_q\Big[Q^{(1)}_{q^4W}-Q^{(3)}_{l^2 q^2W}\Big]\nonumber \\
  & + {\displaystyle\sum_{f\in \{u,d,e\}}}Y_f
  \Big[Q^{(1)}_{l^2f^2W} +
  Q^{(1)}_{q^2f^2W}\Big]\Big\}\delta_{pr}\delta_{st} \;,
\nonumber
\end{flalign}\vspace*{-0.6cm}
\begin{flalign}                           
Q^{(4)}_{\psi^4W}\equiv ({\Psi}^4)^{I\mu\nu}_{WB} \widetilde W_{\mu\nu}^I =&
  \frac{g g'}{2} {\displaystyle \sum_{prst}}
  \Big\{-Y_l\Big[Q^{(2)}_{l^4W}+Q^{(2)}_{l^2 q^2W}\Big] -Y_q\Big[Q^{(2)}_{q^4W}-Q^{(4)}_{l^2 q^2W}\Big] \nonumber\\
  & + {\displaystyle\sum_{f\in \{u,d,e\}}}Y_f
  \Big[Q^{(2)}_{l^2f^2W} +
  Q^{(2)}_{q^2f^2W}\Big]\Big\}\delta_{pr}\delta_{st} \;,
  \nonumber
\end{flalign}\vspace*{-0.6cm}
\begin{flalign}                         
  Q^{(1)}_{\psi^4G}
  \equiv ({\Psi}^4)^{A\mu\nu}_{GG} G_{\mu\nu}^A =&
g_s^2 {\displaystyle \sum_{prst}}
\Big\{2\Big[Q^{(5)}_{q^2u^2G}+Q^{(5)}_{q^2d^2G}+Q^{(5)}_{u^2d^2G}\Big]\delta_{pr}\delta_{st} \label{eq:qclass19}       
\\
      &  - \Big[\frac{1}{2}Q^{(2)}_{q^4G}+\frac{1}{2}Q^{(4)}_{q^4G}  -
        Q^{(2)}_ {u^4G}  - Q^{(2)}_{d^4G}
        \Big]\delta_{pt}\delta_{sr}\Big\} \;,
\nonumber
\end{flalign}\vspace*{-0.6cm}
\begin{flalign}                         
Q^{(2)}_{\psi^4G}\equiv ({\Psi}^4)^{A\mu\nu}_{GG} \widetilde G_{\mu\nu}^A =&
g_s^2 {\displaystyle \sum_{prst}}
\Big\{2\Big[Q^{(6)}_{q^2u^2G}+Q^{(6)}_{q^2d^2G}+Q^{(6)}_{u^2d^2G}\Big]\delta_{pr}\delta_{st} \nonumber \\
      & + \Big[\frac{1}{2}Q^{(1)}_{q^4G}+\frac{1}{2}Q^{(3)}_{q^4G} -
        Q^{(1)}_{u^4G} - Q^{(1)}_{d^4G}
        \Big]\delta_{pt}\delta_{sr}\Big\} \;,
\nonumber
\end{flalign}\vspace*{-0.6cm}
\begin{flalign}                         
Q^{(3)}_{\psi^4G}\equiv ({\Psi}^4)^{A\mu\nu}_{GB} G_{\mu\nu}^A =&
-\, g_s g' {\displaystyle \sum_{prst}}
\Big\{{\displaystyle\sum_{f\in \{lq\}}}{\displaystyle\sum_{f'\in \{q,u,d\}}} Y_f Q^{(1)}_{f^2 {f'}^2G}
+ Y_e\Big[-Q^{(1)}_{q^2e^2G}+Q^{(1)}_{e^2u^2G}+Q^{(1)}_{e^2d^2G}\Big]  \nonumber \\ &
+ Y_u\Big[Q^{(1)}_{u^4G}-Q^{(3)}_{q^2u^2G}+Q^{(1)}_{u^2d^2G}\Big]
+ Y_d\Big[Q^{(1)}_{d^4G} -Q^{(3)}_{q^2d^2G}+Q^{(3)}_{u^2d^2G}\Big]
\Big\}\delta_{pr}\delta_{st} \;,
\nonumber
\end{flalign}\vspace*{-0.6cm}
\begin{flalign}                         
Q^{(4)}_{\psi^4G}\equiv ({\Psi}^4)^{A\mu\nu}_{GB} \widetilde G_{\mu\nu}^A =& 
-\, g_s g' {\displaystyle \sum_{prst}}
\Big\{{\displaystyle\sum_{f\in \{lq\}}}{\displaystyle\sum_{f'\in \{q,u,d\}}} Y_f Q^{(2)}_{f^2 {f'}^2G}
+ Y_e\Big[-Q^{(2)}_{q^2e^2G}+Q^{(2)}_{e^2u^2G}+Q^{(2)}_{e^2d^2G}\Big]\nonumber \\ &
+ Y_u\Big[Q^{(2)}_{u^4G}-Q^{(4)}_{q^2u^2G}+Q^{(2)}_{u^2d^2G}\Big]
+ Y_d\Big[Q^{(2)}_{d^4G} - Q^{(4)}_{q^2d^2G}+Q^{(4)}_{u^2d^2G}\Big]
\Big\}\delta_{pr}\delta_{st} \;.
\nonumber
\end{flalign}
They are all generated by the direct application of the EOM for the
weak and strong gauge bosons in operators in class $X^3 D^2$. In
particular $Q^{(1)}_{\psi^4W}$--$Q^{(4)}_{\psi^4W}$ arise when using
the EOM of $W^{\mu\nu}$ in the operators $R^{(1,2)}_{W^3 D^2}$ in
Eqs.~\eqref{eq:r14}--\eqref{eq:r15}, and $R^{(1,2)}_{BW^2 D^2}$
in~\eqref{eq:r22}--\eqref{eq:r23}, respectively.  Equivalently,
$Q^{(1)}_{\psi^4G}$--$Q^{(4)}_{\psi^4G}$ arise when using the EOM of
$G^{\mu\nu}$ in operators $R^{(1,2)}_{G^3 D^2}$
(\eqref{eq:r18}--\eqref{eq:r19}) and $R^{(1,2)}_{BG^2 D^2}$
(\eqref{eq:r26}--\eqref{eq:r27}).

\item {\bf Class} {\boldmath$20:\psi^4HD$:} there are four universal
  four-fermion operators genrerated by the contraction of the
  gauge-Higgs fermion currents in Eqs.~\eqref{eq:4fhb}
  and~\eqref{eq:4fhw} with the derivative of a Higgs field
\begin{flalign}  
  Q^{(1)}_{\psi^4HD}\equiv & 
  ({\Psi^4})_{BH}^{\mu j} D_\mu H_j =
 -i\, g'{\displaystyle \sum_{prst}}
  \,
    {\displaystyle \sum_{ f \in \{q,l,u,d,e\}}} 
\!\!\!\! Y_f\,\Big\{{ -}
  y^{u\dagger}_{ts}\, Q^{\dagger(1)}_{f^2quHD}
  \, + \,
  \, y^d_{st}\, Q^{(1)}_{f^2qdHD} , + \,   y^e_{st}\,
  Q^{(1)}_{f^2leHD} \Big\}\delta_{pr}\; ,\nonumber
\end{flalign}\vspace*{-0.6cm}
\begin{flalign}                         
  Q^{(2)}_{\psi^4HD}\equiv &
({\Psi^4})_{WH}^{I\mu j}(\tau^I)^k_j D_\mu H_k =   
{ -i}\,\frac{g}{2}\,{\displaystyle \sum_{prst}}
\Big\{\,  +\,
       y^{u\dagger}_{ts}\,
\Big[Q^{\dagger(2)}_{q^3 uHD}+Q^{\dagger(3)}_{l^2quHD}\Big] \label{eq:qclass20}       
\\
&   + \, y^d_{st}\, \Big[Q^{(2)}_{q^3 dHD}+Q^{(3)}_{l^2q d HD}\Big]
     +   y^e_{st}\,\Big[Q^{(2)}_{l^3 eHD}+Q^{(3)}_{leq^2HD}\Big]
\,\Big\}\delta_{pr} \;,
\nonumber
\end{flalign}
and their hermitian conjugates. They are generated by applying the EOM
for the electroweak gauge bosons and the Higgs in the four operators
of class $X H^2 D^2$: $R^{(1,2)}_{BH^2 D^4}$
(\eqref{eq:r81}--\eqref{eq:r82}) and $R^{(1,2)}_{WH^2 D^4}$
(\eqref{eq:r84}--\eqref{eq:r85}).
Notice that, to keep the notation compact, we took the liberty of
reordering the fermion labeling for some operators in  the first equation.
In particular, in the case of $Q^{(1)}_{f^2leHD}$, when $f=q$, the operator
needs to be identified with $Q^{(1)}_{leq^2HD}$  in Table~\ref{tab:uniclass20}. 

\item {\bf Class} {\boldmath$21:\psi^4D^2$:} Finally, there are four
  universal four-fermion operators generated directly by the
  contraction of the derivatives of two fermion currents in
  Eqs.~\eqref{eq:4fd1}--\eqref{eq:4fd4}
\begin{flalign}
  Q^{(1)}_{\psi^4D^2} \equiv  
({\rm D}{\Psi}^4)_{HH}=&
 {\displaystyle \sum_{prst}}\Big\{
 -\!\!\!\!{\displaystyle\sum_{f \in \{u,d\}}}
         y^{f\dagger}_{pt}\,  y^f_{sr}\,
        \big(\frac{1}{6}\,Q^{(1)}_{q^2f^2D^2}
        \,+\, Q^{(3)}_{q^2f^2D^2}\Big)\, -\,
        \frac{1}{2}\, y^{e\dagger}_{pt}\,  y^e_{sr}\,
        Q^{(1)}_{l^2e^2D^2}\nonumber\\
& \;+\;\left[  \,  -\, y^e_{pr}\,  y^u_{st} \, Q^{(1)}_{lequD^2}
\, +\,
   y^u_{pr}\,  y^d_{st} \, Q^{(1)}_{q^2udD^2}
\,+\,
 y^e_{pr}\,  y^{d\dagger}_{st} \, Q^{(1)}_{leqdD^2}\;+\,{\rm h.c.}\right]
                                \Big\}\;,
 \nonumber
\end{flalign}\vspace*{-0.6cm}
\begin{flalign}                         
  Q^{(2)}_{\psi^4D^2} \equiv  
    ({\rm D}{\Psi}^4)_{BB} =& 
    {g'}^2 {\displaystyle \sum_{prst}} \Big\{
    {\displaystyle\sum_{f\in \{q,l,u,d,e\}}}
    Y_f^2 Q^{(1)}_{f^4D^2}
        +2 {\displaystyle\sum_{f\in \{q,l\}}}{\displaystyle\sum_{f\in \{u,d,e\}}}
      Y_f Y_f' Q^{(1)}_{f^2{f'}^2D^2}\nonumber \\
      &+ 2\, Y_q\, Y_l\, Q^{(1)}_{l^2 q^2 D^2}
      +2\Big[ Y_e Y_u Q^{(1)}_{e^2u^2D^2} + Y_e Y_d Q^{(1)}_{e^2d^2D^2}+ Y_u Y_d Q^{(1)}_{u^2d^2D^2}\Big]
        \Big\}\delta_{pr}\delta_{st} \;,
\label{eq:qclass21}       
\end{flalign}\vspace*{-0.6cm}
\begin{flalign}                         
Q^{(3)}_{\psi^4D^2} \equiv  
({\rm D}{\Psi}^4)_{WW}=&
\frac{g^2}{4} {\displaystyle \sum_{prst}}
 \Big\{
\big[Q^{(3)}_{q^4D^2}
  + 2 Q^{(3)}_{l^2q^2D^2}\big]   
\delta_{pr}\delta_{st}+ Q^{(1)}_{l^4D^2}\left(2\delta_{pt}\delta_{sr}-\delta_{pr}\delta_{st}\right)
\Big\} \;,
\nonumber
\end{flalign}\vspace*{-0.6cm}
\begin{flalign}                         
Q^{(4)}_{\psi^4D^2} \equiv  
  ({\rm D}{\Psi}^4)_{GG} =&
\,g_s^2 {\displaystyle \sum_{prst}}
\Big\{2\,\Big[Q^{(3)}_{q^2u^2D^2}+Q^{(3)}_{q^2d^2D^2}+Q^{(3)}_{u^2d^2D^2}\Big]
\delta_{pr}\delta_{st} \nonumber \\
&\hspace*{-2cm}+
\frac{1}{2} \Big[Q^{(1)}_{u^4D^2} +Q^{(1)}_{d^4D^2}  
  \Big]\left(\delta_{pt}\delta_{sr}-\frac{1}{3}\delta_{pr}\delta_{st}\right)
+\frac{1}{2} Q^{(1)}_{q^4D^2}
\left(\frac{1}{2}\delta_{pt}\delta_{sr}
-\frac{1}{3}\delta_{pr}\delta_{st}\right)
+\frac{1}{4}  Q^{(3)}_{q^4D^2} \delta_{pt}\delta_{sr}
\Big\} \;.
\nonumber
\end{flalign}
They,  respectively, originate from applying the EOM for the Higgs
in $R^{(1)}_{H^2D^6}$ (\eqref{eq:r13}), for the hypercharge gauge boson
in $R^{(1)}_{B^2D^4}$ (\eqref{eq:r30}), for the weak gauge boson
in $R^{(1)}_{W^2D^4}$ (\eqref{eq:r31}), and for the gluon
in $R^{(1)}_{G^2D^4}$ (\eqref{eq:r32}).
Notice that, for convenience, in the equation for $Q^{(2)}_{\psi^4D^2}$,
we have  a superscript of $(1)$  in $Q^{(1)}_{e^4 D^2}$ to
to the M8B $Q_{e^4 D^2}$ operator.
We have included this minimal change of labeling when listing the
operators of this class in Table~\ref{tab:uniclass21}.

\end{itemize}

\section{Some simple universal UV completions}
\label{sec:uv}

{

  Here we briefly discuss a few existing models which match onto the
  universal basis. For each model we point out the bosonic operators which are
  generated in the universal basis and specified those which are rotated to the
  fermionic basis in this work.  \medskip

  A straightforward way to construct universal extensions of the SM is
  to enlarge its scalar sector with new scalar fields which do not
  couple to the SM fermions.  The simplest bosonic UV completion is
  therefore that of the SM supplemented by a real singlet scalar $S$
\begin{eqnarray}
  \Delta\mathcal L=\frac{1}{2}(\partial_\mu
  S)^2-\frac{1}{2}M^2S^2-A|H|^2S-\frac{1}{2}k|H|^2S^2-\frac{1}{3!}\mu
  S^3-\frac{1}{4!}\tilde \lambda_S S^4 \;.
  \label{eq:uv2}
\end{eqnarray}
After tree level integration of the heavy $S$, the low energy
effective theory contains the following operators up to dimension
eight~\cite{deBlas:2017xtg,Ellis:2023zim,Corbett:2015lfa}
\begin{eqnarray}
  \Delta {\mathcal L}_{\rm eff}
&=&\frac{A^2}{2M^2}|H|^4+\frac{A^2}{2M^4}\partial_\mu
    |H|^2\partial^\mu
    |H|^2+\frac{A^2}{2M^4}\left(\frac{A\mu}{3M^2}-k\right)|H|^6 
    \nonumber\\
&+&\frac{A^2}{2M^6}\left(k^2-\frac{A^2\tilde\lambda_S}{12M^2}-\frac{A\mu
    k}{M^2}\right)Q_{H^8}+\frac{2A^2}{M^6}\left(\frac{A\mu}{2M^2}-k^2\right)|H|^2\partial_\mu
    |H|^2\partial^\mu |H|^2+\frac{A^2}{2M^6}\Box|H|^2\Box|H|^2
    \;.
\end{eqnarray}
The last two dimension-eight operators can be written in our basis by
the following relations:
\begin{eqnarray}
|H|^2\partial_\mu |H|^2\partial^\mu
  |H|^2&=&R^{(1)}_{H^6D^2}+R^{(2)}_{H^6D^2}+2Q^{(1)}_{H^6} \;,
  \label{eq:opuv1}
  \\
  \Box|H|^2\Box|H|^2&=&4Q_{H^4}^{(3)}+4\left(R^{(5)}_{H^4D^4}+R^{(6)}_{H^4D^4}\right)+R^{(8)}_{H^4D^4}
  +2R^{(9)}_{H^4D^4}+R^{(10)}_{H^4D^4} \;.
\label{eq:opuv2}  
\end{eqnarray}
From the results presented in Appendix~\ref{app:rotbosfer}, we can see
that the rotation of the operator in Eq.~\eqref{eq:opuv1} to M8B only
generates one fermionic universal operator, the real part of
$Q^{(1)}_{\psi^2H^5}$ (see Eqs.~\eqref{eq:r1} and ~\eqref{eq:r2}),
while the operator in Eq.~\eqref{eq:opuv2} generates a linear
combination of five fermionic operators, the real parts of
$Q^{(1)}_{\psi^2H^5}$, $Q^{(1)}_{\psi^2 H^3 D^2}$, and
$Q^{(2)}_{\psi^4 H^2}$, together with $Q^{(1)}_{\psi^4 H^2}$, and
$Q^{(3)}_{\psi^4 H^2}$ ((see Eqs.~\eqref{eq:r5},~\eqref{eq:r6}, and
~\eqref{eq:r8}--~\eqref{eq:r10}).  \medskip

Another possibility is the addition of a scalar $SU(2)_L$ triplet
field $\phi^a$ with Y=0.  In this case, the new terms in
lagrangian read
\begin{eqnarray}
  \Delta{\mathcal L}
  =\frac{1}{2}(D_\mu\phi^a)(D^\mu\phi^a)-\frac{1}{2}M^2(\phi^a)^2+\kappa
  H^\dagger\sigma^a H\phi^a-\lambda_{\phi
    H}(\phi^a)^2|H|^2-\lambda_\phi|\phi^a|^4 \;.
\end{eqnarray}
The low energy effective theory of this model is, up to
dimension-eight operators in terms of our basis can be written as
~\cite{Ellis:2023zim, Corbett:2024evt, Corbett:2021eux},
\begin{eqnarray}
  \Delta{\mathcal L}_{eff}&=&\frac{\kappa^2}{8M^2}|H|^4-\frac{\lambda_{\phi H}\kappa^2}{4M^4}|H|^6+\frac{\kappa^2}{8M^4}\left[(D_\mu H)^\dagger H(D_\mu H)^\dagger H+h.c.\right]+\frac{\kappa^2}{2M^4}|H|^2|D_\mu H|^2-\frac{\kappa^2}{4M^4}|(D_\mu H)^\dagger H|^2\nonumber\\
  &&-\frac{\kappa^2}{16M^6}\left(\frac{\lambda_\phi\kappa^2}{M^2}-8\lambda_{\phi H}^2\right)Q_{H^8}+\frac{\lambda_{\phi H}\kappa^2}{2M^6}\left(R^{(1)}_{H^6D^2}+R^{(2)}_{H^6D^2}+2Q^{(2)}_{H^6}\right)\\
  &&+\frac{\kappa^2}{8M^6}\left(8Q^{(1)}_{H^4}-4Q^{(3)}_{H^4}+4R^{(1)}_{H^4D^4}+4R^{(4)}_{H^4D^4}+4R^{(7)}_{H^4D^4}+R^{(8)}_{H^4D^4}-2R^{(9)}_{H^4D^4}+R^{(10)}_{H^4D^4}\right)\nonumber
  \label{eq:uv3} 
\end{eqnarray}
In this case, from the Appendix~\ref{app:rotbosfer} we note that the
rotation of the operators to M8B generates the following fermionic
operators, the real parts of $Q^{(1)}_{\psi^2H^5}$,
$Q^{(2)}_{\psi^2 H^3 D^2}$, and $Q^{(2)}_{\psi^4 H^2}$; together with
$Q^{(1)}_{\psi^4 H^2}$, and $Q^{(3)}_{\psi^4 H^2}$ (see
Eqs.~\eqref{eq:r1}--~\eqref{eq:r3}, ~\eqref{eq:r6}, and
~\eqref{eq:r9}--~\eqref{eq:r12} ).\medskip

Next we consider a scenario presenting a hidden sector where its
particles are not charged under the SM gauge group. In the kinetic
mixing model, the hiden sector exhibits a $U(1)_X$ gauge symmetry and
possesses a gauge boson $V_\mu$ which interacts with the SM via
\begin{eqnarray}
  \Delta\mathcal{
    L}=-\frac{1}{4}V_{\mu\nu}V^{\mu\nu}+\frac{1}{2}M^2V_\mu
  V^\mu-\frac{k}{2}B_{\mu\nu}V^{\mu\nu} \;.
\end{eqnarray}
%
For heavy $V_{\mu}$ we can integrate it out at tree level and obtain
the following dimension-six and -eight operators~\cite{Hays:2020scx,
  Corbett:2024evt}
\begin{eqnarray}
\Delta {\mathcal L}_{\rm eff} =
  -\frac{k^2}{2M^2}(\partial_\nu B^{\mu\nu})(\partial^\alpha
 B_{\mu\alpha})+\frac{k^2}{2M^4}(\partial_\nu
 B^{\mu\nu})(\partial^2\partial^\alpha B_{\mu\alpha}) \;.
\end{eqnarray}
The dimension eight-operator is identified with the operator
$R^{(1)}_{B^2D^4}$ in our basis, which is associated to seven fermionic
universal operators in M8B:
the real parts of 
$Q^{(1)}_{\psi^2H^5}$ and $Q^{(2)}_{\psi^4 H^2}$,
together with  $Q^{(1)}_{\psi^4 H^2}$, $Q^{(3)}_{\psi^4 H^2}$, $Q^{(4)}_{\psi^4 D^2}$,
$Q^{(1)}_{\psi^2 H^2 D^3}$, and $Q^{(2)}_{\psi^2 H^4,D}$,
as can be seen in Eq.~\eqref{eq:r30}.
\medskip

At large energy, composite Higgs models possess a strongly interacting
sector that is naturally connected to the SM bosons.  Ergo, possible
high-energy resonances can give rise to a plethora of bosonic
low-energy effective operators depending on the spectrum in the UV
region. As an illustration, let us consider the minimal model based on
the coset $SO(5)/SO(4)$~\cite{Kaplan:1983fs, Agashe:2004rs,
  Panico:2015jxa} and consider a vector resonance ($\rho_\mu$) that
transforms in the adjoint of $SO(4)$. In this scenario many operators
are generated and we will list a few of them. \medskip

The formalism develop by Coleman, Callan, Wess and
Zumino~\cite{Callan:1969sn} allow us to write down the allowed
interactions of this resonance~\cite{Giudice:2007fh,
  Contino:2011np}. Denoting by $\Pi = h^a T^a$ where $T^a$ are the
$SO(5)$ broken generators and $h^a$ the Goldstone bosons, we define
$U = \exp(i \Pi)$.  For simplicity, we assume that the SM gauge group
satisfies $G_{\rm SM} \subset SO(4)$.  In order to write down the
$\rho_\mu$ interactions we need the building blocks ${\mathcal D}_\mu$
and ${\mathcal E}_\mu$:
\begin{eqnarray}
  U^\dagger (\partial_\mu + i A_\mu) U \equiv  
i {\mathcal D}_\mu^A T^A +i {\mathcal E}_\mu^a T^a = i {\mathcal D}_\mu
+ i {\mathcal E}_\mu \;,
\end{eqnarray}
with $T^A$ being the unbroken generators and $A_\mu$ the SM gauge
fields. The lowest order terms of ${\mathcal D_\mu}$ and
${\mathcal E}_\mu$ are
\begin{eqnarray}
   {\mathcal D}_\mu &=& D_\mu \Pi - \frac{1}{6} \left[ \Pi,
                        \Pi~\Dfb ~\Pi \right] + ... \;
  \\
   {\mathcal E}_\mu &=& A_\mu - \frac{i}{2} \Pi~\Dfb~\Pi + ... \;.
\end{eqnarray}

The most general two derivative $SO(5)$-invariant action for
the $\rho_\mu$ is
\begin{eqnarray}
m_\rho^4  \Delta {\mathcal L}^{(2)} = m_\rho^2 {\mathcal D}^A_\mu {\mathcal
  D}^\mu_A - \frac{1}{4} \rho_{\mu\nu} \rho^{\mu\nu} + \frac{1}{2} m_\rho^2 (
\rho_\mu - {\mathcal E}_\mu)^2 \;,
\end{eqnarray}
where
$\rho_{\mu\nu} = \partial_\mu \rho_\nu - \partial_\nu \rho_\mu + i
[\rho_\mu,\rho_\nu]$. Assuming that $\rho_\mu$ is heavy we can
integrate it out to obtain~\cite{Giudice:2007fh}
\begin{eqnarray}
\Delta {\mathcal L}_{\rm eff} = - \frac{1}{4 g_\rho^2 } {\mathcal E}_{\mu\nu}
{\mathcal E}^{\mu\nu} - \frac{1}{2 g_\rho^2} D^\mu {\mathcal
  E}_{\mu\nu} \frac{1}{\partial^2+ m_\rho^2} D_\alpha {\mathcal
  E}^{\alpha\nu}
\label{eq:int1}
\end{eqnarray}
where $g_\rho$ is the coupling constant. Terms containing four or more
derivatives and four or more field strength tensors have been
omitted. The first term of the above equation leads to the
dimension-six operators
\begin{eqnarray}
  \left( H^\dagger \tau^I  ~\Dfb_\mu~H \right) D_\nu W^{I,\mu\nu}
  \;\;\;\hbox{ and}\;\;\;
    \left( H^\dagger   ~\Dfb_\mu~H \right) \partial_\nu B^{\mu\nu} \;.
\end{eqnarray}
On the other hand, the second term of Eq.~(\ref{eq:int1}) gives rise
to the following dimension-six and -eight operators
\begin{align}
\hspace*{1cm}&D_\nu W^{I,\mu\nu} D^\alpha W^{I}{\alpha\nu}\;,  & 
& \partial_\nu B^{\mu\nu}  \partial^\alpha B_{\alpha\nu}\;, & 
\\
&R^{(1)}_{W^2D^4}=D_\nu W^{I,\mu\nu}  D^\beta D_\beta    D^\alpha
W^{I}{\alpha\mu}\;, &
& R^{(1)}_{B^2D^4}
=\partial_\nu B^{\mu\nu} \partial^\beta
\partial_\beta  \partial^\alpha B_{\alpha\nu}\;. &
\label{eq:d8compo}
\end{align}
Upon rotation to M8B the operators in Eq.~\eqref{eq:d8compo}
give rise respectively to two linear combinations
involving ten and seven universal operators respectively given in
Eqs.~\eqref{eq:r31} and\eqref{eq:r30}. \medskip

In addition, possible four derivative $\rho$ interactions can produce genuine
anomalous quartic gauge couplings~\cite{Eboli:2006wa}, {\it i.e.}
anomalous quartic couplings that do not have triple couplings
associated to them. For instance, the following $\rho$
self-interaction
\begin{eqnarray}
  \frac{1}{m_\rho^4}~(\rho_{\mu\nu}\rho^{\mu\nu})(\rho_{\alpha\beta}\rho^{\alpha\beta})
\end{eqnarray}
generates the low-energy effective interaction
\begin{eqnarray}
  \frac{1}{m_\rho^4}~({\mathcal E}_{\mu\nu}{\mathcal
    E}^{\mu\nu})({\mathcal E}_{\alpha\beta}{\mathcal E}^{\alpha\beta})
\end{eqnarray}
that contains the operators
\begin{eqnarray}
Q^{(1)}_{W^4} = (W^I_{\mu\nu}W^{I,\mu\nu}) (W^I_{\alpha\beta}W^{I,\alpha\beta}) 
\;\;\;,\;\;\; 
Q^{(1)}_{W^2 B^2} = (W^I_{\mu\nu}W^{I,\mu\nu}) (B_{\alpha\beta}B^{\alpha\beta}) 
\;\;\;,\;\;\; 
Q^{(1)}_{B^4} = (B_{\mu\nu}B^{\mu\nu}) (B_{\alpha\beta}B^{\alpha\beta}) 
\;. 
\end{eqnarray}

}


\section{Final remarks}
\label{sec:rem}

In the absence of a smoking-gun signal for new physics at the LHC,
EFTs, in particular the SMEFT, have become the standard tool for
model-agnostic BSM explorations. Unfortunately, their power is in some
sense also their weakness as, taken in their greatest generality, the
number of parameters ({\sl i.e.} the number of independent Wilson
coefficients) is prohibitively large already at dimension-six.
Identifying physically motivated hypotheses which reduce the EFT
parameter space while still capturing a large class of BSM theories
presents a motivated route to predictability.  Universality, {\sl
  i.e.}  the assumption that the NP couples dominantly to the Standard
Model bosons, is one such hypothesis. At dimension-six, universality
reduces the dimensionality of the SMEFT parameter space from 2499 to
16, allowing for a constrained effective parametrization of NP
effects~\cite{Wells:2015uba}.\medskip


In this work we have taken the next step by constructing the
dimension-eight SMEFT operator basis which at the high energy matching
scale can be used to encode the effects of universal extensions of the
SM.  To do so we have identified a suitable basis of independent
operators formed with the Standard Model bosonic fields at
dimension-eight.  It contains 175 operators -- that is, the assumption
of Universality reduces the number of independent SMEFT coefficients
at dimension eight from 44807 to 175. 89 of these operators are part
of the general SMEFT dimension-eight basis in the literature; see
Table~\ref{tab:murbos}). Our choice of the additional 86 operators is
presented in Eqs.~\eqref{eq:86bos1}--\eqref{eq:86bos10}.  In the
general dimension-eight SMEFT basis in the literature these 86
operators have been traded for combinations of the {89} bosonic
operators in the basis and additional operators involving fermions.
Thus in universal theories, only a subset of fermionic operators are
generated (see Appendix~\ref{app:ferM8B}) and their Wilson coefficients
have well defined relations: they must be linear combinations of the
86 independent couplings of the universal fermionic operators
presented in Section~\ref{sec:ferm}.\medskip

The drastic reduction of independent parameters implied by the
Universality assumption opens up the possibility of employing it for
quantitative phenomenology { because just a few of them contribute to
  a specific reaction}.  For example, the direct effect of the
dimension-eight universal operators can be seen in the production of
multiple $H$, $W^\pm$ and $Z$. Operators in the classes $X^3H^2$ and
$X^3D^2$ modify the triple gauge boson vertices, so contributing to
diboson production at tree level. Moreover, many classes contribute to
modifications of the quartic coupling among the SM gauge bosons, as
well as to vertices with Higgs and gauge bosons.  In addition, an
interesting subset of operators from the rotated basis are those which
include, after rotation, the Murphy basis operators
$Q^{(i)}_{f^2VH^2D}$ for $i\in 1,3,5$ and $Q^{(i)}_{f^2H^2D^3}$ for
$i\in 1,3,4$. These operators introduce novel kinematics to the Higgs
associated production process \cite{Corbett:2023yhk}.  The
$Q^{(i)}_{f^2VH^2D}$ are generated by rotating
$R^{(5,6)}_{B^2H^2D^2}$, $R^{(5,6)}_{BWH^2D}$, $R^{(3)}_{BH^2D^4}$,
$R^{(1)}_{BW^2D^2}$, and $R^{(5,6)}_{W^2H^2D^2}$. Similarly
$Q^{(i)}_{f^2H^2D^3}$ are generated by rotating $R^{(1)}_{B^2D^4}$ and
$R^{(1)}_{W^2D^4}$. We notice in passing that the results in
Sec.~\ref{sec:uv} show that the simple composite Higgs model there
presented generates both $R^{(1)}_{B^2D^4}$ and $R^{(1)}_{W^2D^4}$,
while $R^{(1)}_{B^2D^4}$ also emerges in the minimal $U(1)_X$
kinetic-mixing scenario.  \medskip

Of course, the phenomenological  program first requires a careful accounting
for the relevant field
redefinitions and other finite renormalization effects since some of
the operators give contributions to the definition of the SM
parameters~\cite{Corbett:2023qtg}. 
%
Moreover, even in a scenario where the high energy model is universal, there
will be non-universal effects at the low energy EFT due to the
renormalization group running~\cite{Wells:2015cre} but controlled by
the universal parameters at the matching scale.  They should also be
taken into account in phenomenological studies.  Furthermore, the rich
phenomenology of dimension-eight operators possesses many constraints
originating from the causality and analyticity of the scattering
amplitudes; see, for instance, references~\cite{Bi:2019phv,
  Yamashita:2020gtt, Chala:2021wpj, Chala:2023jyx,
  Chala:2023xjy}. These bounds define the regions of the parameter
space associated with well-defined ultra-violet completions of the SM.
We leave the quantitative exploration of the phenomenology of
universal SMEFT at dimension-eight for future work.

\acknowledgments

OJPE is partially supported by CNPq grant number 305762/2019-2 and
FAPESP grant 2019/04837-9.  This project is funded by USA-NSF grant
PHY-2210533.
It has also received support from the European Union's
Horizon 2020 research and innovation program under the Marie Sk\l
odowska-Curie grant agreement No 860881-HIDDeN, and Horizon Europe
research and innovation programme under the Marie Sk\l odowska-Curie
Staff Exchange grant agreement No 101086085 -- ASYMMETRY''.  It also
receives support from grants PID2019-105614GB-C21, and ``Unit of
Excellence Maria de Maeztu 2020-2023'' award to the ICC-UB
CEX2019-000918-M, funded by MCIN/AEI/10.13039/501100011033, and from
grant 2021-SGR-249 (Generalitat de Catalunya).

\newpage

\appendix

\section{Rotation of the higher derivative bosonic universal operators
  into fermionic universal operators }
\label{app:rotbosfer}

This appendix contains the explicit expression of the 86 bosonic
operators of universal theories not included in M8B in terms of the
bosonic and fermionic operators in M8B as generated by the EOM.  For
the sake of convenience, we present in Appendix~\ref{app:ferM8B} the
definition of all M8B fermionic operators involved.  For some of the
operators the final expressions in terms of operators in M8B basis may
look rather cumbersome.  This is particularly the case when the Higgs
currents introduced by the application of the EOMs results in bosonic
operators which are not included in M8B and which, therefore, need
further simplification by re-application of IBP, BI and EOMs.
\medskip

Notice that the mass term of the Higgs equation of motion
Eq.~\eqref{eq:eom1} leads to the appearance of operators with
dimension less than eight when we apply the EOM.  For these terms we
did not express the resulting operators in terms of any specific
dimension-six operator basis of those in the literature:
HISZ~\cite{Hagiwara:1993ck, Hagiwara:1996kf},
Warsaw~\cite{Grzadkowski:2010es}, EGGM~\cite{Elias-Miro:2013eta}, or
SILH~\cite{Elias-Miro:2013mua}.  For convenience we define the
following combinations involving terms arising from the Higgs mass
part of the EOM and some Higgs operators which appear recurrently in
the expressions
\begin{displaymath}
\begin{array}{lll}  
  \Delta_0\equiv\lambda \left[v^2 (H^\dagger H)-  2(H^\dagger H)^2\right]\; ,
  &&  \Delta_1\equiv\lambda \left[v^2 (H^\dagger H)^3- 2 Q_{H^8}\right]\; ,\\
  \Delta_2\equiv\lambda \left[v^2 (H^\dagger H)^3- 2 Q_{H^8}\right]+
    2 Q^{(1)}_{H^6} + Q^{(2)}_{H^6}\;, 
  &&
  \Delta_3\equiv\lambda \left[v^2 (H^\dagger H)^3- 2 Q_{H^8}\right]
  +5 Q^{(1)}_{H^6}\;, \\
\Delta_4\equiv\lambda \left[v^2 (H^\dagger H)^3- 2 Q_{H^8}\right]
+4 Q^{(1)}_{H^6}+Q^{(2)}_{H^6}\;, &&
\Delta_5\equiv\lambda \left[v^2 (H^\dagger H)^3- 2 Q_{H^8}\right]
  +3 Q^{(1)}_{H^6} + 2 Q^{(2)}_{H^6}\;.   
\end{array}
\end{displaymath}

  \noindent\underline{\bf\large Operators in the class $H^6 D^2$}
\begin{flalign}
       R^{(1)}_{H^6 D^2}&= \Delta_1 - Q^{(1)}_{\psi^2H^5}\;, 
\label{eq:r1}
\end{flalign}\vspace*{-0.8cm}
\begin{flalign} 
      R^{(2)}_{H^6 D^2}& = \Delta_1 - Q^{\dagger(1)}_{\psi^2H^5}\;.
      \label{eq:r2}\end{flalign}

\noindent\underline{\bf\large Operators in the class $H^4 D^4$}
\begin{flalign}
  R^{(1)}_{H^4 D^4}=& \lambda v^2 (H^\dagger \tau^I H) (D^\mu H^\dagger) \tau^I (D_\mu H) - 2\lambda Q^{(2)}_{H^6}- Q^{(2)}_{\psi^2H^3D^2}\;,
\label{eq:r3}\end{flalign}\vspace*{-0.8cm}
\begin{flalign}   
     R^{(2)}_{H^4 D^4}=& \lambda \Delta_2 +\lambda v^2 
      (H^\dagger( D^\mu H))^2 
      - \lambda Q^{\dagger(1)}_{\psi^2H^5} - Q^{(3)}_{\psi^2H^3D^2}\;,  
\label{eq:r4}\end{flalign}\vspace*{-0.8cm}
\begin{flalign} 
      R^{(3)}_{H^4 D^4}=& \lambda \Delta_2 +\lambda v^2
      ( D^\mu H^\dagger H)^2 -\lambda Q^{(1)}_{\psi^2H^5} -
      Q^{\dagger(3)}_{\psi^2H^3D^2} \;,
\label{eq:r5}\end{flalign}\vspace*{-0.8cm}
\begin{flalign} 
      R^{(4)}_{H^4 D^4}=& \lambda v^2 (H^\dagger \tau^I H) (D^\mu H^\dagger) \tau^I (D_\mu H) - 2\lambda Q^{(2)}_{H^6} - Q^{\dagger(2)}_{\psi^2H^3D^2}\;,
\label{eq:r6}\end{flalign}\vspace*{-0.8cm}
\begin{flalign}
      R^{(5)}_{H^4 D^4}=& \lambda v^2 (H^\dagger H)(D_\mu H^\dagger)( D^\mu H) - 2\lambda Q^{(1)}_{H^6} - Q^{(1)}_{\psi^2H^3D^2} \;,
\label{eq:r7}\end{flalign}\vspace*{-0.8cm}
\begin{flalign}  
      R^{(6)}_{H^4 D^4}=& \lambda v^2(H^\dagger H)(D_\mu H^\dagger)( D^\mu H) - 2\lambda Q^{(1)}_{H^6} - Q^{\dagger(1)}_{\psi^2H^3D^2}\;,
\label{eq:r8}\end{flalign}\vspace*{-0.8cm}
\begin{flalign} 
      R^{(7)}_{H^4 D^4}=& (\Delta_0)^2
      -\lambda v^2 \left(H^\dagger J^\dagger_H+{\rm h.c.}\right) (H^\dagger H) 
+2\lambda\left( Q^{(1)}_{\psi^2H^5} + Q^{\dagger(1)}_{\psi^2H^5} \right) + Q^{(1)}_{\psi^4H^2} \;,
\label{eq:r9}\end{flalign}\vspace*{-0.8cm}
\begin{flalign} 
      R^{(8)}_{H^4 D^4}=& (\Delta_0)^2
      -2\lambda v^2 (H^\dagger H) H^\dagger J^\dagger_H
      +4\lambda Q^{(1)}_{\psi^2H^5} + Q^{(2)}_{\psi^4H^2}\;  ,
\label{eq:r10}\end{flalign}\vspace*{-0.8cm}
\begin{flalign}       
      R^{(9)}_{H^4 D^4}=& (\Delta_0)^2
      -\lambda v^2 \left(J_H H +{\rm h.c.}\right) (H^\dagger H)
      +2\lambda\left(Q^{(1)}_{\psi^2H^5} + Q^{\dagger(1)}_{\psi^2H^5}\right)
      + Q^{(3)}_{\psi^4H^2} + \frac{1}{2}Q^{(1)}_{\psi^4H^2} \; ,
\label{eq:r11}\end{flalign}\vspace*{-0.8cm}
\begin{flalign} 
  R^{(10)}_{H^4 D^4}=& (\Delta_0)^2
  -2\lambda v^2 (H^\dagger H) J_HH 
      +4\lambda Q^{\dagger(1)}_{\psi^2H^5} +
      Q^{\dagger(2)}_{\psi^4H^2} \;.
\label{eq:r12}\end{flalign}

\noindent\underline{\bf\large Operator in the class $H^6 D^2$}
\begin{flalign}
  R^{(1)}_{H^2 D^6}=&-8\lambda^2\Delta_1 +\lambda^2 v^4 (D^\mu H^\dagger D_\mu H)
  -2\lambda^2 v^2 D^\mu(H^\dagger H) D_\mu(H^\dagger H)
  -4\lambda^2 v^2 (H^\dagger H)(D^\mu H^\dagger D_\mu H) \nonumber \\
  & 
  -\lambda v^2 (D^\mu H^\dagger D_\mu J_H^\dagger+ {\rm h.c.})
 -6\lambda^2 v^2 (H^\dagger H)(H^\dagger J_H^\dagger+ {\rm h.c.})
  -4\lambda^2 Q^{(1)}_{H^6}\nonumber  \\
  &
-2\lambda\left( 3 Q^{(1)}_{\psi^2H^3D^2}   +
Q^{(2)}_{\psi^2H^3 D^2} +2 Q^{(3)}_{\psi^2H^3D^2}  + {\rm h.c.}\right)
  +16 \lambda^2\left(Q^{(1)}_{\psi^2H^5} + Q^{\dagger(1)}_{\psi^2H^5}\right)\nonumber  \\
  &
  +2\lambda\left( 3 Q^{(1)}_{\psi^4H^2}   
+Q^{(2)}_{\psi^4H^2}
+Q^{\dagger (2)}_{\psi^4H^2}+ 2 Q^{(3)}_{\psi^4H^2} \right)
+Q^{(1)}_{\psi^4 D^2}  \;.
\label{eq:r13}\end{flalign}

\noindent\underline{\bf\large Operators in class $X^3D^2$}
\begin{flalign}
  R^{(1)}_{W^3 D^2}=&
  -\frac{g^3}{4} \Delta_3 + 2 i g^2  Q^{(1)}_{WH^4D^2} +
\frac{g^2}{4}\left( g'Q^{(1)}_{WBH^4} + gQ^{(1)}_{W^2H^4} \right) 
\nonumber \\&
+\frac{g^3}{8}\left(Q^{(1)}_{\psi^2 H^5}+Q^{\dagger (1)}_{\psi^2 H^5}\right)
-\frac{g^2}{4} \left(  Q^{(2)}_{\psi^2H^4D} + Q^{(4)}_{\psi^2H^4D}\right)
-g Q^{(11)}_{\psi^2WH^2D} + Q^{(1)}_{\psi^4 W}
\;,
\label{eq:r14}\end{flalign}\vspace*{-0.8cm}
\begin{flalign} 
R^{(2)}_{W^3 D^2}=& 2i  g^2  Q^{(2)}_{WH^4D^2} +
\frac{g^2}{4}\left( g'Q^{(2)}_{WBH^4} + gQ^{(2)}_{W^2H^4} \right)
-g Q^{(12)}_{\psi^2WH^2D} + Q^{(2)}_{\psi^4 W} 
\;,\label{eq:r15}\end{flalign}\vspace*{-0.85cm}
\begin{flalign} 
R^{(3)}_{W^3 D^2}=& - g  Q^{(4)}_{W^2H^2D^2}
-\frac{g}{4}\left(g' Q^{(1)}_{W^2BH^2} +g Q^{(1)}_{W^3H^2}\right)
-\frac{1}{2}Q^{(1)}_{\psi^2W^2H(D)} 
\;,
\label{eq:r16}\end{flalign}\vspace*{-0.85cm}
\begin{flalign} 
R^{(4)}_{W^3 D^2}=& -  g Q^{(6)}_{W^2H^2D^2}
-\frac{g}{4}\left(-g' Q^{(2)}_{W^2BH^2} +2 g Q^{(2)}_{W^3H^2} \right)
-Q^{(2)}_{\psi^2W^2H(D)} 
\;,
\label{eq:r17}\end{flalign}\vspace*{-0.85cm}
\begin{flalign}
R^{(1)}_{G^3 D^2}=&Q^{(1)}_{\psi^4G}
\;,
\label{eq:r18}
\end{flalign}\vspace*{-0.85cm}
\begin{flalign} 
R^{(2)}_{G^3 D^2}=& Q^{(2)}_{\psi^4G}
\;,
\label{eq:r19}\end{flalign}\vspace*{-0.85cm}
\begin{flalign} 
R^{(3)}_{G^3 D^2}=&-\frac{1}{2}Q^{(1)}_{\psi^2G^2H(D)}
\;,
\label{eq:r20}\end{flalign}\vspace*{-0.85cm}
\begin{flalign} 
R^{(4)}_{G^3 D^2}=& - Q^{(2)}_{\psi^2G^2H(D)}
\;,
\label{eq:r21}\end{flalign}\vspace*{-0.85cm}
\begin{flalign}
  R^{(1)}_{BW^2 D^2}=&
  i\frac{gg'}{2}Q^{(3)}_{WH^4D^2}  
  -\frac{g'}{2}Q^{(7)}_{\psi^2WH^2D} + \frac{g}{2}Q^{(3)}_{\psi^2WH^2D}
-\frac{gg'}{8}  Q^{(3)}_{\psi^2 H^4 D}
  + Q^{(3)}_{\psi^4W}
  \;,
\label{eq:r22}\end{flalign}\vspace*{-0.85cm}
\begin{flalign}   
R^{(2)}_{BW^2 D^2}=&
i\frac{gg'}{2}Q^{(4)}_{WH^4D^2} 
    -\frac{g'}{2}Q^{(8)}_{\psi^2WH^2D} + \frac{g}{2}Q^{(4)}_{\psi^2WH^2D} + Q^{(4)}_{\psi^4W}
    \;,
\label{eq:r23}\end{flalign}\vspace*{-0.85cm}
\begin{flalign}     
R^{(3)}_{BW^2 D^2}=&-g Q^{(3)}_{WBH^2D^2} + Q^{(1)}_{\psi^2WBH(D)}
\;,
\label{eq:r24}\end{flalign}\vspace*{-0.85cm}
\begin{flalign} 
R^{(4)}_{BW^2 D^2}=& -g Q^{(5)}_{WBH^2D^2} + Q^{(2)}_{\psi^2WBH(D)}
\;,
\label{eq:r25}\end{flalign}\vspace*{-0.85cm}
\begin{flalign}
R^{(1)}_{BG^2 D^2}=& -\frac{g'}{2}Q^{(3)}_{\psi^2GH^2D} + Q^{(3)}_{\psi^4G}
\;,
\label{eq:r26}\end{flalign}\vspace*{-0.85cm}
\begin{flalign} 
R^{(2)}_{BG^2 D^2}=& -\frac{g'}{2}Q^{(4)}_{\psi^2GH^2D} + Q^{(4)}_{\psi^4G}
\;,
\label{eq:r27}\end{flalign}\vspace*{-0.85cm}
\begin{flalign} 
R^{(3)}_{BG^2 D^2}=& Q^{(1)}_{\psi^2GBH(D)}
\;,
\label{eq:r28}\end{flalign}\vspace*{-0.85cm}
\begin{flalign} 
R^{(4)}_{BG^2 D^2}=& Q^{(2)}_{\psi^2GBH(D)}
\;.
\label{eq:r29}\end{flalign}

\noindent\underline{\bf\large Operators in class $X^2D^4$}
\begin{flalign}
  R^{(1)}_{B^2 D^4}=&
  \frac{g^2{g'}^2}{4} \Delta_4 +
  {g'}^2\left(Q^{(1)}_{H^4}-Q^{(2)}_{H^4}\right)
  +\frac{{g'}^2}{8}\left({g'}^2 Q^{(1)}_{B^2H^4}-g^2 Q^{(1)}_{W^2H^4}
  \right)
  +i {g'}^2\left(g' Q^{(1)}_{BH^4D^2}-g Q^{(1)}_{WH^4D^2}\right) \nonumber \\ &
  -\frac{g'}{2} Q^{(1)}_{\psi^2 H^2 D^3}
  +\frac{{g'}^2 g}{4} Q^{(2)}_{\psi^2H^4D}
  -\frac{g^2 {g'}^2}{8}\left(Q^{(1)}_{\psi^2H^5}+Q^{\dagger (1)}_{\psi^2H^5}\right)
  \nonumber \\&
  -\frac{3 {g'}^2}{4}
  \left(- Q^{(1)}_{\psi^4 H^2}
    +Q^{(2)}_{\psi^4 H^2}+Q^{\dagger (2)}_{\psi^4 H^2}
    -2Q^{(3)}_{\psi^4 H^2}\right)
    + Q^{(2)}_{\psi^4 D^2}  
    \;,
 \label{eq:r30}\end{flalign}\vspace*{-0.7cm}
\begin{flalign}    
  R^{(1)}_{W^2 D^4}=& \frac{g^4}{4}\Delta_3+\frac{g^2 {g'}^2}{2}\Delta_5
    -g^2\left(Q^{(1)}_{H^4}+Q^{(2)}_{H^4}-2Q^{(3)}_{H^4}\right) \nonumber \\&
  -\frac{{g}^2}{8}\left(3{g'}^2 Q^{(1)}_{B^2H^4}+g^2 Q^{(1)}_{W^2H^4}
  +4 g g' Q^{(1)}_{WBH^4}\right) 
  -  i g^2 \left(3 g' Q^{(1)}_{BH^4D^2}+g Q^{(1)}_{WH^4D^2}\right)
\nonumber\\ &
-\frac{g}{2} Q^{(2)}_{\psi^2 H^2 D^3}-g^2Q^{(11)}_{\psi^2 WH^2 D}
+\frac{g^2}{4}\left(g Q^{(2)}_{\psi^2 H^4 D}+g Q^{(4)}_{\psi^2 H^4 D}+4g' Q^{(1)}_{\psi^2 H^4 D}
  \right)     
\nonumber   \\&
  -\frac{g^2(g^2+2{g'}^2)}{8} \left(Q^{(1)}_{\psi^2H^5}+Q^{\dagger (1)}_{\psi^2H^5}\right)
  -\frac{3 {g}^2}{4}
  \left(- 3 Q^{(1)}_{\psi^4 H^2}
    +Q^{(2)}_{\psi^4 H^2}+Q^{\dagger (2)}_{\psi^4 H^2}
    + 2Q^{(3)}_{\psi^4 H^2}\right)
    + Q^{(3)}_{\psi^4 D^2}  
\label{eq:r31}
\end{flalign}\vspace*{-0.8cm}
\begin{flalign}     
R^{(1)}_{G^2 D^4}=& Q^{(4)}_{\psi^4D^2} \;.
\label{eq:r32}
\end{flalign}

\noindent\underline{\bf\large Operators in class $X^2H^2D^2$}
\begin{flalign}
R^{(1)}_{B^2 H^2 D^2}=&\lambda v^2 B_{\mu\nu} B^{\mu\nu} (H^\dagger H) - 2 \lambda Q^{(1)}_{B^2H^4} - Q^{(1)}_{\psi^2B^2H}
\;,
\label{eq:r33}\end{flalign}\vspace*{-0.8cm}
\begin{flalign} 
R^{(2)}_{B^2 H^2 D^2}=&
\lambda v^2 B_{\mu\nu} B^{\mu\nu} (H^\dagger H) - 2 \lambda Q^{(1)}_{B^2H^4} - Q^{\dagger(1)}_{\psi^2B^2H}
\;,
\label{eq:r34}\end{flalign}\vspace*{-0.8cm}
\begin{flalign} 
R^{(3)}_{B^2 H^2 D^2}=&\lambda v^2 B_{\mu\nu} \widetilde{B}^{\mu\nu} (H^\dagger H) - 2 \lambda Q^{(2)}_{B^2H^4} - Q^{(2)}_{\psi^2B^2H}
\;,
\label{eq:r35}\end{flalign}\vspace*{-0.8cm}
\begin{flalign} 
R^{(4)}_{B^2 H^2 D^2}=&\lambda v^2 B_{\mu\nu} \widetilde{B}^{\mu\nu} (H^\dagger H) - 2 \lambda Q^{(2)}_{B^2H^4} - Q^{\dagger(2)}_{\psi^2B^2H}
\;,
\label{eq:r36}\end{flalign}\vspace*{-0.8cm}
\begin{flalign} 
  R^{(5)}_{B^2 H^2 D^2}=&-\frac{{g'}^2}{8} \Delta_5
+i \frac{g'}{2}
  Q^{(1)}_{BH^4D^2} +\frac{g'}{8}\left( g' Q^{(1)}_{B^2 H^4}+ g Q^{(1)}_{WB H^4}\right)\nonumber \\
&- \frac{1}{2}\left(Q^{(1)}_{\psi^2BH^2D} + iQ^{(3)}_{\psi^2BH^2D} \right)
-\frac{g'}{4} Q^{(1)}_{\psi^2 H^4 D}
+\frac{{g'}^2}{16} \left(Q^{(1)}_{\psi^2H^5}+Q^{\dagger (1)}_{\psi^2H^5}
\right)
\;,
\label{eq:r37}\end{flalign}\vspace*{-0.7cm}
\begin{flalign} 
R^{(6)}_{B^2 H^2 D^2}=&
-\frac{{g'}^2}{8} \Delta_5 +
i\frac{g'}{2}
  Q^{(1)}_{BH^4D^2} + \frac{g'}{8}\left(g' Q^{(1)}_{B^2 H^4}+ g Q^{(1)}_{WB H^4}\right)\nonumber \\
&- \frac{1}{2}\left(Q^{(1)}_{\psi^2BH^2D} - i Q^{(3)}_{\psi^2BH^2D} \right)
-\frac{g'}{4} Q^{(1)}_{\psi^2 H^4 D}
+\frac{{g'}^2}{16} \left(Q^{(1)}_{\psi^2H^5}+Q^{\dagger (1)}_{\psi^2H^5}
\right)
\;,
\label{eq:r38}\end{flalign}\vspace*{-0.7cm}
\begin{flalign} 
R^{(7)}_{B^2 H^2 D^2}=&
i\frac{g'}{2}
  Q^{(2)}_{BH^4D^2} +\frac{g'}{8}\left(g' Q^{(2)}_{B^2 H^4}+ g Q^{(2)}_{WB H^4}\right)
- \frac{1}{2}\left(Q^{(2)}_{\psi^2BH^2D} + i Q^{(4)}_{\psi^2BH^2D} \right)
\;,
\label{eq:r39}\end{flalign}\vspace*{-0.8cm}
\begin{flalign} 
R^{(8)}_{B^2 H^2 D^2}=&
i \frac{g'}{2}
  Q^{(2)}_{BH^4D^2} +\frac{g'}{8}\left(g' Q^{(2)}_{B^2 H^4}+ g Q^{(2)}_{WB H^4}\right)
- \frac{1}{2}\left(Q^{(2)}_{\psi^2BH^2D} - i Q^{(4)}_{\psi^2BH^2D} \right)
\;,
\label{eq:r40}\end{flalign}\vspace*{-0.8cm}
\begin{flalign} 
R^{(9)}_{B^2 H^2 D^2}=&
\frac{{g'}^2}{4} \Delta_5 
-\frac{{g'}^2}{8} \left(Q^{(1)}_{\psi^2H^5}+Q^{\dagger (1)}_{\psi^2H^5}\right)
+g'Q^{(1)}_{\psi^2H^4D} + Q^{(4)}_{\psi^4H^2}
\;,
\label{eq:r41}\end{flalign}\vspace*{-0.8cm}
\begin{flalign}
R^{(1)}_{W^2 H^2 D^2}=&\lambda v^2 W^I_{\mu\nu} W^{I,\mu\nu} (H^\dagger H) - 2 \lambda Q^{(1)}_{W^2H^4} - Q^{(1)}_{\psi^2W^2H}
\;,
\label{eq:r42}\end{flalign}\vspace*{-0.8cm}
\begin{flalign} 
R^{(2)}_{W^2 H^2 D^2}=&\lambda v^2 W^I_{\mu\nu} W^{I,\mu\nu} (H^\dagger H) - 2 \lambda Q^{(1)}_{W^2H^4} - Q^{\dagger(1)}_{\psi^2W^2H}
\;,
\label{eq:r43}\end{flalign}\vspace*{-0.8cm}
\begin{flalign} 
R^{(3)}_{W^2 H^2 D^2}=& \lambda v^2 W^I_{\mu\nu} \widetilde{W}^{I,\mu\nu} (H^\dagger H) - 2 \lambda Q^{(2)}_{W^2H^4} - Q^{(2)}_{\psi^2W^2H}
\;,
\label{eq:r44}\end{flalign}\vspace*{-0.8cm}
\begin{flalign} 
R^{(4)}_{W^2 H^2 D^2}=&  \lambda v^2 W^I_{\mu\nu} \widetilde{W}^{I,\mu\nu} (H^\dagger H) - 2 \lambda Q^{(2)}_{W^2H^4} - Q^{\dagger(2)}_{\psi^2W^2H}
\;,
\label{eq:r45}\end{flalign}\vspace*{-0.8cm}
\begin{flalign}
R^{(5)}_{W^2 H^2 D^2}=&
-\frac{{g}^2}{8} \Delta_3 +\frac{g}{2}\left(
  i Q^{(1)}_{WH^4D^2} - Q^{(3)}_{WH^4D^2}\right)+\frac{g}{8}\left(
  g'Q^{(1)}_{WB H^4}+  gQ^{(1)}_{W^2 H^4}\right)
\label{eq:r46}
    \\& 
    - \frac{1}{2}\left(Q^{(5)}_{\psi^2WH^2D} + iQ^{(7)}_{\psi^2WH^2D} \right)
    -\frac{g}{8} \left( Q^{(2)}_{\psi^2 H^4 D}+Q^{(4)}_{\psi^2 H^4 D}
-i Q^{(3)}_{\psi^2 H^4 D}
\right)
+\frac{g^2}{16} \left(Q^{(1)}_{\psi^2H^5}+Q^{\dagger (1)}_{\psi^2H^5}
\right)
\;, \nonumber
\end{flalign}\vspace*{-0.65cm}
\begin{flalign} 
R^{(6)}_{W^2 H^2 D^2}=&
-\frac{{g}^2}{8} \Delta_3 +\frac{g}{2}\left(
  i Q^{(1)}_{WH^4D^2} + Q^{(3)}_{WH^4D^2}\right)+\frac{g}{8}
  \left( g' Q^{(1)}_{WB H^4}+ g Q^{(1)}_{W^2 H^4}\right)\label{eq:r47}
    \\
&- \frac{1}{2}\left(Q^{(5)}_{\psi^2WH^2D} - iQ^{(7)}_{\psi^2WH^2D} \right)
-\frac{g}{8} \left( Q^{(2)}_{\psi^2 H^4 D}+Q^{(4)}_{\psi^2 H^4 D}
+i Q^{(3)}_{\psi^2 H^4 D}
\right)
+\frac{g^2}{16} \left(Q^{(1)}_{\psi^2H^5}+Q^{\dagger (1)}_{\psi^2H^5}
\right)
\;,
\nonumber
\end{flalign}\vspace*{-0.65cm}
\begin{flalign} 
R^{(7)}_{W^2 H^2 D^2}=&\frac{g}{2}\left(
  iQ^{(2)}_{WH^4D^2} - Q^{(4)}_{WH^4D^2}\right)+\frac{g}{8}\left(
  g' Q^{(2)}_{WB H^4} + g Q^{(2)}_{W^2 H^4}\right)
- \frac{1}{2}\left(Q^{(6)}_{\psi^2WH^2D} + iQ^{(8)}_{\psi^2WH^2D} \right)
\;,
\label{eq:r48}\end{flalign}\vspace*{-0.8cm}
\begin{flalign} 
R^{(8)}_{W^2 H^2 D^2}=&\frac{g}{2}\left(
  i Q^{(2)}_{WH^4D^2} + Q^{(4)}_{WH^4D^2}\right)+\frac{g}{8}\left(
  g' Q^{(2)}_{WB H^4} + g Q^{(2)}_{W^2 H^4}\right)
- \frac{1}{2}\left(Q^{(6)}_{\psi^2WH^2D} -i Q^{(8)}_{\psi^2WH^2D} \right)
\;,
\label{eq:r49}\end{flalign}\vspace*{-0.8cm}
\begin{flalign}
R^{(9)}_{W^2 H^2 D^2}=&
\frac{g^2}{4} \Delta_3
  + \frac{g}{2}\left(Q^{(2)}_{\psi^2H^4D} + Q^{(4)}_{\psi^2H^4D} \right)
  -\frac{g^2}{8} \left(Q^{(1)}_{\psi^2H^5}+Q^{\dagger (1)}_{\psi^2H^5}\right)
  + Q^{(5)}_{\psi^4H^2}  
  \;,
  \label{eq:r50}\end{flalign}\vspace*{-0.8cm}
\begin{flalign} 
R^{(10)}_{W^2 H^2 D^2}=&
i\frac{g^2}{4} \Delta_3
- i \frac{g}{2}\left(4
   i Q^{(1)}_{WH^4D^2} - Q^{(3)}_{WH^4D^2}\right) -i\frac{g}{4}\left(
   g' Q^{(1)}_{WB H^4}+ g Q^{(1)}_{W^2 H^4}\right)
   \label{eq:r51}
\\
&+ \frac{1}{2}\left(Q^{(9)}_{\psi^2WH^2D} + iQ^{(11)}_{\psi^2WH^2D} \right)
+i\frac{g}{4} \left( Q^{(2)}_{\psi^2 H^4 D}+Q^{(4)}_{\psi^2 H^4 D}
-i Q^{(3)}_{\psi^2 H^4 D}
\right)
-i\frac{g^2}{8} \left(Q^{(1)}_{\psi^2H^5}+Q^{\dagger (1)}_{\psi^2H^5}
\right)
\;,\nonumber
\end{flalign}\vspace*{-0.65cm}
\begin{flalign} 
R^{(11)}_{W^2 H^2 D^2}=&
- i \frac{g^2}{4} \Delta_3 +i\frac{g}{2}\left(4 i
  Q^{(1)}_{WH^4D^2} + Q^{(3)}_{WH^4D^2}\right)+i\frac{g}{4}\left(
  g' Q^{(1)}_{WB H^4}+ g Q^{(1)}_{W^2 H^4}\right)
  \label{eq:r52}
\\
&+ \frac{1}{2}\left(Q^{(9)}_{\psi^2WH^2D} - iQ^{(11)}_{\psi^2WH^2D} \right)
-i\frac{g}{4} \left( Q^{(2)}_{\psi^2 H^4 D}+Q^{(4)}_{\psi^2 H^4 D}
+i Q^{(3)}_{\psi^2 H^4 D}
\right)
+i\frac{g^2}{8} \left(Q^{(1)}_{\psi^2H^5}+Q^{\dagger (1)}_{\psi^2H^5}
\right)
\;,
\nonumber\end{flalign}\vspace*{-0.8cm}
\begin{flalign} 
R^{(12)}_{W^2 H^2 D^2}=& \frac{g}{2}\left(4 
  Q^{(2)}_{WH^4D^2} +i Q^{(4)}_{WH^4D^2}\right)-i\frac{g}{4}\left(
  g' Q^{(2)}_{WB H^4}+ g Q^{(2)}_{W^2 H^4}\right)
+ \frac{1}{2}\left(Q^{(10)}_{\psi^2WH^2D} + iQ^{(12)}_{\psi^2WH^2D} \right)
\;,
\label{eq:r53}\end{flalign}\vspace*{-0.8cm}
\begin{flalign} 
R^{(13)}_{W^2 H^2 D^2}=&
\frac{g}{2}\left(-4 
  Q^{(2)}_{WH^4D^2} + i Q^{(4)}_{WH^4D^2}\right)+i \frac{g}{4}\left(
  g' Q^{(2)}_{WB H^4}+ g Q^{(2)}_{W^2 H^4}\right)
+ \frac{1}{2}\left(Q^{(10)}_{\psi^2WH^2D} - iQ^{(12)}_{\psi^2WH^2D} \right)
\;,
\label{eq:r54}\end{flalign}\vspace*{-0.8cm}
\begin{flalign}
R^{(1)}_{G^2 H^2 D^2}=&\lambda v^2 G^A_{\mu\nu} G^{A,\mu\nu} (H^\dagger H) - 2 \lambda Q^{(1)}_{G^2H^4} - Q^{(1)}_{\psi^2G^2H}
\;,
\label{eq:r55}\end{flalign}\vspace*{-0.8cm}
\begin{flalign} 
R^{(2)}_{G^2 H^2 D^2}=& \lambda v^2 G^A_{\mu\nu} G^{A,\mu\nu} (H^\dagger H) - 2 \lambda Q^{(1)}_{G^2H^4} - Q^{\dagger(1)}_{\psi^2G^2H}
\;,
\label{eq:r56}\end{flalign}\vspace*{-0.8cm}
\begin{flalign} 
R^{(3)}_{G^2 H^2 D^2}=& \lambda v^2 G^A_{\mu\nu} \widetilde{G}^{A,\mu\nu} (H^\dagger H) - 2 \lambda Q^{(2)}_{G^2H^4} - Q^{(2)}_{\psi^2G^2H}
\;,
\label{eq:r57}\end{flalign}\vspace*{-0.8cm}
\begin{flalign} 
R^{(4)}_{G^2 H^2 D^2}=& \lambda v^2 G^A_{\mu\nu} \widetilde{G}^{A,\mu\nu} (H^\dagger H) - 2 \lambda Q^{(2)}_{G^2H^4} - Q^{\dagger(2)}_{\psi^2G^2H}
\;,
\label{eq:r58}\end{flalign}\vspace*{-0.8cm}
\begin{flalign}
R^{(5)}_{G^2 H^2 D^2}=& -\frac{1}{2}\left(Q^{(1)}_{\psi^2GH^2D} +\, { i }\,Q^{(3)}_{\psi^2GH^2D}\right)
\;,
\label{eq:r59}\end{flalign}\vspace*{-0.8cm}
\begin{flalign} 
R^{(6)}_{G^2 H^2 D^2}=& -\frac{1}{2}\left(Q^{(1)}_{\psi^2GH^2D} - \,{  i}\, Q^{(3)}_{\psi^2GH^2D}\right)
\;,
\label{eq:r60}\end{flalign}\vspace*{-0.8cm}
\begin{flalign} 
R^{(7)}_{G^2 H^2 D^2}=&  -\frac{1}{2}\left(Q^{(2)}_{\psi^2GH^2D} + \,{ i}\,
Q^{(4)}_{\psi^2GH^2D}\right)
\;,
\label{eq:r61}\end{flalign}\vspace*{-0.8cm}
\begin{flalign} 
R^{(8)}_{G^2 H^2 D^2}=& -\frac{1}{2}\left(Q^{(2)}_{\psi^2GH^2D} - \,{  i}\,Q^{(4)}_{\psi^2GH^2D}\right)
\;,
\label{eq:r62}\end{flalign}\vspace*{-0.8cm}
\begin{flalign} 
R^{(9)}_{G^2 H^2 D^2}=&Q^{(6)}_{\psi^4H^2}
\;,
\label{eq:r63}\end{flalign}\vspace*{-0.8cm}
\begin{flalign}
R^{(1)}_{B W H^2 D^2}=& \lambda v^2 B_{\mu\nu} W^{I,\mu\nu} (H^\dagger \tau^I H) - 2 \lambda Q^{(1)}_{WBH^4} - Q^{(1)}_{\psi^2WBH}
\;,
\label{eq:r64}\end{flalign}\vspace*{-0.8cm}
\begin{flalign} 
R^{(2)}_{B W H^2 D^2}=&\lambda v^2 B_{\mu\nu} W^{I,\mu\nu} (H^\dagger \tau^I H) - 2 \lambda Q^{(1)}_{WBH^4} - Q^{\dagger(1)}_{\psi^2WBH}
\;,
\label{eq:r65}\end{flalign}\vspace*{-0.8cm}
\begin{flalign} 
R^{(3)}_{B W H^2 D^2}=& \lambda v^2 B_{\mu\nu} \widetilde{W}^{I,\mu\nu} (H^\dagger \tau^I H) - 2 \lambda Q^{(2)}_{WBH^4} - Q^{(2)}_{\psi^2WBH}
\;,
\label{eq:r66}\end{flalign}\vspace*{-0.8cm}
\begin{flalign} 
R^{(4)}_{B W H^2 D^2}=&\lambda v^2 B_{\mu\nu} \widetilde{W}^{I,\mu\nu} (H^\dagger \tau^I H) - 2 \lambda Q^{(2)}_{WBH^4} - Q^{\dagger(2)}_{\psi^2WBH}
\;,
\label{eq:r67}\end{flalign}\vspace*{-0.8cm}
\begin{flalign}
R^{(5)}_{B W H^2 D^2}=&-\frac{g g'}{8}\left( Q_{H^6}^{(1)}-Q_{H^6}^{(2)}\right)
+\frac{g'}{2}\left(  i
  Q^{(1)}_{WH^4D^2} -   Q^{(3)}_{WH^4D^2}\right)+\frac{gg'}{16}\left(
   Q^{(1)}_{W^2 H^4}-  Q^{(3)}_{W^2 H^4}\right)\nonumber \\
&+\frac{1}{2}\left(Q^{(1)}_{\psi^2WH^2D} + iQ^{(3)}_{\psi^2WH^2D} \right)
-\frac{g'}{8}\left(Q^{(4)}_{\psi^2H^4 D}
-i Q^{(3)}_{\psi^2H^4 D}\right)
\;,
\label{eq:r68}\end{flalign}\vspace*{-0.65cm}
\begin{flalign} 
R^{(6)}_{B W H^2 D^2}=&
-\frac{g g'}{8}\left( Q_{H^6}^{(1)}-Q_{H^6}^{(2)}\right)
+\frac{g'}{2}\left(  i
  Q^{(1)}_{WH^4D^2} +   Q^{(3)}_{WH^4D^2}\right) +\frac{gg'}{16}
  \left( Q^{(1)}_{W^2 H^4}- Q^{(3)}_{W^2 H^4}\right)\nonumber \\
&+\frac{1}{2}\left(Q^{(1)}_{\psi^2WH^2D} - iQ^{(3)}_{\psi^2WH^2D} \right)
-\frac{g'}{8}\left(Q^{(4)}_{\psi^2H^4 D}
+i Q^{(3)}_{\psi^2H^4 D}\right)
\;,
\label{eq:r69}\end{flalign}\vspace*{-0.65cm}
\begin{flalign} 
R^{(7)}_{B W H^2 D^2}=&
\frac{g'}{2}\left(  i
  Q^{(2)}_{WH^4D^2} -   Q^{(4)}_{WH^4D^2}\right)+\frac{gg'}{16}
  \left( Q^{(2)}_{W^2 H^4}-  Q^{(4)}_{W^2 H^4}\right)
+\frac{1}{2}\left(Q^{(2)}_{\psi^2WH^2D} + iQ^{(4)}_{\psi^2WH^2D} \right)
\;,
\label{eq:r70}\end{flalign}\vspace*{-0.8cm}
\begin{flalign} 
R^{(8)}_{B W H^2 D^2}=&
\frac{g'}{2}\left(  i
  Q^{(2)}_{WH^4D^2} +   Q^{(4)}_{WH^4D^2}\right)+\frac{gg'}{16}
  \left( Q^{(2)}_{W^2 H^4}-  Q^{(4)}_{W^2 H^4}\right)
+\frac{1}{2}\left(Q^{(2)}_{\psi^2WH^2D} - iQ^{(4)}_{\psi^2WH^2D} \right)
\;,
\label{eq:r71}\end{flalign}\vspace*{-0.8cm}
\begin{flalign}
R^{(9)}_{B W H^2 D^2}=&
-\frac{g g'}{8}\Delta_5
+
i\frac{3 g}{2} Q^{(1)}_{BH^4D^2}+ \frac{g}{8}\left(
 g Q^{(1)}_{WBH^4} +g' Q^{(1)}_{B^2H^4}\right)
\nonumber\\
&+\frac{1}{2}\left(Q^{(5)}_{\psi^2BH^2D} + iQ^{(7)}_{\psi^2BH^2D}\right)
-\frac{g}{4} Q^{(1)}_{\psi^2H^4D} 
+\frac{gg'}{16} \left(Q^{(1)}_{\psi^2H^5}+Q^{\dagger (1)}_{\psi^2H^5}\right)
\;,
\label{eq:r72}\end{flalign}\vspace*{-0.65cm}
\begin{flalign} 
R^{(10)}_{B W H^2 D^2}=&
-\frac{g g'}{8}\Delta_5
+i \frac{ 3 g}{2} Q^{(1)}_{BH^4D^2} +
\frac{g}{8}\left(
g Q^{(1)}_{WBH^4} +g' Q^{(1)}_{B^2H^4}\right)
\nonumber\\
&+\frac{1}{2}\left(Q^{(5)}_{\psi^2BH^2D} - iQ^{(7)}_{\psi^2BH^2D}\right)
-\frac{g}{4} Q^{(1)}_{\psi^2H^4D} 
+\frac{gg'}{16} \left(Q^{(1)}_{\psi^2H^5}+Q^{\dagger (1)}_{\psi^2H^5}\right)
\;,
\label{eq:r73}\end{flalign}\vspace*{-0.8cm}
\begin{flalign} 
R^{(11)}_{B W H^2 D^2}=&
i\frac{3 g}{2}
Q^{(2)}_{BH^4D^2} +
\frac{g}{8}\left(
g Q^{(2)}_{WBH^4} +g' Q^{(2)}_{B^2H^4}\right)
+\frac{1}{2}\left(Q^{(6)}_{\psi^2BH^2D} + iQ^{(8)}_{\psi^2BH^2D}\right)
\;,
\label{eq:r74}\end{flalign}\vspace*{-0.8cm}
\begin{flalign} 
R^{(12)}_{B W H^2 D^2}=&
i\frac{3 g}{2}
  Q^{(2)}_{BH^4D^2} +
\frac{g}{8}\left(g Q^{(2)}_{WBH^4} +g' Q^{(2)}_{B^2H^4}\right)
+\frac{1}{2}\left(Q^{(6)}_{\psi^2BH^2D}  - i Q^{(8)}_{\psi^2BH^2D}\right)
\;,
\label{eq:r75}\end{flalign}\vspace*{-0.8cm}
\begin{flalign} 
R^{(13)}_{B W H^2 D^2}=&
\frac{gg'}{4}\Delta_5 -\frac{gg'}{8} \left(Q^{(1)}_{\psi^2H^5}+Q^{\dagger (1)}_{\psi^2H^5}\right)
  +\frac{g}{2} Q^{(1)}_{\psi^2H^4D} 
  +\frac{g'}{4}\left(Q^{(2)}_{\psi^2H^4D} - Q^{(4)}_{\psi^2H^4D}\right)
  + Q^{(7)}_{\psi^4H^2}
\;.
\label{eq:r76}
\end{flalign}

\noindent\underline{\bf\large Operators in class $X H^4D^2$} 
\begin{flalign}
  R^{(1)}_{B  H^4 D^2}=& i\frac{g'}{2}\Delta_5 +iQ^{(1)}_{\psi^2H^4D}
  -i\frac{g'}{4}\left(Q^{(1)}_{\psi^2 H^5}+Q^{\dagger (1)}_{\psi^2 H^5}\right)\;,
\label{eq:r77}\end{flalign}\vspace*{-0.7cm}
\begin{flalign}  
R^{(1)}_{W H^4 D^2}=&
 i\frac{g}{2}\Delta_3+i\frac{1}{2}\left(Q^{(2)}_{\psi^2H^4D} + Q^{(4)}_{\psi^2H^4D}\right)
- i\frac{g}{4}
 \left(Q^{(1)}_{\psi^2 H^5}+Q^{\dagger (1)}_{\psi^2 H^5}\right)
 \;,
 \label{eq:r78}\end{flalign}\vspace*{-0.7cm}
\begin{flalign} 
R^{(2)}_{W H^4 D^2}=& -\frac{g}{2}\Big( Q^{(1)}_{H^6} - Q^{(2)}_{H^6} \Big) - \frac{1}{2}\left(Q^{(4)}_{\psi^2H^4D} - iQ^{(3)}_{\psi^2H^4D} \right)
\;,
\label{eq:r79}\end{flalign}\vspace*{-0.7cm}
\begin{flalign} 
R^{(3)}_{W H^4 D^2}=& -\frac{g}{2}\Big( Q^{(1)}_{H^6} - Q^{(2)}_{H^6} \Big) - \frac{1}{2}\left(Q^{(4)}_{\psi^2H^4D} + iQ^{(3)}_{\psi^2H^4D} \right)
\;.
\label{eq:r80}\end{flalign}

\noindent\underline{\bf\large Operators in class $X H^2D^4$} 
\begin{flalign}
  R^{(1)}_{B H^2 D^4}=&-i\frac{g'}{2}
  \left[\lambda \Delta_5 -
    \lambda v^2 
    (D^\nu H^\dagger H)(H^\dagger \overleftrightarrow{D}_\nu H)\right]
  +\lambda v^2 ( D_\nu H^\dagger H) J^\nu_B
      \label{eq:r81} \\
&  -i\frac{g'}{4}\left(Q^{\dagger(1)}_{\psi^2H^3D^2} + Q^{\dagger(2)}_{\psi^2H^3D^2} -
  2 Q^{\dagger(3)}_{\psi^2H^3D^2}\right) -i \lambda Q^{(1)}_{\psi^2 H^4 D}
  +i\frac{g'\lambda}{4}
  \left(Q^{(1)}_{\psi^2 H^5}+Q^{\dagger(1)}_{\psi^2 H^5}\right)
    - Q^{\dagger(1)}_{\psi^4 HD}
    \;,\nonumber
\end{flalign}\vspace*{-0.65cm}
\begin{flalign}
R^{(2)}_{B H^2 D^4}=
& i\frac{g'}{2}
\left[\lambda\Delta_5 + \lambda v^2
    (H^\dagger D^\nu H)(H^\dagger \overleftrightarrow{D}_\nu H)
\right]
  +\lambda v^2 ( H^\dagger D_\nu H) J^\nu_B
\label{eq:r82}  
  \\
  &  +i\frac{g'}{4}\left(Q^{(1)}_{\psi^2H^3D^2} + Q^{(2)}_{\psi^2H^3D^2}
  - 2 Q^{(3)}_{\psi^2H^3D^2}
  \right)
  +i \lambda Q^{(1)}_{\psi^2 H^4 D}
-i\frac{g'\lambda}{4}\left(Q^{(1)}_{\psi^2 H^5}+Q^{\dagger(1)}_{\psi^2 H^5}\right)
  - Q^{(1)}_{\psi^4 HD}
  \;,
  \nonumber
\end{flalign}\vspace*{-0.65cm}
\begin{flalign}
R^{(3)}_{B H^2 D^4}=&
i\frac{g' {g}^2}{8}
\Delta_4
  + i g' \left(Q^{(1)}_{H^4}-Q^{(2)}_{H^4}\right)
-\frac{g'}{2}\left( g' Q^{(1)}_{BH^4D^2}-g Q^{(1)}_{WH^4D^2}\right)
\nonumber \\
&
- i\frac{g g'}{16}\left( g Q^{(1)}_{W^2H^4}+g Q^{(3)}_{W^2H^4}+2 g'  Q^{(1)}_{WBH^4}
  \right)
 -i \frac{1}{4} Q^{(1)}_{\psi^2 H^2 D^3}
+i \frac{1}{2}\left(g' Q^{(1)}_{\psi^2 BH^2 D} + g Q^{(1)}_{\psi^2 WH^2 D}\right)\nonumber \\&
-i \frac{g^2+{g'}^2}{4} Q^{(1)}_{\psi^2H^4 D}
+i \frac{gg'}{8} Q^{(2)}_{\psi^2H^4 D}
    -i\frac{g^2 {g'}}{ 16}  \left(Q^{(1)}_{\psi^2 H^5}+Q^{\dagger (1)}_{\psi^2 H^5}
    \right)
\nonumber
\\&
    -i\frac{1}{2}\left[g'\left(-Q^{(1)}_{\psi^4 H^2}+
    Q^{(2)}_{\psi^4 H^2}+Q^{\dagger (2)}_{\psi^4 H^2}
    -2Q^{(3)}_{\psi^4 H^2}
    +Q^{(4)}_{\psi^4 H^2}\right)+g Q^{(7)}_{\psi^4 H^2}\right]
  \;,
\label{eq:r83}\end{flalign}\vspace*{-0.65cm}
\begin{flalign}  
  R^{(1)}_{W H^2 D^4}=&
  -i\frac{g}{2}
  \left[  \lambda \Delta_3 -
\lambda v^2 
(D^\nu H^\dagger\tau^I H)(H^\dagger \overleftrightarrow{D^I}_\nu H)\right]
    +\lambda v^2 ( D_\nu H^\dagger\tau^I H) J^{I\nu}_W
    \nonumber \\
    &  -i\frac{g}{4}\left(3 Q^{\dagger(1)}_{\psi^2H^3D^2} - Q^{\dagger(2)}_{\psi^2H^3D^2} - 2 Q^{\dagger(3)}_{\psi^2H^3D^2}\right)
-i \frac{\lambda}{2}\left( Q^{(2)}_{\psi^2 H^4 D}+Q^{(4)}_{\psi^2 H^4 D}-i Q^{(3)}_{\psi^2 H^4 D}\right)\nonumber
    \\&
    +i\frac{g\lambda}{4}\left(Q^{(1)}_{\psi^2 H^5}+Q^{\dagger(1)}_{\psi^2 H^5}\right)
 - Q^{\dagger(2)}_{\psi^4 HD}
 \;,
\label{eq:r84}\end{flalign}\vspace*{-0.65cm}
\begin{flalign}
R^{(2)}_{W H^2 D^4}=&
i\frac{g}{2}
\left[ \lambda\Delta_3+
  \lambda v^2\
  (H^\dagger\tau^I D^\nu H)(H^\dagger \overleftrightarrow{D^I}_\nu H)\right]
  +\lambda v^2 ( H^\dagger\tau^I D_\nu H) J^{I\nu}_W
  \nonumber \\
  &  +i \frac{g}{4}\left(3 Q^{(1)}_{\psi^2H^3D^2} - Q^{(2)}_{\psi^2H^3D^2} - 2 Q^{(3)}_{\psi^2H^3D^2}\right)
  +i \frac{\lambda}{2}\left( Q^{(2)}_{\psi^2 H^4 D}+Q^{(4)}_{\psi^2 H^4 D}+i Q^{(3)}_{\psi^2 H^4 D}\right)
  \nonumber \\&
  - i\frac{g\lambda}{4}\left(Q^{(1)}_{\psi^2 H^5}+Q^{\dagger(1)}_{\psi^2 H^5}\right)
 - Q^{(2)}_{\psi^4 HD}
 \;,
\label{eq:r85}\end{flalign}\vspace*{-0.65cm}
\begin{flalign} 
R^{(3)}_{W H^2 D^4}=&
i\frac{g {g'}^2}{4}\Delta_5
- i g \left(Q^{(1)}_{H^4}+Q^{(2)}_{H^4}-2 Q^{(3)}_{H^4}
\right)
 -i\frac{gg'}{4}\left( g' Q^{(1)}_{B^2H^4}+g Q^{(1)}_{WBH^4}\right)
-\frac{g}{2}\left(g Q^{(1)}_{WH^4D^2}
-3 g' Q^{(1)}_{BH^4D^2}\right) \nonumber \\ &
-i \frac{1}{4} Q^{(2)}_{\psi^2 H^2 D^3}
+i \frac{1}{2}\left(g' Q^{(5)}_{\psi^2 BH^2 D} + g Q^{(5)}_{\psi^2 WH^2 D}
-2 g Q^{(11)}_{\psi^2 WH^2 D}
\right) \nonumber \\&
-i \frac{1}{8}\left((g^2+{g'}^2) Q^{(2)}_{\psi^2H^4 D}+(g^2-{g'}^2) Q^{(4)}_{\psi^2H^4 D}
- 4 g g' Q^{(1)}_{\psi^2 H^4 D}
\right)
    -i\frac{g {g'}^2}{8}  \left(Q^{(1)}_{\psi^2 H^5}+Q^{\dagger (1)}_{\psi^2 H^5}
    \right)\nonumber \\&
  -i\frac{1}{2}\left[g\left(-3 Q^{(1)}_{\psi^4 H^2}
    +Q^{(2)}_{\psi^4 H^2}+Q^{\dagger (2)}_{\psi^4 H^2}
    + 2Q^{(3)}_{\psi^4 H^2}+Q^{(5)}_{\psi^4 H^2}\right)+g' Q^{(7)}_{\psi^4 H^2}\right]
  \;.
  \label{eq:r86}
\end{flalign}

\section{Useful relations}
\label{sec:relations}

One useful tool to simplify the intermediate expressions are the Fierz
transformations~\cite{Nishi:2004st}. We made extensive use of the following
relations:
\begin{itemize}
\item $SU(2)$ identities:
  \begin{eqnarray}
    &&  { (\tau^I)_i^l\,(\tau^J)_l^j} = \delta^{IJ}
       (\delta)^j_i \,+\,i\,\epsilon^{IJK} (\tau^K)_i^j\;,
    \\
    && (\tau^I)_j^l\,   (\tau^I)_m^n = 2\delta_j^n\,\delta^l_m-\delta_j^l\,\delta _m^n
       \;,
       \label{eq:fierz2}
    \\   
&&    2 (\tau^I)_m^l \,\delta _j^n -(\tau^I)_m^n\,\delta_j^l-(\tau^I)_j^l\delta_m^n
   ={ +} i\, \epsilon^{IJK} (\tau^J)_j^l \, (\tau^K)_m^n\;,
    \\
&&    (\tau^I)_m^n \,\delta _j^l -(\tau^I)_j^l\,\delta_m^n
    ={ +} i\, \epsilon^{IJK} (\tau^J)_j^n \, (\tau^K)_m^l\; .    
  \end{eqnarray}      

\item $SU(3)$ identities:
  \begin{eqnarray}
  && (T^A)_{a}^b \, (T^B)_b^c= \frac{1}{6}\,\delta^{AB}\, \delta_a^c \,+\,\frac{1}{2}
     \left( d^{ABC}\,+\,i\, f^{ABC}\right) (T^C)_{a}^c \;,
    \\
&&  (T^A)_{a}^b \, (T^A)_c^d=
  \frac{1}{2}\, \delta_a^d\delta_c^b \,-\,\frac{1}{6}\, \delta_a^b\delta_c^d \;,\\
&&  (T^A)_a^d \, \delta_c^b \,-\,
  \frac{1}{3}\left( (T^A)_a^b\, \delta_c^d\,+\,
  (T^A)_c^d\, \delta_a^b\right)\,- \,d^{ABC}\, (T^B)_a^b\,  (T^C)_c^d\,
  =\,-\,i\, f^{ABC} \, (T^B)_a^b \, (T^C)_c^d \;.
\end{eqnarray}

\item Relations with $\gamma$ matrices:
 \begin{eqnarray}
&&\gamma^\mu\,\gamma^\nu= g^{\mu\nu} \,-\, i \,\sigma^{\mu\nu} \;,\\
   &&   \gamma^\mu\, \gamma^\alpha\gamma^\rho= g^{\mu\alpha}\,\gamma^\rho \,+\, 
   g^{\alpha\rho}\, \gamma^\mu\,-\,g^{\mu\rho}\,
   \gamma^\alpha \,-\, i\, \epsilon^{\mu\alpha\rho\sigma}\, \gamma_\sigma\gamma_5 \;,\\
&& \sigma^{\mu\nu}= -\,i\,\frac{1}{2}\,\epsilon^{\mu\nu\rho\eta}\,\sigma_{\rho\eta}\,\gamma^5 \;,\\
   &&    D^\mu (\overline{f} \gamma^\nu  M f)
    -D^\nu (\overline{f} \gamma^{ \mu}  M f)\,=\,
    i\,\left[ \overline{f}\sigma^{\mu\nu} M\, (\slash\!\!\!\! Df)\,
      -\,(\overline{ \slash\!\!\!\! Df}) \sigma^{\mu\nu} M f
    \pm \epsilon^{\mu\nu\rho\eta}\, \overline{f}\gamma_{ \rho} M \,
    \overleftrightarrow{D}_{ \eta}  f\right]\;.
 \end{eqnarray}
 where $M$ can be the identity, a Pauli matrix $\tau^I$ or a $T^A$
 matrix and the upper (lower) sign corresponds to right (left)-handed
 fermions.
 
\item Lorentz scalar fermionic Fierz identities:
  \begin{eqnarray}
  &&    (\overline{l}_a, e_b)\, (\overline{e}_c\, {l_d})
    \,=\,-\, \frac{1}{2}\,
    (\overline{l}_a\,\gamma^\mu {l_d})\, (\overline{e}_c\, \gamma_\mu\, e_b)\;,\\
    && (\overline{l}^j_a, e_b)\, (\overline{e}_c\, {l_d}_k)
    \,=\,-\, \frac{1}{4}\, \left[
      (\overline{l}_a\,\gamma^\mu {l_d})\, (\overline{e}_c\, \gamma_\mu\, e_b)\,
      \delta^j_k\,
      +\, (\overline{l}_a\,\gamma^\mu\,\tau^I {l_d})\,
      (\overline{e}_c\, \gamma_\mu \,e_b)\,(\tau^I)^{j}_k\right] \;,     
    \\
  &&(\overline{q}_a\, u_b)\, (\overline{u}_c\, {q_d})   
    =-\frac{1}{6}\,
    (\overline{q}_a\,\gamma^\mu {q_d})\, (\overline{u}_c\, \gamma_\mu\, u_b)\,
    -\;
    (\overline{q}_a\,\gamma^\mu T^A\, {q_d})\,
    (\overline{u}_c\, \gamma_\mu\;T^A\, u_b)\;,\\    
  &&(\overline{q}_{ a}^j\, u_b)\, (\overline{u}_c\, {q_d}_k)   
    =-\,\frac{\delta^j_k} {2}\,\left[\frac{1}{6}
    (\overline{q}_a\,\gamma^\mu {q_d})\, (\overline{u}_c\,\gamma_\mu  u_b)\,
    +\;
    (\overline{q}_a\,\gamma^\mu T^A\, {q_d})\, (\overline{u}_c\,
     \gamma_\mu\;T^{ A} u_b)\,\right]\;,\nonumber\\
    &&\hspace*{2.5cm}-\,\frac{(\tau)^j_k}{2}\,\left[\frac{1}{6}
    (\overline{q}_a\,\gamma^\mu\tau^I {q_d})\, (\overline{u}_c\,\gamma_\mu\,  u_b)\,
    +\;
    (\overline{q}_a\,\gamma^\mu T^A\,\tau^I\, {q_d})\,
       (\overline{u}_c\, \gamma_\mu\;T^A \, u_b)\,\right]\, \;,
  \end{eqnarray}  
  \begin{eqnarray}    
      &&  (\overline{f}_a\gamma^\mu f_a)\,   (\overline{f}_b\gamma_\mu f_b)=
    (\overline{f}_a\gamma^\mu f_b)\,   (\overline{f}_b\gamma_\mu f_a)\;,
    \hspace*{6cm}
  {\rm for}\; f=q,l,u,d,e,\\
  &&  (\bar l_a\gamma^\mu\tau^I l_a)\,   (\bar l_b\gamma_\mu\tau^I l_b)=
  2\; (\bar l_a\gamma^\mu l_b)\,   (\bar l_b\gamma_\mu l_a) \,-\,
  (\bar l_a\gamma^\mu l_a)\,   (\bar l_b\gamma_\mu l_b) \;,\\
  &&  (\overline{f}_a\gamma^\mu T^A f_a)\,   (\overline{f}_b\gamma_\mu T^A f_b)\,=
  \,-\,\frac{1}{6}\,(\overline{f}_a\gamma^\mu f_a)\,   (\overline{f}_b\gamma_\mu f_b)
\,+\,\frac{1}{2}\,(\overline{f}_a\gamma^\mu f_b)\,   (\overline{f}_b\gamma_\mu f_a)\;,
  \hspace*{1cm}
     {\rm for}\; f=q,u,d \;.
  \end{eqnarray}
  \\

\item Lorentz tensor fermionic Fierz identities: In the expressions
  below, $S^{\mu\nu}$ represent a piece which is symmetric under the
  exchange $\mu\leftrightarrow \nu$ and which will not contribute to the
  operators for which these relations are being used, so for
  simplicity we have not included their explicit forms.  They can be
  found in Ref.~\cite{Liao:2012uj}.
  \begin{eqnarray}  
  &&  (\overline{e}_a\gamma^\mu e_a)\,   (\overline{e}_b\gamma^\nu e_b)\,
    \,=\, S_e^{\mu\nu} \,{ +}\,\frac{1}{2}\, i\, \epsilon^{\mu\nu\rho\sigma}\,
    (\overline{e}_a\gamma^\rho e_b)\,   (\overline{e}_b\gamma^\sigma e_a)\;, \\  
&&  (\bar l_a\gamma^\mu  l_a)\,
(\bar  l_b\gamma^\nu l_b)\,
      \,=\, S_l^{\mu\nu}\,{ -}\,
      \,\frac{1}{2}\, i\, \epsilon^{\mu\nu\rho\sigma}\,
      (\overline{l}_a\gamma^\rho  l_b)\,   (\bar l_b\gamma^\sigma  l_a)\;,\\
 &&  (\overline{u}_a\gamma^\mu u_a)\,   (\overline{u}_b\gamma^\nu u_b)\,
    \,=\, S_u^{\mu\nu} \,{ +}\,\frac{1}{2}\, i\, \epsilon^{\mu\nu\rho\sigma}\,
    \left[\frac{1}{3}\,(\overline{u}_a\gamma^\rho u_b)\,   (\overline{u}_b\gamma^\sigma u_b)
\,+\,2\,(\overline{u}_a\gamma^\rho T^Au_b)\,   (\overline{u}_b\gamma^\sigma T^A u_a)\right]\;,\\
 &&  (\overline{q}_a\gamma^\mu q_a)\,   (\overline{q}_b\gamma^\nu q_b)\,
    \,=\, S_q^{\mu\nu} \,{ -}\,\frac{1}{2}\, i\, \epsilon^{\mu\nu\rho\sigma}\,
    \left[\frac{1}{3}\,(\overline{q}_a\gamma^\rho q_b)\,   (\overline{q}_b\gamma^\sigma q_a)
      \,+\,2\,(\overline{q}_a\gamma^\rho T^A q_b)\,
      (\overline{q}_b\gamma^\sigma T^A q_a)\right]\;,\\
&&  (\bar l_a\gamma^\mu \tau^I l_a)\,
(\bar  l_b\gamma^\nu\tau^J l_b)\,\epsilon^{IJK}
      \,=\, {S^K_l}^{\mu\nu}\!{ +}\,
      \, \epsilon^{\mu\nu\rho\sigma}\,
      (\overline{l}_a\gamma^\rho \tau^K l_b)\,   (\bar l_b\gamma^\sigma  l_b)\;,\\
 &&  (\overline{q}_a\gamma^\mu\tau^I q_a)\,   (\overline{q}_b\gamma^\nu \tau^Jq_b)\,\epsilon^{IJK}\!\!
    = {S^K_q}^{\mu\nu} \!\!{ +}\, \epsilon^{\mu\nu\rho\sigma}
    \left[\frac{1}{3}\,(\overline{q}_a\gamma^\rho \tau^Kq_b)\,   (\overline{q}_b\gamma^\sigma q_b)
      \,+\,2\,(\overline{q}_a\gamma^\rho T^A \tau^K q_b)\,   (\overline{q}_b\gamma^\sigma T^A q_b)\right]\;,\\
    &&    (\overline{u}_a\gamma^\mu T^A u_a)\,   (\overline{u}_b\gamma^\nu T^B u_b)\, f^{ABC}
    =\, {S_u^C}^{\mu\nu}\,
    { -}\,\frac{1}{2}\,\epsilon^{\mu\nu\rho\sigma}\,
(\overline{u}_a\gamma^\rho T^C u_b)\,   (\overline{u}_b\gamma^\sigma  u_a)\;,\\
&&    (\overline{q}_a\gamma^\mu T^A q_a)\,   (\overline{q}_b\gamma^\nu T^B q_b)\, f^{ABC}
    =\, {S_q^C}^{\mu\nu}\,
    { +}\,\frac{1}{2}\,\epsilon^{\mu\nu\rho\sigma}\,
(\overline{q}_a\gamma^\rho T^C q_b)\,   (\overline{q}_b\gamma^\sigma  q_a)\;.
\end{eqnarray}
\end{itemize}

\section{Relevant fermionic operators in M8B}
\label{app:ferM8B}

For convenience and reference, we here list the operators in M8B that
contain fermion and are in the rotation of the bosonic universal
operators and, therefore, appear in
Eqs.~\eqref{eq:qclass9}--\eqref{eq:qclass21}.

\begin{table}[h]
\begin{center}
\begin{adjustbox}{width=0.95\textwidth,center}
\small
\begin{minipage}[t]{5.cm}
\renewcommand{\arraystretch}{1.1}
\begin{tabular}[t]{|l|l|}
\hline  
\multicolumn{2}{|c|}{\boldmath$9:\psi^2X^2H + \hc$} \\
\hline
$Q_{leG^2H}^{(1)}$  &  $(\bar l_p e_r) H G^A_{\mu\nu} G^{A\mu\nu}$ \\
$Q_{leG^2H}^{(2)}$  &  $(
\bar l_p e_r) H \widetilde G^A_{\mu\nu} G^{A\mu\nu}$ \\
$Q_{leW^2H}^{(1)}$  &  $(\bar l_p e_r) H W^I_{\mu\nu} W^{I\mu\nu} $ \\
$Q_{leW^2H}^{(2)}$  &  $(\bar l_p e_r) H \widetilde W^I_{\mu\nu} W^{I\mu\nu}$ \\
$Q_{leW^2H}^{(3)}$  &  $\epsilon^{IJK} (\bar l_p \sigma^{\mu\nu} e_r) \tau^I H W_{\mu\rho}^J W_\nu^{K\rho}$ \\
$Q_{quG^2H}^{(1)}$  &  $(\overline{q}_p u_r) \widetilde H G^A_{\mu\nu} G^{A\mu\nu} $ \\
$Q_{quG^2H}^{(2)}$  &  $(\overline{q}_p u_r) \widetilde H \widetilde G^A_{\mu\nu} G^{A\mu\nu} $ \\
$Q_{quG^2H}^{(5)}$  &  $f^{ABC} (\overline{q}_p \sigma^{\mu\nu} T^A u_r) \widetilde H G^B_{\mu\rho} G_{\nu}^{C\rho} $ \\
$Q_{quGBH}^{(3)}$  &  $(\overline{q}_p \sigma^{\mu\nu} T^A u_r) \widetilde H G_{\mu\rho}^A B_\nu^{\,\,\,\rho}$ \\
$Q_{quW^2H}^{(1)}$  &  $(\overline{q}_p u_r) \widetilde H W^I_{\mu\nu} W^{I\mu\nu} $ \\
$Q_{quW^2H}^{(2)}$  &  $(\overline{q}_p u_r) \widetilde H \widetilde W^I_{\mu\nu} W^{I\mu\nu} $ \\
$Q_{quW^2H}^{(3)}$  &  $\epsilon^{IJK} (\overline{q}_p \sigma^{\mu\nu} u_r) \tau^I \widetilde H W_{\mu\rho}^J W_\nu^{K\rho}$ \\
$Q_{quWBH}^{(1)}$  &  $(\overline{q}_p u_r) \tau^I \widetilde H W^I_{\mu\nu} B^{\mu\nu}$ \\
$Q_{quWBH}^{(2)}$  &  $(\overline{q}_p u_r) \tau^I \widetilde H \widetilde W^I_{\mu\nu} B^{\mu\nu} $ \\
$Q_{quWBH}^{(3)}$  &  $(\overline{q}_p \sigma^{\mu\nu} u_r) \tau^I \widetilde H W_{\mu\rho}^I B_\nu^{\,\,\,\rho}$ \\
$Q_{quB^2H}^{(1)}$  &  $(\overline{q}_p u_r) \widetilde H B_{\mu\nu} B^{\mu\nu} $ \\
$Q_{quB^2H}^{(2)}$  &  $(\overline{q}_p u_r) \widetilde H \widetilde B_{\mu\nu} B^{\mu\nu}$
\\\hline
\end{tabular}
\end{minipage}
\hspace{0.5cm}
\begin{minipage}[t]{5.cm}
\renewcommand{\arraystretch}{1.1}
\begin{tabular}[t]{|l|l|}
\hline  
\multicolumn{2}{|c|}{\boldmath$9:\psi^2X^2H + \hc$} \\
\hline
$Q_{leWBH}^{(1)}$  &  $(\bar l_p e_r) \tau^I H W^I_{\mu\nu} B^{\mu\nu} $ \\
$Q_{leWBH}^{(2)}$  &  $(\bar l_p e_r) \tau^I H \widetilde W^I_{\mu\nu} B^{\mu\nu} $ \\
$Q_{leWBH}^{(3)}$  &  $(\bar l_p \sigma^{\mu\nu} e_r) \tau^I H W_{\mu\rho}^I B_\nu^{\,\,\,\rho}$ \\
$Q_{leB^2H}^{(1)}$  &  $(\bar l_p e_r) H B_{\mu\nu} B^{\mu\nu} $ \\
$Q_{leB^2H}^{(2)}$  &  $(\bar l_p e_r) H \widetilde B_{\mu\nu} B^{\mu\nu} $ \\
$Q_{qdG^2H}^{(1)}$  &  $(\overline{q}_p d_r) H G^A_{\mu\nu} G^{A\mu\nu} $ \\
$Q_{qdG^2H}^{(2)}$  &  $(\overline{q}_p d_r) H \widetilde G^A_{\mu\nu} G^{A\mu\nu} $ \\
$Q_{qdG^2H}^{(5)}$  &  $f^{ABC} (\overline{q}_p \sigma^{\mu\nu} T^A d_r) H G^B_{\mu\rho} G_{\nu}^{C\rho}$ \\
$Q_{qdGBH}^{(3)}$  &  $(\overline{q}_p \sigma^{\mu\nu} T^A d_r) H G_{\mu\rho}^A B_\nu^{\,\,\,\rho}$ \\
$Q_{qdW^2H}^{(1)}$  &  $(\overline{q}_p d_r) H W^I_{\mu\nu} W^{I\mu\nu} $ \\
$Q_{qdW^2H}^{(2)}$  &  $(\overline{q}_p d_r) H \widetilde W^I_{\mu\nu} W^{I\mu\nu} $ \\
$Q_{qdW^2H}^{(3)}$  &  $\epsilon^{IJK} (\overline{q}_p \sigma^{\mu\nu} d_r) \tau^I H W_{\mu\rho}^J W_\nu^{K\rho}$ \\
$Q_{qdWBH}^{(1)}$  &  $(\overline{q}_p d_r) \tau^I H W^I_{\mu\nu} B^{\mu\nu} $ \\
$Q_{qdWBH}^{(2)}$  &  $(\overline{q}_p d_r) \tau^I H \widetilde W^I_{\mu\nu} B^{\mu\nu} $ \\
$Q_{qdWBH}^{(3)}$  &  $(\overline{q}_p \sigma^{\mu\nu} d_r) \tau^I H W_{\mu\rho}^I B_\nu^{\,\,\,\rho}$ \\
$Q_{qdB^2H}^{(1)}$  &  $(\overline{q}_p d_r) H B_{\mu\nu} B^{\mu\nu} $ \\
$Q_{qdB^2H}^{(2)}$  &  $(\overline{q}_p d_r) H \widetilde B_{\mu\nu} B^{\mu\nu} $\\\hline
\end{tabular}
\end{minipage}
\hspace*{0.5cm}
\begin{minipage}[t]{5cm}
\renewcommand{\arraystretch}{1.1}
\begin{tabular}[t]{|l|l|}
\hline  
\multicolumn{2}{|c|}{\boldmath$11:\psi^2H^2D^3$} \\
\hline
$Q_{l^2H^2D^3}^{(1)}$  &  $i (\bar{l}_p \gamma^{\mu} D^{\nu} l_r) (D_{(\mu}D_{\nu)}H^{\dag} H)$ \\
$Q_{l^2H^2D^3}^{(2)}$  &  $i (\bar{l}_p \gamma^{\mu} D^{\nu} l_r) (H^{\dag} D_{(\mu}D_{\nu)} H)$ \\
$Q_{l^2H^2D^3}^{(3)}$  &  $i (\bar{l}_p \gamma^{\mu} \tau^I D^{\nu} l_r) (D_{(\mu}D_{\nu)}H^{\dag} \tau^I H)$ \\
$Q_{l^2H^2D^3}^{(4)}$  &  $i (\bar{l}_p \gamma^{\mu} \tau^I D^{\nu} l_r) (H^{\dag} \tau^I D_{(\mu}D_{\nu)} H)$ \\
$Q_{e^2H^2D^3}^{(1)}$  &  $i (\bar{e}_p \gamma^{\mu} D^{\nu} e_r) (D_{(\mu}D_{\nu)}H^{\dag} H)$ \\
$Q_{e^2H^2D^3}^{(2)}$  &  $i (\bar{e}_p \gamma^{\mu} D^{\nu} e_r) (H^{\dag} D_{(\mu}D_{\nu)} H)$ \\
$Q_{q^2H^2D^3}^{(1)}$  &  $i (\bar{q}_p \gamma^{\mu} D^{\nu} q_r) (D_{(\mu}D_{\nu)}H^{\dag} H)$ \\
$Q_{q^2H^2D^3}^{(2)}$  &  $i (\bar{q}_p \gamma^{\mu} D^{\nu} q_r) (H^{\dag} D_{(\mu}D_{\nu)} H)$ \\
$Q_{q^2H^2D^3}^{(3)}$  &  $i (\bar{q}_p \gamma^{\mu} \tau^I D^{\nu} q_r) (D_{(\mu}D_{\nu)}H^{\dag} \tau^I H)$ \\
$Q_{q^2H^2D^2}^{(4)}$  &  $i (\bar{q}_p \gamma^{\mu} \tau^I D^{\nu} q_r) (H^{\dag} \tau^I D_{(\mu}D_{\nu)} H)$ \\
$Q_{u^2H^2D^3}^{(1)}$  &  $i (\bar{u}_p \gamma^{\mu} D^{\nu} u_r) (D_{(\mu}D_{\nu)}H^{\dag} H)$ \\
$Q_{u^2H^2D^3}^{(2)}$  &  $i (\bar{u}_p \gamma^{\mu} D^{\nu} u_r) (H^{\dag} D_{(\mu}D_{\nu)} H)$ \\
$Q_{d^2H^2D^3}^{(1)}$  &  $i (\bar{d}_p \gamma^{\mu} D^{\nu} d_r) (D_{(\mu}D_{\nu)}H^{\dag} H)$ \\
$Q_{d^2H^2D^3}^{(2)}$  &  $i (\bar{d}_p \gamma^{\mu} D^{\nu} d_r) (H^{\dag} D_{(\mu}D_{\nu)} H)$ \\\hline
\end{tabular}
\end{minipage}
\end{adjustbox}
\end{center}
\caption{
  The dimension-eight operators in the M8B with particle content
  $\psi^2X^2H$    and $\psi^2H^2 D^3$
generated in universal theories. 
For the operators in the first two columns their  hermitian conjugates
are a priori independent operators.  For operators
$\psi^2H^2 D^3$ their hermitian conjugates  are not independent
operators. 
The subscripts $p, r$ are weak-eigenstate indices.
}
\label{tab:uniclass9-11}
\end{table}

\begin{table}[h]
\begin{center}
\begin{adjustbox}{width=0.97\textwidth,center}
\small
\begin{minipage}[t]{3.7cm}
\renewcommand{\arraystretch}{1.1}
\begin{tabular}[t]{|l|l|}
\hline  
\multicolumn{2}{|c|}{\boldmath$12:\psi^2H^5 + \hc$} \\
\hline
$Q_{leH^5}$  & $(H^\dag H)^2 (\bar l_p e_r H)$ \\
$Q_{quH^5}$  & $(H^\dag H)^2 (\overline{q}_p u_r \widetilde H )$ \\
$Q_{qdH^5}$  & $(H^\dag H)^2 (\overline{q}_p d_r H)$
\\
\hline
\hline
\multicolumn{2}{|c|}{\boldmath$17:\psi^2H^3D^2 + \hc$} \\
\hline
$Q_{leH^3D^2}^{(1)}$  & $(D_\mu H^\dag D^\mu H) (\bar l_p e_r H)$ \\
$Q_{leH^3D^2}^{(2)}$  & $(D_\mu H^\dag \tau^I D^\mu H) (\bar l_p e_r \tau^I H)$ \\
$Q_{leH^3D^2}^{(5)}$  & $(H^\dag D_\mu H) (\bar l_p e_r D^\mu H)$ \\
$Q_{quH^3D^2}^{(1)}$  & $(D_\mu H^\dag D^\mu H) (\overline{q}_p u_r \widetilde H)$ \\
$Q_{quH^3D^2}^{(2)}$  & $(D_\mu H^\dag \tau^I D^\mu H) (\overline{q}_p u_r \tau^I \widetilde H)$ \\
$Q_{quH^3D^2}^{(5)}$  & $(D_\mu H^\dag H) (\overline{q}_p u_r D^\mu \widetilde H)$ \\
$Q_{qdH^3D^2}^{(1)}$  & $(D_\mu H^\dag D^\mu H) (\overline{q}_p d_r H)$ \\
$Q_{qdH^3D^2}^{(2)}$  & $(D_\mu H^\dag \tau^I D^\mu H) (\overline{q}_p d_r \tau^I H)$ \\
$Q_{qdH^3D^2}^{(5)}$  & $(H^\dag D_\mu H) (\overline{q}_p d_r D^\mu H)$ \\\hline
\end{tabular}
\end{minipage}
\hspace{1.1cm}
\begin{minipage}[t]{9.9cm}
\renewcommand{\arraystretch}{1.1}
\begin{tabular}[t]{|l|l|}
\hline    
\multicolumn{2}{|c|}{\boldmath$13:\psi^2H^4D$} \\
\hline
$Q_{l^2H^4D}^{(1)}$  &  $ i ({ \bar{l}}_p \gamma^{\mu} l_r) (H^{\dag} \overleftrightarrow{D}_{\mu} H) (H^{\dag} H)$ \\
$Q_{l^2H^4D}^{(2)}$  &  $ i ({ \bar{l}}_p \gamma^{\mu} \tau^I l_r) [(H^{\dag} \overleftrightarrow{D}_{\mu}^I H) (H^{\dag} H) + (H^{\dag} \overleftrightarrow{D}_{\mu} H) (H^{\dag} \tau^I H)]$ \\
$Q_{l^2H^4D}^{(3)}$  &  $ i \epsilon^{IJK} ({ \bar{l}}_p \gamma^{\mu} \tau^I l_r) (H^{\dag} \overleftrightarrow{D}_{\mu}^J H) (H^{\dag} \tau^K H)$ \\
$Q_{l^2H^4D}^{(4)}$  &  $ \epsilon^{IJK} ({ \bar{l}}_p \gamma^{\mu} \tau^I l_r) (H^{\dag} \tau^J H) D_{\mu} (H^{\dag} \tau^K H)$ \\
${ Q^{(1)}_{e^2H^4D}}$  &  $ i ({ \bar{e}}_p \gamma^{\mu} e_r) (H^{\dag} \overleftrightarrow{D}_{\mu} H) (H^{\dag} H)$ \\
$Q_{q^2H^4D}^{(1)}$  &  $ i ({ \bar{q}}_p \gamma^{\mu} q_r) (H^{\dag} \overleftrightarrow{D}_{\mu} H) (H^{\dag} H)$ \\
$Q_{q^2H^4D}^{(2)}$  &  $ i ({ \bar{q}}_p \gamma^{\mu} \tau^I q_r) [(H^{\dag} \overleftrightarrow{D}_{\mu}^I H) (H^{\dag} H) + (H^{\dag} \overleftrightarrow{D}_{\mu} H) (H^{\dag} \tau^I H)]$ \\
$Q_{q^2H^4D}^{(3)}$  &  $ i \epsilon^{IJK} ({ \bar{q}}_p \gamma^{\mu} \tau^I q_r) (H^{\dag} \overleftrightarrow{D}_{\mu}^J H) (H^{\dag} \tau^K H)$ \\
$Q_{q^2H^4D}^{(4)}$  &  $ \epsilon^{IJK} ({ \bar{q}}_p \gamma^{\mu} \tau^I q_r) (H^{\dag} \tau^J H) D_{\mu} (H^{\dag} \tau^K H)$ \\
${ Q^{(1)}_{u^2H^4D}}$  &  $ i ({ \bar{u}}_p \gamma^{\mu} u_r) (H^{\dag} \overleftrightarrow{D}_{\mu} H) (H^{\dag} H)$ \\
${Q^{(1)}_{d^2H^4D}}$  &  $ i ({ \bar{d}}_p \gamma^{\mu} d_r) (H^{\dag} \overleftrightarrow{D}_{\mu} H) (H^{\dag} H)$ \\\hline
%
\end{tabular}
\end{minipage}
\end{adjustbox}
\end{center}
\caption{The dimension-eight operators in the M8B with particle
  content $\psi^2H^5$, $\psi^2H^4 D$ and $\psi^2H^3 D^2$ that are
  generated in universal theories.  For the operators in the first
  column their hermitian conjugates are a priori independent
  operators. Operators in class 13 are hermitian.
  For operators $Q^{(1)}_{f^2 H^4 D}$, where $f = u,d,e$, 
  we have added a superscript of $(1)$ to the M8B operators. The subscripts $p, r$ are
  weak-eigenstate indices.}
\label{tab:uniclass12-13-17}
\end{table}

\begin{table}[h]
\begin{center}
\begin{adjustbox}{width=0.98\textwidth,center}
\small
\begin{minipage}[t]{5.cm}
\renewcommand{\arraystretch}{1.1}
\begin{tabular}[t]{|l|l|}
\hline  
\multicolumn{2}{|c|}{\boldmath$15:(\bar R R)XH^2D$} \\
\hline
$Q_{e^2WH^2D}^{(1)}$  &  $(\bar{e}_p \gamma^\nu e_r) D^{\mu} (H^\dag \tau^I H) W_{\mu\nu}^I$ \\
$Q_{e^2WH^2D}^{(2)}$  &  $(\bar{e}_p \gamma^\nu e_r) D^{\mu} (H^\dag \tau^I H) \widetilde W_{\mu\nu}^I$ \\
$Q_{e^2WH^2D}^{(3)}$  &  $(\bar{e}_p \gamma^\nu e_r) (H^\dag \overleftrightarrow{D}^{I\mu} H) W_{\mu\nu}^I$ \\
$Q_{e^2WH^2D}^{(4)}$  &  $(\bar{e}_p \gamma^\nu e_r) (H^\dag \overleftrightarrow{D}^{I\mu} H) \widetilde W_{\mu\nu}^I$ \\
$Q_{e^2BH^2D}^{(1)}$  &  $(\bar{e}_p \gamma^\nu e_r) D^{\mu} (H^\dag H) B_{\mu\nu}$ \\
$Q_{e^2BH^2D}^{(2)}$  &  $(\bar{e}_p \gamma^\nu e_r) D^{\mu} (H^\dag H) \widetilde B_{\mu\nu}$ \\
$Q_{e^2BH^2D}^{(3)}$  &  $(\bar{e}_p \gamma^\nu e_r) (H^\dag \overleftrightarrow{D}^\mu H) B_{\mu\nu}$ \\
$Q_{e^2BH^2D}^{(4)}$  &  $(\bar{e}_p \gamma^\nu e_r) (H^\dag \overleftrightarrow{D}^\mu H) \widetilde B_{\mu\nu}$ \\
$Q_{u^2GH^2D}^{(1)}$  &  $(\bar{u}_p \gamma^\nu T^A u_r) D^{\mu} (H^\dag H) G_{\mu\nu}^A$ \\
$Q_{u^2GH^2D}^{(2)}$  &  $(\bar{u}_p \gamma^\nu T^A u_r) D^{\mu} (H^\dag H) \widetilde G_{\mu\nu}^A$ \\
$Q_{u^2GH^2D}^{(3)}$  &  $(\bar{u}_p \gamma^\nu T^A u_r) (H^\dag \overleftrightarrow{D}^\mu H) G_{\mu\nu}^A$ \\
$Q_{u^2GH^2D}^{(4)}$  &  $(\bar{u}_p \gamma^\nu T^A u_r) (H^\dag \overleftrightarrow{D}^\mu H) \widetilde G_{\mu\nu}^A$ \\
$Q_{u^2WH^2D}^{(1)}$  &  $(\bar{u}_p \gamma^\nu u_r) D^{\mu} (H^\dag \tau^I H) W_{\mu\nu}^I$ \\
$Q_{u^2WH^2D}^{(2)}$  &  $(\bar{u}_p \gamma^\nu u_r) D^{\mu} (H^\dag \tau^I H) \widetilde W_{\mu\nu}^I$ \\
$Q_{u^2WH^2D}^{(3)}$  &  $(\bar{u}_p \gamma^\nu u_r) (H^\dag \overleftrightarrow{D}^{I\mu} H) W_{\mu\nu}^I$ \\
$Q_{u^2WH^2D}^{(4)}$  &  $(\bar{u}_p \gamma^\nu u_r) (H^\dag \overleftrightarrow{D}^{I\mu} H) \widetilde W_{\mu\nu}^I$ \\
$Q_{u^2BH^2D}^{(1)}$  &  $(\bar{u}_p \gamma^\nu u_r) D^{\mu} (H^\dag H) B_{\mu\nu}$ \\
$Q_{u^2BH^2D}^{(2)}$  &  $(\bar{u}_p \gamma^\nu u_r) D^{\mu} (H^\dag H) \widetilde B_{\mu\nu}$ \\
$Q_{u^2BH^2D}^{(3)}$  &  $(\bar{u}_p \gamma^\nu u_r) (H^\dag \overleftrightarrow{D}^\mu H) B_{\mu\nu}$ \\
$Q_{u^2BH^2D}^{(4)}$  &  $(\bar{u}_p \gamma^\nu u_r) (H^\dag \overleftrightarrow{D}^\mu H) \widetilde B_{\mu\nu}$\\
$Q_{d^2GH^2D}^{(1)}$  &  $(\bar{d}_p \gamma^\nu T^A d_r) D^{\mu} (H^\dag H) G_{\mu\nu}^A$ \\
$Q_{d^2GH^2D}^{(2)}$  &  $(\bar{d}_p \gamma^\nu T^A d_r) D^{\mu} (H^\dag H) \widetilde G_{\mu\nu}^A$ \\
$Q_{d^2GH^2D}^{(3)}$  &  $(\bar{d}_p \gamma^\nu T^A d_r) (H^\dag \overleftrightarrow{D}^\mu H) G_{\mu\nu}^A$ \\
$Q_{d^2GH^2D}^{(4)}$  &  $(\bar{d}_p \gamma^\nu T^A d_r) (H^\dag \overleftrightarrow{D}^\mu H) \widetilde G_{\mu\nu}^A$ \\
$Q_{d^2WH^2D}^{(1)}$  &  $(\bar{d}_p \gamma^\nu d_r) D^{\mu} (H^\dag \tau^I H) W_{\mu\nu}^I$ \\
$Q_{d^2WH^2D}^{(2)}$  &  $(\bar{d}_p \gamma^\nu d_r) D^{\mu} (H^\dag \tau^I H) \widetilde W_{\mu\nu}^I$ \\
$Q_{d^2WH^2D}^{(3)}$  &  $(\bar{d}_p \gamma^\nu d_r) (H^\dag \overleftrightarrow{D}^{I\mu} H) W_{\mu\nu}^I$ \\
$Q_{d^2WH^2D}^{(4)}$  &  $(\bar{d}_p \gamma^\nu d_r) (H^\dag \overleftrightarrow{D}^{I\mu} H) \widetilde W_{\mu\nu}^I$ \\
$Q_{d^2BH^2D}^{(1)}$  &  $(\bar{d}_p \gamma^\nu d_r) D^{\mu} (H^\dag H) B_{\mu\nu}$ \\
$Q_{d^2BH^2D}^{(2)}$  &  $(\bar{d}_p \gamma^\nu d_r) D^{\mu} (H^\dag H) \widetilde B_{\mu\nu}$ \\
$Q_{d^2BH^2D}^{(3)}$  &  $(\bar{d}_p \gamma^\nu d_r) (H^\dag \overleftrightarrow{D}^\mu H) B_{\mu\nu}$ \\
$Q_{d^2BH^2D}^{(4)}$  &  $(\bar{d}_p \gamma^\nu d_r) (H^\dag \overleftrightarrow{D}^\mu H) \widetilde B_{\mu\nu}$ \\
%
%
%
%
%
%
\hline
\end{tabular}
\end{minipage}
\hspace*{0.5cm}
\begin{minipage}[t]{5.5cm}
\renewcommand{\arraystretch}{1.1}
\begin{tabular}[t]{|l|l|}
\hline  
\multicolumn{2}{|c|}{\boldmath$15:(\bar L L)XH^2D$} \\
\hline
$Q_{l^2WH^2D}^{(1)}$  &  $(\bar{l}_p \gamma^\nu l_r) D^{\mu} (H^\dag \tau^I H) W_{\mu\nu}^I$ \\
$Q_{l^2WH^2D}^{(2)}$  &  $(\bar{l}_p \gamma^\nu l_r) D^{\mu} (H^\dag \tau^I H) \widetilde W_{\mu\nu}^I$ \\
$Q_{l^2WH^2D}^{(3)}$  &  $(\bar{l}_p \gamma^\nu l_r) (H^\dag \overleftrightarrow{D}^{I\mu} H) W_{\mu\nu}^I$ \\
$Q_{l^2WH^2D}^{(4)}$  &  $(\bar{l}_p \gamma^\nu l_r) (H^\dag \overleftrightarrow{D}^{I\mu} H) \widetilde W_{\mu\nu}^I$ \\
$Q_{l^2WH^2D}^{(5)}$  &  $(\bar{l}_p \gamma^\nu \tau^I l_r) D^{\mu} (H^\dag H) W_{\mu\nu}^I$ \\
$Q_{l^2WH^2D}^{(6)}$  &  $(\bar{l}_p \gamma^\nu \tau^I l_r) D^{\mu} (H^\dag H) \widetilde W_{\mu\nu}^I$ \\
$Q_{l^2WH^2D}^{(7)}$  &  $(\bar{l}_p \gamma^\nu \tau^I l_r) (H^\dag \overleftrightarrow{D}^\mu H) W_{\mu\nu}^I$ \\
$Q_{l^2WH^2D}^{(8)}$  &  $(\bar{l}_p \gamma^\nu \tau^I l_r) (H^\dag \overleftrightarrow{D}^\mu H) \widetilde W_{\mu\nu}^I$ \\
$Q_{l^2WH^2D}^{(9)}$  &  $\epsilon^{IJK} (\bar{l}_p \gamma^\nu \tau^I l_r) D^{\mu} (H^\dag \tau^J H) W_{\mu\nu}^K$ \\
$Q_{l^2WH^2D}^{(10)}$  &  $\epsilon^{IJK} (\bar{l}_p \gamma^\nu \tau^I l_r) D^{\mu} (H^\dag \tau^J H) \widetilde W_{\mu\nu}^K$ \\
$Q_{l^2WH^2D}^{(11)}$  &  $\epsilon^{IJK} (\bar{l}_p \gamma^\nu \tau^I l_r) (H^\dag \overleftrightarrow{D}^{J\mu} H) W_{\mu\nu}^K$ \\
$Q_{l^2WH^2D}^{(12)}$  &  $\epsilon^{IJK} (\bar{l}_p \gamma^\nu \tau^I l_r) (H^\dag \overleftrightarrow{D}^{J\mu} H) \widetilde W_{\mu\nu}^K$ \\
$Q_{l^2BH^2D}^{(1)}$  &  $(\bar{l}_p \gamma^\nu \tau^I l_r) D^{\mu} (H^\dag \tau^I H) B_{\mu\nu}$ \\
$Q_{l^2BH^2D}^{(2)}$  &  $(\bar{l}_p \gamma^\nu \tau^I l_r) D^{\mu} (H^\dag \tau^I H) \widetilde B_{\mu\nu}$ \\
$Q_{l^2BH^2D}^{(3)}$  &  $(\bar{l}_p \gamma^\nu \tau^I l_r) (H^\dag \overleftrightarrow{D}^{I\mu} H) B_{\mu\nu}$ \\
$Q_{l^2BH^2D}^{(4)}$  &  $(\bar{l}_p \gamma^\nu \tau^I l_r) (H^\dag \overleftrightarrow{D}^{I\mu} H) \widetilde B_{\mu\nu}$ \\
$Q_{l^2BH^2D}^{(5)}$  &  $(\bar{l}_p \gamma^\nu l_r) D^{\mu} (H^\dag H) B_{\mu\nu}$ \\
$Q_{l^2BH^2D}^{(6)}$  &  $(\bar{l}_p \gamma^\nu l_r) D^{\mu} (H^\dag H) \widetilde B_{\mu\nu}$ \\
$Q_{l^2BH^2D}^{(7)}$  &  $(\bar{l}_p \gamma^\nu l_r) (H^\dag \overleftrightarrow{D}^\mu H) B_{\mu\nu}$ \\
$Q_{l^2BH^2D}^{(8)}$  &  $(\bar{l}_p \gamma^\nu l_r) (H^\dag \overleftrightarrow{D}^\mu H) \widetilde B_{\mu\nu}$ \\\hline
\end{tabular}
\end{minipage}
\hspace{0.5cm}
\begin{minipage}[t]{5.5cm}
\renewcommand{\arraystretch}{1.1}
\begin{tabular}[t]{|l|l|}
\hline
  \multicolumn{2}{|c|}{\boldmath$15:(\bar L L)XH^2D$} \\
\hline
%
%
%
%
$Q_{q^2GH^2D}^{(5)}$  &  $(\bar{q}_p \gamma^\nu T^A q_r) D^{\mu} (H^\dag H) G_{\mu\nu}^A$ \\
$Q_{q^2GH^2D}^{(6)}$  &  $(\bar{q}_p \gamma^\nu T^A q_r) D^{\mu} (H^\dag H) \widetilde G_{\mu\nu}^A$ \\
$Q_{q^2GH^2D}^{(7)}$  &  $(\bar{q}_p \gamma^\nu T^A q_r) (H^\dag \overleftrightarrow{D}^\mu H) G_{\mu\nu}^A$ \\
$Q_{q^2GH^2D}^{(8)}$  &  $(\bar{q}_p \gamma^\nu T^A q_r) (H^\dag \overleftrightarrow{D}^\mu H) \widetilde G_{\mu\nu}^A$ \\
$Q_{q^2WH^2D}^{(1)}$  &  $(\bar{q}_p \gamma^\nu q_r) D^{\mu} (H^\dag \tau^I H) W_{\mu\nu}^I$ \\
$Q_{q^2WH^2D}^{(2)}$  &  $(\bar{q}_p \gamma^\nu q_r) D^{\mu} (H^\dag \tau^I H) \widetilde W_{\mu\nu}^I$ \\
$Q_{q^2WH^2D}^{(3)}$  &  $(\bar{q}_p \gamma^\nu q_r) (H^\dag \overleftrightarrow{D}^{I\mu} H) W_{\mu\nu}^I$ \\
$Q_{q^2WH^2D}^{(4)}$  &  $(\bar{q}_p \gamma^\nu q_r) (H^\dag \overleftrightarrow{D}^{I\mu} H) \widetilde W_{\mu\nu}^I$ \\
$Q_{q^2WH^2D}^{(5)}$  &  $(\bar{q}_p \gamma^\nu \tau^I q_r) D^{\mu} (H^\dag H) W_{\mu\nu}^I$ \\
$Q_{q^2WH^2D}^{(6)}$  &  $(\bar{q}_p \gamma^\nu \tau^I q_r) D^{\mu} (H^\dag H) \widetilde W_{\mu\nu}^I$ \\
$Q_{q^2WH^2D}^{(7)}$  &  $(\bar{q}_p \gamma^\nu \tau^I q_r) (H^\dag \overleftrightarrow{D}^\mu H) W_{\mu\nu}^I$ \\
$Q_{q^2WH^2D}^{(8)}$  &  $(\bar{q}_p \gamma^\nu \tau^I q_r) (H^\dag \overleftrightarrow{D}^\mu H) \widetilde W_{\mu\nu}^I$ \\
$Q_{q^2WH^2D}^{(9)}$  &  $\epsilon^{IJK} (\bar{q}_p \gamma^\nu \tau^I q_r) D^{\mu} (H^\dag \tau^J H) W_{\mu\nu}^K$ \\
$Q_{q^2WH^2D}^{(10)}$  &  $\epsilon^{IJK} (\bar{q}_p \gamma^\nu \tau^I q_r) D^{\mu} (H^\dag \tau^J H) \widetilde W_{\mu\nu}^K$ \\
$Q_{q^2WH^2D}^{(11)}$  &  $\epsilon^{IJK} (\bar{q}_p \gamma^\nu \tau^I q_r) (H^\dag \overleftrightarrow{D}^{J\mu} H) W_{\mu\nu}^K$ \\
$Q_{q^2WH^2D}^{(12)}$  &  $\epsilon^{IJK} (\bar{q}_p \gamma^\nu \tau^I q_r) (H^\dag \overleftrightarrow{D}^{J\mu} H) \widetilde W_{\mu\nu}^K$ \\
$Q_{q^2BH^2D}^{(1)}$  &  $(\bar{q}_p \gamma^\nu \tau^I q_r) D^{\mu} (H^\dag \tau^I H) B_{\mu\nu}$ \\
$Q_{q^2BH^2D}^{(2)}$  &  $(\bar{q}_p \gamma^\nu \tau^I q_r) D^{\mu} (H^\dag \tau^I H) \widetilde B_{\mu\nu}$ \\
$Q_{q^2BH^2D}^{(3)}$  &  $(\bar{q}_p \gamma^\nu \tau^I q_r) (H^\dag \overleftrightarrow{D}^{I\mu} H) B_{\mu\nu}$ \\
$Q_{q^2BH^2D}^{(4)}$  &  $(\bar{q}_p \gamma^\nu \tau^I q_r) (H^\dag \overleftrightarrow{D}^{I\mu} H) \widetilde B_{\mu\nu}$ \\
$Q_{q^2BH^2D}^{(5)}$  &  $(\bar{q}_p \gamma^\nu q_r) D^{\mu} (H^\dag H) B_{\mu\nu}$ \\
$Q_{q^2BH^2D}^{(6)}$  &  $(\bar{q}_p \gamma^\nu q_r) D^{\mu} (H^\dag H) \widetilde B_{\mu\nu}$ \\
$Q_{q^2BH^2D}^{(7)}$  &  $(\bar{q}_p \gamma^\nu q_r) (H^\dag \overleftrightarrow{D}^\mu H) B_{\mu\nu}$ \\
$Q_{q^2BH^2D}^{(8)}$  &  $(\bar{q}_p \gamma^\nu q_r) (H^\dag \overleftrightarrow{D}^\mu H) \widetilde B_{\mu\nu}$ \\\hline
\end{tabular}
\end{minipage}
\end{adjustbox}
\end{center}
\caption{The dimension-eight operators in the M8B with particle
  content $\psi^2X H^2 D$ generated in universal theories.  All
  operators are {either hermitian or anti-hermitian}. Once again,
  the subscripts $p, r$ are weak-eigenstate indices.}
\label{tab:uniclass15}
\end{table}

\begin{table}[h]
\begin{center}
\begin{adjustbox}{width=0.98\textwidth,center}
\small
\begin{minipage}[t]{5.9cm}
\renewcommand{\arraystretch}{1.1}
\begin{tabular}[t]{|l|l|}
\hline
  \multicolumn{2}{|c|}{\boldmath$14:\psi^2X^2D$} \\
\hline
$Q_{l^2W^2D}^{(2)}$  &  $\epsilon^{IJK} (\bar l_p \gamma^\mu \tau^I \overleftrightarrow{D}^\nu l_r) W_{\mu\rho}^J W_{\nu}^{K\rho}$ \\
$Q_{l^2W^2D}^{(4)}$  &  $\epsilon^{IJK} (\bar l_p \gamma^\mu \tau^I \overleftrightarrow{D}^\nu l_r) (W_{\mu\rho}^J \widetilde{W}_{\nu}^{K\rho} + \widetilde{W}_{\mu\rho}^J W_{\nu}^{K\rho})$ \\
$Q_{q^2W^2D}^{(2)}$  &  $\epsilon^{IJK} (\overline{q}_p \gamma^\mu \tau^I \overleftrightarrow{D}^\nu q_r) W_{\mu\rho}^J W_{\nu}^{K\rho}$ \\
$Q_{q^2G^2D}^{(2)}$  &  $f^{ABC} (\overline{q}_p \gamma^\mu T^A \overleftrightarrow{D}^\nu q_r) G_{\mu\rho}^B G_{\nu}^{C\rho}$ \\
$Q_{q^2W^2D}^{(4)}$  &  $\epsilon^{IJK} (\overline{q}_p \gamma^\mu \tau^I \overleftrightarrow{D}^\nu q_r) (W_{\mu\rho}^J \widetilde{W}_{\nu}^{K\rho} + \widetilde{W}_{\mu\rho}^J W_{\nu}^{K\rho})$ \\
$Q_{q^2G^2D}^{(5)}$  &  $f^{ABC} (\overline{q}_p \gamma^\mu T^A \overleftrightarrow{D}^\nu q_r) (G_{\mu\rho}^B \widetilde{G}_{\nu}^{C\rho} + \widetilde{G}_{\mu\rho}^B G_\nu^{C \rho})$ \\
$Q_{u^2G^2D}^{(2)}$  &  $f^{ABC} (\bar u_p \gamma^\mu T^A \overleftrightarrow{D}^\nu u_r) G_{\mu\rho}^B G_{\nu}^{C\rho}$ \\
$Q_{u^2G^2D}^{(5)}$  &  $f^{ABC} (\bar u_p \gamma^\mu T^A \overleftrightarrow{D}^\nu u_r) (G_{\mu\rho}^B \widetilde{G}_{\nu}^{C\rho} + \widetilde{G}_{\mu\rho}^B G_{\nu}^{C\rho})$ \\
$Q_{d^2G^2D}^{(2)}$  &  $f^{ABC} (\bar d_p \gamma^\mu T^A \overleftrightarrow{D}^\nu d_r) G_{\mu\rho}^B G_{\nu}^{C\rho}$ \\
$Q_{d^2G^2D}^{(5)}$  &  $f^{ABC} (\bar d_p \gamma^\mu T^A \overleftrightarrow{D}^\nu d_r) (G_{\mu\rho}^B \widetilde{G}_{\nu}^{C\rho} + \widetilde{G}_{\mu\rho}^B G_{\nu}^{C\rho})$ \\\hline
\end{tabular}
\end{minipage}
\hspace{1.3cm}
\begin{minipage}[t]{7.4cm}
\renewcommand{\arraystretch}{1.1}
\begin{tabular}[t]{|l|l|}
\hline  
\multicolumn{2}{|c|}{\boldmath$14:\psi^2X^2D$} \\
\hline
$Q_{l^2WBD}^{(1)}$  &  $(\bar l_p \gamma^\mu \tau^I \overleftrightarrow{D}^\nu l_r) (B_{\mu\rho} W_{\nu}^{I\rho} - B_{\nu\rho} W_{\mu}^{I\rho})$ \\
$Q_{l^2WBD}^{(3)}$  &  $(\bar l_p \gamma^\mu \tau^I \overleftrightarrow{D}^\nu l_r) (B_{\mu\rho}\widetilde{W}_\nu^{I \rho} - B_{\nu\rho} \widetilde W_\mu^{I \rho})$ \\
$Q_{q^2WBD}^{(1)}$  &  $(\overline{q}_p \gamma^\mu \tau^I \overleftrightarrow{D}^\nu q_r) (B_{\mu\rho} W_{\nu}^{I\rho} - B_{\nu\rho} W_{\mu}^{I\rho})$ \\
$Q_{q^2WBD}^{(3)}$  &  $(\overline{q}_p \gamma^\mu \tau^I \overleftrightarrow{D}^\nu q_r) (B_{\mu\rho}\widetilde{W}_\nu^{I \rho} - B_{\nu\rho} \widetilde W_\mu^{I \rho})$ \\
$Q_{q^2GBD}^{(1)}$  &  $(\overline{q}_p \gamma^\mu T^A \overleftrightarrow{D}^\nu q_r) (B_{\mu\rho} G_{\nu}^{A\rho} - B_{\nu\rho} G_{\mu}^{A\rho})$ \\
$Q_{q^2GBD}^{(3)}$  &  $(\overline{q}_p \gamma^\mu T^A \overleftrightarrow{D}^\nu q_r) (B_{\mu\rho} \widetilde{G}_\nu^{A \rho} - B_{\nu\rho} \widetilde{G}_\mu^{A \rho})$ \\
$Q_{u^2GBD}^{(1)}$  &  $(\bar u_p \gamma^\mu T^A \overleftrightarrow{D}^\nu u_r) (B_{\mu\rho} G_{\nu}^{A\rho} - B_{\nu\rho} G_{\mu}^{A\rho})$ \\
$Q_{u^2GBD}^{(3)}$  &  $(\bar u_p \gamma^\mu T^A \overleftrightarrow{D}^\nu u_r) (B_{\mu\rho} \widetilde{G}_\nu^{A \rho} - B_{\nu\rho} \widetilde{G}_\mu^{A \rho})$ \\
$Q_{d^2GBD}^{(1)}$  &  $(\bar d_p \gamma^\mu T^A \overleftrightarrow{D}^\nu d_r) (B_{\mu\rho} G_{\nu}^{A\rho} - B_{\nu\rho} G_{\mu}^{A\rho})$ \\
$Q_{d^2GBD}^{(3)}$  &  $(\bar d_p \gamma^\mu T^A \overleftrightarrow{D}^\nu d_r) (B_{\mu\rho} \widetilde{G}_\nu^{A \rho} - B_{\nu\rho} \widetilde{G}_\mu^{A \rho})$ \\\hline
\end{tabular}
\end{minipage}
\end{adjustbox}
\end{center}
\caption{The dimension-eight operators in the M8B with particle
  content $\psi^2X^2 D$ generated in universal theories.  All
  operators are {anti-hermitian}. As before, the subscripts $p, r$
  are weak-eigenstate indices.}
\label{tab:uniclass14}
\end{table}

\begin{table}[h]
\begin{center}
\begin{adjustbox}{width=0.98\textwidth,center}
\small
\begin{minipage}[t]{5.5cm}
  \renewcommand{\arraystretch}{1.1}
  \begin{tabular}[t]{|l|l|}
\hline    
\multicolumn{2}{|c|}{\boldmath$18:(\bar L L)(\bar L L)H^2$} \\
\hline
$Q_{l^4H^2}^{(1)}$  &  $(\bar l_p \gamma^\mu l_r) (\bar l_s \gamma_\mu l_t) (H^\dag H)$ \\
$Q_{l^4H^2}^{(2)}$  &  $(\bar l_p \gamma^\mu l_r) (\bar l_s \gamma_\mu \tau^I l_t) (H^\dag \tau^I H)$ \\
$Q_{q^4H^2}^{(1)}$  & $(\overline{q}_p \gamma^\mu q_r) (\overline{q}_s \gamma_\mu q_t) (H^\dag H)$ \\
$Q_{q^4H^2}^{(2)}$  & $(\overline{q}_p \gamma^\mu q_r) (\overline{q}_s \gamma_\mu \tau^I q_t) (H^\dag \tau^I H)$ \\
$Q_{q^4H^2}^{(3)}$  & $(\overline{q}_p \gamma^\mu \tau^I q_r) (\overline{q}_s \gamma_\mu \tau^I q_t) (H^\dag H)$ \\
$Q_{l^2q^2H^2}^{(1)}$  & $(\bar l_p \gamma^\mu l_r) (\overline{q}_s \gamma_\mu q_t) (H^\dag H)$ \\
$Q_{l^2q^2H^2}^{(2)}$  & $(\bar l_p \gamma^\mu \tau^I l_r) (\overline{q}_s \gamma_\mu q_t) (H^\dag \tau^I H)$ \\
$Q_{l^2q^2H^2}^{(3)}$  & $(\bar l_p \gamma^\mu \tau^I l_r) (\overline{q}_s \gamma_\mu \tau^I q_t) (H^\dag H)$ \\
$Q_{l^2q^2H^2}^{(4)}$  & $(\bar l_p \gamma^\mu l_r) (\overline{q}_s \gamma_\mu \tau^I q_t) (H^\dag \tau^I H)$ \\\hline\hline
\multicolumn{2}{|c|}{\boldmath$18:(\bar R R)(\bar R R)H^2$} \\
\hline
$Q_{e^4H^2}^{(1)}$  &  $(\bar e_p \gamma^\mu e_r) (\bar e_s \gamma_\mu e_t) (H^\dag H)$ \\
$Q_{u^4H^2}^{(1)}$  &  $(\bar u_p \gamma^\mu u_r) (\bar u_s \gamma_\mu u_t) (H^\dag H)$ \\
$Q_{d^4H^2}^{(1)}$  &  $(\bar d_p \gamma^\mu d_r) (\bar d_s \gamma_\mu d_t) (H^\dag H)$ \\
$Q_{e^2u^2H^2}^{(1)}$  &  $(\bar e_p \gamma^\mu e_r) (\bar u_s \gamma_\mu u_t) (H^\dag H)$ \\
$Q_{e^2d^2H^2}^{(1)}$  &  $(\bar e_p \gamma^\mu e_r) (\bar d_s\gamma_\mu d_t) (H^\dag H)$ \\
$Q_{u^2d^2H^2}^{(1)}$  &  $(\bar u_p \gamma^\mu u_r) (\bar d_s \gamma_\mu d_t) (H^\dag H)$ \\
$Q_{u^2d^2H^2}^{(2)}$  &  $(\bar u_p \gamma^\mu T^A u_r) (\bar d_s \gamma_\mu T^A d_t) (H^\dag H)$\\\hline
\end{tabular}
\end{minipage}
\hspace*{0.3cm}
\begin{minipage}[t]{5.5cm}
\renewcommand{\arraystretch}{1.1}
\begin{tabular}[t]{|l|l|}
\hline  
\multicolumn{2}{|c|}{\boldmath$18:(\bar L L)(\bar R R)H^2$} \\
\hline
$Q_{l^2e^2H^2}^{(1)}$  &  $(\bar l_p \gamma^\mu l_r) (\bar e_s \gamma_\mu e_t) (H^\dag H)$ \\
$Q_{l^2e^2H^2}^{(2)}$  &  $(\bar l_p \gamma^\mu \tau^I l_r) (\bar e_s \gamma_\mu e_t) (H^\dag \tau^I H)$ \\
$Q_{l^2u^2H^2}^{(1)}$  &  $(\bar l_p \gamma^\mu l_r) (\bar u_s \gamma_\mu u_t) (H^\dag H)$ \\
$Q_{l^2u^2H^2}^{(2)}$  &  $(\bar l_p \gamma^\mu \tau^I l_r) (\bar u_s \gamma_\mu u_t) (H^\dag \tau^I H)$ \\
$Q_{l^2d^2H^2}^{(1)}$  &  $(\bar l_p \gamma^\mu l_r) (\bar d_s \gamma_\mu d_t) (H^\dag H)$ \\
$Q_{l^2d^2H^2}^{(2)}$  &  $(\bar l_p \gamma^\mu \tau^I l_r) (\bar d_s \gamma_\mu d_t) (H^\dag \tau^I H)$ \\
$Q_{q^2e^2H^2}^{(1)}$  &  $(\overline{q}_p \gamma^\mu q_r) (\bar e_s \gamma_\mu e_t) (H^\dag H)$ \\
$Q_{q^2e^2H^2}^{(2)}$  &  $(\overline{q}_p \gamma^\mu \tau^I q_r) (\bar e_s \gamma_\mu e_t) (H^\dag \tau^I H)$ \\
$Q_{q^2u^2H^2}^{(1)}$  &  $(\overline{q}_p \gamma^\mu q_r) (\bar u_s \gamma_\mu u_t) (H^\dag H)$ \\ 
$Q_{q^2u^2H^2}^{(2)}$  &  $(\overline{q}_p \gamma^\mu \tau^I q_r) (\bar u_s \gamma_\mu u_t) (H^\dag \tau^I H)$ \\ 
$Q_{q^2u^2H^2}^{(3)}$  &  $(\overline{q}_p \gamma^\mu T^A q_r) (\bar u_s \gamma_\mu T^A u_t) (H^\dag H)$ \\ 
$Q_{q^2u^2H^2}^{(4)}$  &  $(\overline{q}_p \gamma^\mu T^A \tau^I q_r) (\bar u_s \gamma_\mu T^A u_t) (H^\dag \tau^I H)$ \\ 
$Q_{q^2d^2H^2}^{(1)}$  &  $(\overline{q}_p \gamma^\mu q_r) (\bar d_s \gamma_\mu d_t) (H^\dag H)$ \\
$Q_{q^2d^2H^2}^{(2)}$  &  $(\overline{q}_p \gamma^\mu \tau^I q_r) (\bar d_s \gamma_\mu d_t) (H^\dag \tau^I H)$ \\
$Q_{q^2d^2H^2}^{(3)}$  &  $(\overline{q}_p \gamma^\mu T^A q_r) (\bar d_s \gamma_\mu T^A d_t) (H^\dag H)$ \\
$Q_{q^2d^2H^2}^{(4)}$  &  $(\overline{q}_p \gamma^\mu T^A \tau^I q_r) (\bar d_s \gamma_\mu T^A d_t) (H^\dag \tau^I H)$ \\\hline
\end{tabular}
\end{minipage}
\hspace{0.6cm}
\begin{minipage}[t]{6.2cm}
\renewcommand{\arraystretch}{1.1}
\begin{tabular}[t]{|l|l|}
\hline  
\multicolumn{2}{|c|}{\boldmath$18:(\bar L R)(\bar L R)H^2 + \hc$} \\
\hline
$Q_{q^2udH^2}^{(1)}$  &  $(\overline{q}_p^j u_r) \epsilon_{jk} (\overline{q}_s^k d_t) (H^\dag H)$ \\
$Q_{q^2udH^2}^{(2)}$  &  $(\overline{q}_p^j u_r) (\tau^I\epsilon)_{jk} (\overline{q}_s^k d_t) (H^\dag \tau^I H)$ \\
%
%
%
$Q_{lequH^2}^{(1)}$  &  $(\bar l_p^j e_r) \epsilon_{jk} (\overline{q}_s^k u_t) (H^\dag H)$ \\
$Q_{lequH^2}^{(2)}$  &  $(\bar l_p^j e_r) (\tau^I\epsilon)_{jk} (\overline{q}_s^k u_t) (H^\dag \tau^I H)$ \\
%
%
%
$Q_{l^2e^2H^2}^{(3)}$  &  $(\bar l_p e_r H) (\bar l_s e_t H)$ \\
$Q_{leqdH^2}^{(3)}$  &  $(\bar l_p e_r H) (\overline{q}_s d_t H)$ \\
%
%
$Q_{q^2u^2H^2}^{(5)}$  &  $(\overline{q}_p u_r \widetilde H) (\overline{q}_s u_t \widetilde H)$ \\
%
%
$Q_{q^2d^2H^2}^{(5)}$  &  $(\overline{q}_p d_r H) (\overline{q}_s d_t H)$ \\
\hline\hline
\multicolumn{2}{|c|}{\boldmath$18:(\bar L R)(\bar R L)H^2 + \hc$} \\
\hline
$Q_{leqdH^2}^{(1)}$  &  $(\bar l_p^j e_r) (\bar d_s q_{tj}) (H^\dag H)$ \\
$Q_{leqdH^2}^{(2)}$  &  $(\bar l_p e_r) \tau^I (\bar d_s q_t) (H^\dag \tau^I H)$ \\
%
%
$Q_{lequH^2}^{(5)}$  &  $(\bar l_p e_r H) (\widetilde H^\dag \bar u_s q_t)$ \\
$Q_{q^2udH^2}^{(5)}$  &  $(\overline{q}_p d_r H) (\widetilde H^\dag \bar u_s q_t)$ \\
\hline
%
\end{tabular}
\end{minipage}
\end{adjustbox}
\end{center}
\caption{The dimension-eight operators in the M8B with particle
  content $\psi^4 H^2$ generated in universal theories.  All operators
  are {either hermitian or anti-hermitian}. { For operators $Q^{(1)}_{f^4 H^2}$, 
  where $f = u,d,e$; and for $Q^{(1)}_{e^2u^2H^2}$ and $Q^{(1)}_{e^2d^2H^2}$ we have added 
  a superscript of $(1)$ to the M8B operators.}
  The subscripts $p, r,s,t$ are weak-eigenstate indices.}
\label{tab:uniclass18}
\end{table}

\begin{table}[h]
\begin{center}
\begin{adjustbox}{width=0.98\textwidth,center}
\small
\begin{minipage}[t]{5.6cm}
\renewcommand{\arraystretch}{1.1}
\begin{tabular}[t]{|l|l|}
\hline  
\multicolumn{2}{|c|}{\boldmath$19:(\bar L L)(\bar L L)X$} \\
\hline
$Q_{l^4W}^{(1)}$  &  $(\bar l_p \gamma^\mu l_r) (\bar l_s \gamma^\nu \tau^I l_t) W_{\mu\nu}^I$ \\
$Q_{l^4W}^{(2)}$  &  $(\bar l_p \gamma^\mu l_r) (\bar l_s \gamma^\nu \tau^I l_t) \widetilde W_{\mu\nu}^I$ \\
$Q_{q^4G}^{(1)}$  & $(\overline{q}_p \gamma^\mu q_r) (\overline{q}_s \gamma^\nu T^A q_t) G^A_{\mu\nu}$ \\
$Q_{q^4G}^{(2)}$  & $(\overline{q}_p \gamma^\mu q_r) (\overline{q}_s \gamma^\nu T^A q_t) \widetilde G^A_{\mu\nu}$ \\
$Q_{q^4G}^{(3)}$  & $(\overline{q}_p \gamma^\mu \tau^I q_r) (\overline{q}_s \gamma^\nu T^A \tau^I q_t) G^A_{\mu\nu}$ \\
$Q_{q^4G}^{(4)}$  & $(\overline{q}_p \gamma^\mu \tau^I q_r) (\overline{q}_s \gamma^\nu T^A \tau^I q_t) \widetilde G^A_{\mu\nu}$ \\
$Q_{q^4W}^{(1)}$  & $(\overline{q}_p \gamma^\mu q_r) (\overline{q}_s \gamma^\nu \tau^I q_t) W^I_{\mu\nu}$ \\
$Q_{q^4W}^{(2)}$  & $(\overline{q}_p \gamma^\mu q_r) (\overline{q}_s \gamma^\nu \tau^I q_t) \widetilde W^I_{\mu\nu}$ \\
$Q_{q^4W}^{(3)}$  & $(\overline{q}_p \gamma^\mu T^A q_r) (\overline{q}_s \gamma^\nu
T^A \tau^I q_t) W^I_{\mu\nu}$ \\
$Q_{q^4W}^{(4)}$  & $(\overline{q}_p \gamma^\mu T^A q_r) (\overline{q}_s \gamma^\nu T^A \tau^I q_t) \widetilde W^I_{\mu\nu}$ \\
$Q_{l^2q^2G}^{(1)}$  & $(\bar l_p \gamma^\mu l_r) (\overline{q}_s \gamma^\nu T^A q_t) G^A_{\mu\nu}$ \\
$Q_{l^2q^2G}^{(2)}$  & $(\bar l_p \gamma^\mu l_r) (\overline{q}_s \gamma^\nu T^A q_t) \widetilde G^A_{\mu\nu}$ \\
%
%
%
$Q_{l^2q^2W}^{(1)}$  &  $(\bar l_p \gamma^\mu l_r)(\overline{q}_s \gamma^\nu \tau^I q_t) W^I_{\mu\nu}$ \\
$Q_{l^2q^2W}^{(2)}$  &  $(\bar l_p \gamma^\mu l_r)(\overline{q}_s \gamma^\nu \tau^I q_t) \widetilde W^I_{\mu\nu}$ \\
$Q_{l^2q^2W}^{(3)}$  &  $(\bar l_p \gamma^\mu \tau^I l_r)(\overline{q}_s \gamma^\nu q_t) W^I_{\mu\nu}$ \\
$Q_{l^2q^2W}^{(4)}$  &  $(\bar l_p \gamma^\mu \tau^I l_r)(\overline{q}_s \gamma^\nu q_t) \widetilde W^I_{\mu\nu}$ \\
$Q_{l^2q^2W}^{(5)}$  &  $\epsilon^{IJK} (\bar l_p \gamma^\mu \tau^I l_r)(\overline{q}_s \gamma^\nu \tau^J q_t) W^K_{\mu\nu}$ \\
$Q_{l^2q^2W}^{(6)}$  &  $\epsilon^{IJK} (\bar l_p \gamma^\mu \tau^I l_r)(\overline{q}_s \gamma^\nu \tau^J q_t) \widetilde W^K_{\mu\nu}$ \\
\hline
\end{tabular}
\end{minipage}
\begin{minipage}[t]{6.15cm}
\renewcommand{\arraystretch}{1.1}
\begin{tabular}[t]{|l|l|}
\hline
\multicolumn{2}{|c|}{\boldmath$19:(\bar R R)(\bar R R)X$} \\
\hline
$Q_{u^4G}^{(1)}$  &  $(\bar u_p \gamma^\mu u_r) (\bar u_s \gamma^\nu T^A u_t) G^A_{\mu\nu}$ \\
$Q_{u^4G}^{(2)}$  &  $(\bar u_p \gamma^\mu u_r) (\bar u_s \gamma^\nu T^A u_t) \widetilde G^A_{\mu\nu}$ \\
$Q_{d^4G}^{(1)}$  &  $(\bar d_p \gamma^\mu d_r) (\bar d_s \gamma^\nu T^A d_t) G^A_{\mu\nu}$ \\
$Q_{d^4G}^{(2)}$  &  $(\bar d_p \gamma^\mu d_r) (\bar d_s \gamma^\nu T^A d_t) \widetilde G^A_{\mu\nu}$ \\
$Q_{e^2u^2G}^{(1)}$  &  $(\bar e_p \gamma^\mu e_r)(\bar u_s \gamma^\nu T^A u_t) G_{\mu\nu}^A$ \\
$Q_{e^2u^2G}^{(2)}$  &  $(\bar e_p \gamma^\mu e_r)(\bar u_s \gamma^\nu T^A u_t) \widetilde G_{\mu\nu}^A$ \\
%
%
%
$Q_{e^2d^2G}^{(1)}$  &  $(\bar e_p \gamma^\mu e_r)(\bar d_s \gamma^\nu T^A d_t) G_{\mu\nu}^A$ \\
$Q_{e^2d^2G}^{(2)}$  &  $(\bar e_p \gamma^\mu e_r)(\bar d_s \gamma^\nu T^A d_t) \widetilde G_{\mu\nu}^A$ \\
%
%
%
$Q_{u^2d^2G}^{(1)}$  &  $(\bar u_p \gamma^\mu u_r)(\bar d_s \gamma^\nu T^A d_t) G_{\mu\nu}^A$ \\
$Q_{u^2d^2G}^{(2)}$  &  $(\bar u_p \gamma^\mu u_r)(\bar d_s \gamma^\nu T^A d_t) \widetilde G_{\mu\nu}^A$ \\
$Q_{u^2d^2G}^{(3)}$  &  $(\bar u_p \gamma^\mu T^A u_r)(\bar d_s \gamma^\nu d_t) G_{\mu\nu}^A$ \\
$Q_{u^2d^2G}^{(4)}$  &  $(\bar u_p \gamma^\mu T^A u_r)(\bar d_s \gamma^\nu d_t) \widetilde G_{\mu\nu}^A$ \\
$Q_{u^2d^2G}^{(5)}$  &  $f^{ABC} (\bar u_p \gamma^\mu T^A u_r)(\bar d_s \gamma^\nu T^B d_t) G_{\mu\nu}^C$ \\
$Q_{u^2d^2G}^{(6)}$  &  $f^{ABC} (\bar u_p \gamma^\mu T^A u_r)(\bar d_s \gamma^\nu T^B d_t) \widetilde G_{\mu\nu}^C$ \\\hline
\hline
\multicolumn{2}{|c|}{\boldmath$19:(\bar L L)(\bar R R)X$} \\
\hline
$Q_{l^2e^2W}^{(1)}$  &  $(\bar l_p \gamma^\mu \tau^I l_r) (\bar e_s \gamma^\nu e_t) W^I_{\mu\nu}$ \\
$Q_{l^2e^2W}^{(2)}$  &  $(\bar l_p \gamma^\mu \tau^I l_r) (\bar e_s \gamma^\nu e_t) \widetilde W^I_{\mu\nu}$ \\
%
%
%
$Q_{l^2u^2G}^{(1)}$  &  $(\bar l_p \gamma^\mu l_r) (\bar u_s \gamma^\nu T^A u_t) G^A_{\mu\nu}$ \\
$Q_{l^2u^2G}^{(2)}$  &  $(\bar l_p \gamma^\mu l_r) (\bar u_s \gamma^\nu T^A u_t) \widetilde G^A_{\mu\nu}$ \\
$Q_{l^2u^2W}^{(1)}$  &  $(\bar l_p \gamma^\mu \tau^I l_r) (\bar u_s \gamma^\nu u_t) W^I_{\mu\nu}$ \\
$Q_{l^2u^2W}^{(2)}$  &  $(\bar l_p \gamma^\mu \tau^I l_r) (\bar u_s \gamma^\nu u_t) \widetilde W^I_{\mu\nu}$ \\
\hline
\end{tabular}
\end{minipage}
\begin{minipage}[t]{6cm}
\renewcommand{\arraystretch}{1.1}
\begin{tabular}[t]{|l|l|}
\hline
\multicolumn{2}{|c|}{\boldmath$19:(\bar L L)(\bar R R)X$} \\
\hline
%
%
%
$Q_{l^2d^2G}^{(1)}$  &  $(\bar l_p \gamma^\mu l_r) (\bar d_s \gamma^\nu T^A d_t) G^A_{\mu\nu}$ \\
$Q_{l^2d^2G}^{(2)}$  &  $(\bar l_p \gamma^\mu l_r) (\bar d_s \gamma^\nu T^A d_t) \widetilde G^A_{\mu\nu}$ \\
$Q_{l^2d^2W}^{(1)}$  &  $(\bar l_p \gamma^\mu \tau^I l_r) (\bar d_s \gamma^\nu d_t) W^I_{\mu\nu}$ \\
$Q_{l^2d^2W}^{(2)}$  &  $(\bar l_p \gamma^\mu \tau^I l_r) (\bar d_s \gamma^\nu d_t) \widetilde W^I_{\mu\nu}$ \\
%
%
%
$Q_{q^2e^2G}^{(1)}$  &  $(\overline{q}_p \gamma^\mu T^A q_r) (\bar e_s \gamma^\nu e_t) G^A_{\mu\nu}$ \\
$Q_{q^2e^2G}^{(2)}$  &  $(\overline{q}_p \gamma^\mu T^A q_r) (\bar e_s \gamma^\nu e_t) \widetilde G^A_{\mu\nu}$ \\
$Q_{q^2e^2W}^{(1)}$  &  $(\overline{q}_p \gamma^\mu \tau^I q_r) (\bar e_s \gamma^\nu e_t) W^I_{\mu\nu}$ \\
$Q_{q^2e^2W}^{(2)}$  &  $(\overline{q}_p \gamma^\mu \tau^I q_r) (\bar e_s \gamma^\nu e_t) \widetilde W^I_{\mu\nu}$ \\
%
%
%
$Q_{q^2u^2G}^{(1)}$  &  $(\overline{q}_p \gamma^\mu q_r) (\bar u_s \gamma^\nu T^A u_t) G^A_{\mu\nu}$ \\ 
$Q_{q^2u^2G}^{(2)}$  &  $(\overline{q}_p \gamma^\mu q_r) (\bar u_s \gamma^\nu T^A u_t) \widetilde G^A_{\mu\nu}$ \\ 
$Q_{q^2u^2G}^{(3)}$  &  $(\overline{q}_p \gamma^\mu T^A q_r) (\bar u_s \gamma^\nu u_t) G^A_{\mu\nu}$ \\ 
$Q_{q^2u^2G}^{(4)}$  &  $(\overline{q}_p \gamma^\mu T^A q_r) (\bar u_s \gamma^\nu u_t) \widetilde G^A_{\mu\nu}$ \\
$Q_{q^2u^2G}^{(5)}$  &  $f^{ABC} (\overline{q}_p \gamma^\mu T^A q_r) (\bar u_s \gamma^\nu T^B u_t) G^C_{\mu\nu}$ \\ 
$Q_{q^2u^2G}^{(6)}$  &  $f^{ABC} (\overline{q}_p \gamma^\mu T^A q_r) (\bar u_s \gamma^\nu T^B u_t) \widetilde G^C_{\mu\nu}$ \\ 
%
%
%
$Q_{q^2u^2W}^{(1)}$  &  $(\overline{q}_p \gamma^\mu \tau^I q_r) (\bar u_s \gamma^\nu u_t) W^I_{\mu\nu}$ \\ 
$Q_{q^2u^2W}^{(2)}$  &  $(\overline{q}_p \gamma^\mu \tau^I q_r) (\bar u_s \gamma^\nu u_t) \widetilde W^I_{\mu\nu}$ \\ 
%
%
%
%
%
%
%
$Q_{q^2d^2G}^{(1)}$  &  $(\overline{q}_p \gamma^\mu q_r) (\bar d_s \gamma^\nu T^A d_t) G^A_{\mu\nu}$ \\ 
$Q_{q^2d^2G}^{(2)}$  &  $(\overline{q}_p \gamma^\mu q_r) (\bar d_s \gamma^\nu T^A d_t) \widetilde G^A_{\mu\nu}$ \\ 
$Q_{q^2d^2G}^{(3)}$  &  $(\overline{q}_p \gamma^\mu T^A q_r) (\bar d_s \gamma^\nu d_t) G^A_{\mu\nu}$ \\ 
$Q_{q^2d^2G}^{(4)}$  &  $(\overline{q}_p \gamma^\mu T^A q_r) (\bar d_s \gamma^\nu d_t) \widetilde G^A_{\mu\nu}$ \\ 
$Q_{q^2d^2G}^{(5)}$  &  $f^{ABC} (\overline{q}_p \gamma^\mu T^A q_r) (\bar d_s \gamma^\nu T^B d_t) G^C_{\mu\nu}$ \\ 
$Q_{q^2d^2G}^{(6)}$  &  $f^{ABC} (\overline{q}_p \gamma^\mu T^A q_r) (\bar d_s \gamma^\nu T^B d_t) \widetilde G^C_{\mu\nu}$ \\ 
%
%
%
$Q_{q^2d^2W}^{(1)}$  &  $(\overline{q}_p \gamma^\mu \tau^I q_r) (\bar d_s \gamma^\nu d_t) W^I_{\mu\nu}$ \\ 
$Q_{q^2d^2W}^{(2)}$  &  $(\overline{q}_p \gamma^\mu \tau^I q_r) (\bar d_s \gamma^\nu d_t) \widetilde W^I_{\mu\nu}$ \\ 
\hline
\end{tabular}
\end{minipage}
\end{adjustbox}
\end{center}
\caption{The dimension-eight operators in the M8B with particle
  content $\psi^4 X$ generated in universal theories.  All operators
  are {either hermitian or anti-hermitian}.  The subscripts
  $p, r,s,t$ are weak-eigenstate indices.}
\label{tab:uniclass19}
\end{table}

\begin{table}[h]
\begin{center}
\begin{adjustbox}{width=\textwidth,center}
\small
\begin{minipage}[t]{5.15cm}
\renewcommand{\arraystretch}{1.1}
\begin{tabular}[t]{|l|l|}
\hline  
\multicolumn{2}{|c|}{\boldmath$20:\psi^4HD + \hc$} \\
\hline
$Q_{l^3eHD}^{(1)}$  &  $i (\bar l_p \gamma^\mu l_r) [(\bar l_s e_t) D_\mu H]$ \\
$Q_{l^3eHD}^{(2)}$  &  $i (\bar l_p \gamma^\mu \tau^I l_r) [(\bar l_s e_t) \tau^I D_\mu H]$ \\
%
%
$Q_{le^3HD}^{(1)}$  &  $i (\bar e_p \gamma^\mu e_r) [(\bar l_s D_\mu e_t) H]$ \\
$Q_{leq^2HD}^{(1)}$  &  $i (\overline{q}_p \gamma^\mu q_r) [(\bar l_s e_t) D_\mu H]$ \\
%
%
$Q_{leq^2HD}^{(3)}$  &  $i (\overline{q}_p \gamma^\mu \tau^I q_r) [(\bar l_s e_t) \tau^I D_\mu H]$ \\
$Q_{leu^2HD}^{(1)}$  &  $i (\bar u_p \gamma^\mu u_r) [(\bar l_s e_t) D_\mu H]$ \\
$Q_{led^2HD}^{(1)}$  &  $i (\bar d_p \gamma^\mu d_r) [(\bar l_s e_t) D_\mu H]$ \\
\hline
\end{tabular}
\end{minipage}
\begin{minipage}[t]{5.1cm}
\renewcommand{\arraystretch}{1.1}
\begin{tabular}[t]{|l|l|}
\hline  
\multicolumn{2}{|c|}{\boldmath$20:\psi^4HD + \hc$} \\
\hline
$Q_{l^2quHD}^{(1)}$  &  $i (\bar l_p \gamma^\mu l_r) [(\overline{q}_s u_t) D_\mu \widetilde H]$ \\
$Q_{l^2quHD}^{(3)}$  &  $i (\bar l_p \gamma^\mu \tau^I l_r) [(\overline{q}_s u_t) \tau^I D_\mu \widetilde H]$ \\
$Q_{e^2quHD}^{(1)}$  &  $i (\bar e_p \gamma^\mu e_r) [(\overline{q}_s u_t) D_\mu \widetilde H]$ \\
$Q_{q^3uHD}^{(1)}$  &  $i (\overline{q}_p \gamma^\mu q_r) [(\overline{q}_s u_t) D_\mu \widetilde H]$ \\
$Q_{q^3uHD}^{(2)}$  &  $i (\overline{q}_p \gamma^\mu \tau^I q_r) [(\overline{q}_s u_t) \tau^I D_\mu \widetilde H]$ \\
$Q_{qu^3HD}^{(1)}$  &  $i (\bar u_p \gamma^\mu u_r) [(\overline{q}_s u_t) D_\mu \widetilde H]$ \\
$Q_{qud^2HD}^{(1)}$  &  $i (\bar d_p \gamma^\mu d_r) [(\overline{q}_s u_t) D_\mu \widetilde H]$ \\\hline
\end{tabular}
\end{minipage}
\hspace{0.1cm}
\begin{minipage}[t]{5cm}
  \renewcommand{\arraystretch}{1.1}
\begin{tabular}[t]{|l|l|}
\hline  
  \multicolumn{2}{|c|}{\boldmath$20:\psi^4HD + \hc$} \\
\hline
$Q_{l^2qdHD}^{(1)}$  &  $i (\bar l_p \gamma^\mu l_r) [(\overline{q}_s d_t) D_\mu H]$ \\
$Q_{l^2qdHD}^{(3)}$  &  $i (\bar l_p \gamma^\mu \tau^I l_r) [(\overline{q}_s d_t) \tau^I D_\mu H]$ \\
$Q_{e^2qdHD}^{(1)}$  &  $i (\bar e_p \gamma^\mu e_r) [(\overline{q}_s d_t) D_\mu H]$ \\
$Q_{q^3dHD}^{(1)}$  &  $i (\overline{q}_p \gamma^\mu q_r) [(\overline{q}_s d_t) D_\mu H]$ \\
$Q_{q^3dHD}^{(2)}$  &  $i (\overline{q}_p \gamma^\mu \tau^I q_r) [(\overline{q}_s d_t) \tau^I D_\mu H]$ \\
$Q_{qu^2dHD}^{(1)}$  &  $i (\bar u_p \gamma^\mu u_r) [(\overline{q}_s d_t) D_\mu H]$ \\
$Q_{qd^3HD}^{(1)}$  &  $i (\bar d_p \gamma^\mu d_r) [(\overline{q}_s d_t) D_\mu H]$ \\\hline
\end{tabular}
\end{minipage}
\end{adjustbox}
\end{center}
\caption{The dimension-eight operators in the M8B with particle
  content $\psi^4 HD$ generated in universal theories.  For all
  operators their hermitian conjugates are a priori independent
  operators.  The subscripts $p, r,s,t$ are weak-eigenstate indices.}
\label{tab:uniclass20}
\end{table}

\begin{table}[h]
\begin{center}
\begin{adjustbox}{width=\textwidth,center}
\small
\begin{minipage}[t]{5.2cm}
\renewcommand{\arraystretch}{1.1}
\begin{tabular}[t]{|l|l|}
\hline  
\multicolumn{2}{|c|}{\boldmath$21:(\bar L L)(\bar L L)D^2$} \\
\hline
$Q_{l^4D^2}^{(1)}$  &  $D^\nu (\bar l_p \gamma^\mu l_r) D_\nu (\bar l_s \gamma_\mu l_t)$  \\
%
%
$Q_{q^4D^2}^{(1)}$  &  $D^\nu (\overline{q}_p \gamma^\mu q_r) D_\nu (\overline{q}_s \gamma_\mu q_t)$  \\
%
%
$Q_{q^4D^2}^{(3)}$  &  $D^\nu (\overline{q}_p \gamma^\mu \tau^I q_r) D_\nu (\overline{q}_s \gamma_\mu \tau^I q_t)$  \\
%
%
$Q_{l^2q^2D^2}^{(1)}$  &  $D^\nu (\bar l_p \gamma^\mu l_r) D_\nu (\overline{q}_s \gamma_\mu q_t)$  \\
$Q_{l^2q^2D^2}^{(3)}$  &  $D^\nu (\bar l_p \gamma^\mu \tau^I l_r) D_\nu (\overline{q}_s \gamma_\mu \tau^I q_t)$  
\\\hline\hline
\multicolumn{2}{|c|}{\boldmath$21:(\bar L R)(\bar L R)D^2 + \hc$} \\
\hline
$Q_{q^2udD^2}^{(1)}$  &  $D_\mu (\overline{q}_p^j u_r) \epsilon_{jk} D^\mu (\overline{q}_s^k d_t)$  \\
%
%
%
$Q_{lequD^2}^{(1)}$  &  $D_\mu (\bar l_p^j e_r) \epsilon_{jk} D^\mu (\overline{q}_s^k u_t)$ \\
\hline
\end{tabular}
\end{minipage}
\hspace*{0.1cm}
\begin{minipage}[t]{5.3cm}
\renewcommand{\arraystretch}{1.1}
\begin{tabular}[t]{|l|l|}
\hline  
  \multicolumn{2}{|c|}{\boldmath$21:(\bar R R)(\bar R R)D^2$} \\
\hline
$Q_{e^4D^2}^{ (1)}$  &  $D^\nu (\bar e_p \gamma^\mu e_r) D_\nu (\bar e_s \gamma_\mu e_t)$  \\
$Q_{u^4D^2}^{(1)}$  &  $D^\nu (\bar u_p \gamma^\mu u_r) D_\nu (\bar u_s \gamma_\mu u_t)$  \\
%
%
$Q_{d^4D^2}^{(1)}$  &  $D^\nu (\bar d_p \gamma^\mu d_r) D_\nu (\bar d_s \gamma_\mu d_t)$  \\
%
%
$Q_{e^2u^2D^2}^{(1)}$  &  $D^\nu (\bar e_p \gamma^\mu e_r) D_\nu (\bar u_s \gamma_\mu u_t)$  \\
%
%
$Q_{e^2d^2D^2}^{(1)}$  &  $D^\nu (\bar e_p \gamma^\mu e_r) D_\nu (\bar d_s \gamma_\mu d_t)$  \\
%
%
$Q_{u^2d^2D^2}^{(1)}$  &  $D^\nu (\bar u_p \gamma^\mu u_r) D_\nu (\bar d_s \gamma_\mu d_t)$  \\
%
%
$Q_{u^2d^2D^2}^{(3)}$  &  $D^\nu (\bar u_p \gamma^\mu T^A u_r) D_\nu (\bar d_s \gamma_\mu T^A d_t)$  \\
%
%
\hline
\hline
\multicolumn{2}{|c|}{\boldmath$21:(\bar L R)(\bar R L)D^2 + \hc$} \\
\hline
$Q_{leqdD^2}^{(1)}$  &  $D_\mu (\overline{l}_p^j e_r) \epsilon_{jk} D^\mu (\overline{d}_s^k q_t)$  \\\hline
\end{tabular}
\end{minipage}
\hspace*{0.2cm}
\begin{minipage}[t]{6cm}
\renewcommand{\arraystretch}{1.1}
\begin{tabular}[t]{|l|l|}
\hline  
\multicolumn{2}{|c|}{\boldmath$21:(\bar L L)(\bar R R)D^2$} \\
\hline
$Q_{l^2e^2D^2}^{(1)}$  &  $D^\nu (\bar l_p \gamma^\mu l_r) D_\nu (\bar e_s \gamma_\mu e_t)$  \\
%
%
$Q_{l^2u^2D^2}^{(1)}$  &  $D^\nu (\bar l_p \gamma^\mu l_r) D_\nu (\bar u_s \gamma_\mu u_t)$  \\
%
%
$Q_{l^2d^2D^2}^{(1)}$  &  $D^\nu (\bar l_p \gamma^\mu l_r) D_\nu (\bar d_s \gamma_\mu d_t)$  \\
%
%
$Q_{q^2e^2D^2}^{(1)}$  &  $D^\nu (\overline{q}_p \gamma^\mu q_r) D_\nu (\bar e_s \gamma_\mu e_t)$  \\
%
%
$Q_{q^2u^2D^2}^{(1)}$  &  $D^\nu (\overline{q}_p \gamma^\mu q_r) D_\nu (\bar u_s \gamma_\mu u_t)$  \\
%
%
$Q_{q^2u^2D^2}^{(3)}$  &  $D^\nu (\overline{q}_p \gamma^\mu T^A q_r) D_\nu (\bar u_s \gamma_\mu T^A u_t)$  \\
%
%
$Q_{q^2d^2D^2}^{(1)}$  &  $D^\nu (\overline{q}_p \gamma^\mu q_r) D_\nu (\bar d_s \gamma_\mu d_t)$  \\
%
%
$Q_{q^2d^2D^2}^{(3)}$  &  $D^\nu (\overline{q}_p \gamma^\mu T^A q_r) D_\nu (\bar d_s \gamma_\mu T^A d_t)$  \\
\hline
\end{tabular}
\end{minipage}
\end{adjustbox}
\end{center}
\caption{The dimension-eight operators in the M8B with particle
  content $\psi^4 HD$ generated in universal theories.  All operators
  are {either hermitian or anti-hermitian}. {For the operator $Q^{(1)}_{e^4 D^2}$, 
  we have added a superscript of $(1)$ to the M8B operator.}
  The subscripts $p, r,s,t$ are weak-eigenstate indices.}
\label{tab:uniclass21}
\end{table}
\clearpage
\bibliography{references}

\end{document}